\documentclass[smallextended]{svjour3}       
\smartqed  
\usepackage{graphicx}
\usepackage{amsmath}
\usepackage{xspace}
\usepackage{txfonts}
\usepackage{txfonts}
\usepackage{array}
\usepackage{color}
\usepackage{url}
\usepackage{enumitem}
\usepackage{bm}
\usepackage{xspace}
\usepackage{siunitx}
\usepackage[dvipsnames]{xcolor}
\usepackage[colorlinks = true,
            linkcolor = blue,
            urlcolor  = blue,
            citecolor = blue,
            anchorcolor = blue]{hyperref}
\usepackage{threeparttablex} 
\usepackage{longtable}
\usepackage{aas_macros}
\usepackage{siunitx}
\usepackage[left=2.5cm, right=2.5cm, top=2.2cm, bottom=2.3cm]{geometry}

\newcommand{\Msun}{M$_{\odot}$}

\newcommand{\sweet}{SWEET-Cat}
\newcommand{\angstrom}{\textup{\AA}}
\newcommand{\teff}{T$_\mathrm{eff}$}
\newcommand{\logg}{$\log(g)$}
\newcommand{\logRHK}{$\log(R^{\prime}_{\rm HK})$~}
\newcommand{\feh}{[Fe/H]}
\newcommand{\abratio}[2]{\ensuremath{[\mathrm{#1}/\mathrm{#2}]}\xspace}

\def\gtsima{$\; \buildrel > \over \sim \;$}
\def\ltsima{$\; \buildrel < \over \sim \;$}
\def\gtrsim{\lower.5ex\hbox{\gtsima}}
\def\lesssim{\lower.5ex\hbox{\ltsima}}

\journalname{Experimental Astronomy}
\begin{document}

\title{The homogeneous characterisation of Ariel host stars
}


\author{Camilla Danielski \and
        Anna Brucalassi \and
        Serena Benatti \and 
        Tiago Campante \and 
        Elisa Delgado-Mena \and
        Monica Rainer \and 
        Germano Sacco \and 
        Vardan Adibekyan \and
        Katia Biazzo \and
        Diego Bossini  \and
        Giovanni Bruno  \and
        Giada Casali  \and
        Petr Kabath  \and
        Laura Magrini  \and
        Giusi Micela  \and
        Giuseppe Morello  \and
        Pietro Palladino \and
        Nicoletta Sanna \and
        Subhajit Sarkar  \and
        S\'ergio Sousa  \and
        Maria Tsantaki  \and
        Diego Turrini \and
        Mathieu Van der Swaelmen
}

\authorrunning{Danielski et al.} 

\institute{Danielski C.\at
              UCL Centre for Space Exochemistry Data,
              Atlas Building,  Fermi Avenue, Harwell Campus,  Didcot,  OX11 0QR (UK)\\
             \emph{Present address:} Instituto de Astrof\'isica de Andaluc\'ia (IAA-CSIC), 
            Glorieta de la Astronom\'ia s/n, 18008 Granada, Spain  \\
              \email{cdanielski@iaa.es} 
           \and
           Brucalassi A., Casali G., Magrini L., Rainer M., Sacco G., 
           Sanna N.,
           Tsantaki M., Van der Swaelmen M. \at 
           INAF - Osservatorio Astrofisico di Arcetri, Largo E. Fermi 5, 50125 Firenze, Italy
           \and
           Casali G., \at
           Dipartimento di Fisica e Astronomia Augusto Righi, Universit\'a degli Studi di Bologna, 
           Via Gobetti 93/2, I-40129 Bologna, Italy
           \and
           Benatti S., Micela G. \at 
           INAF - Osservatorio Astronomico di Palermo,
           Piazza del Parlamento 1, 90134 Palermo, Italy
           \and
    	   Campante T., Delgado-Mena E., Adibekyan V., Bossini D., Sousa S. \at 
           Instituto de Astrof\'{\i}sica e Ci\^{e}ncias do Espa\c{c}o, Universidade do Porto, CAUP, Rua das Estrelas, 4150-762 Porto, Portugal
           \and
           Campante T., Adibekyan V. \at 
           Departamento de F\'{\i}sica e Astronomia, Faculdade de Ci\^{e}ncias da Universidade do Porto, Rua do Campo Alegre, s/n, 4169-007 Porto, Portugal
           \and
           Biazzo K. \at 
           INAF - Osservatorio Astronomico di Roma
           Via di Frascati, 33, I-00044, Monte Porzio Catone, Italy
           \and
           Bruno G.\at
           INAF - Osservatorio Astrofisico di Catania, 
           Via S. Sofia 78, 95123 Catania, Italy
           \and
           Kabath P. \at 
           Astronomical Institute, Czech Academy of Science, \\ Fri\u{c}ova 298
           251 65 Ond\u{r}ejov \u{C}eská republika
           \and
           Morello G.\at
           AIM, CEA, CNRS, Universit\'e Paris-Saclay, Universit\'e Paris Diderot, Sorbonne Paris Cit\'e, F-91191 Gif-sur-Yvette, France
           \and
           Palladino P. \at 
           Universita' degli Studi di Palermo, Dipartimento di Fisica e Chimica Emilio Segr\'{e}, Viale delle Scienze, Ed. 17, 90128 Palermo, Italy
           \and
           Sarkar S.\at
           School of Physics and Astronomy, Cardiff University, The Parade, Cardiff, CF24 3AA, UK
           \and
           Turrini D.\at
           Institute of Space Astrophysics and Planetology INAF-IAPS, Via Fosso del Cavaliere 100, Rome, Italy\\
           INAF - Osservatorio Astrofisico di Torino, Via Osservatorio 20, Pino Torinese I-10025, Italy
}

\date{Received: date / Accepted: date}

\maketitle

\begin{abstract}
The Ariel mission will characterise the chemical and thermal properties
of the atmospheres of about a thousand exoplanets transiting their host star(s). 
The observation of such a large sample of planets will allow to deepen our understanding of planetary and atmospheric formation at the early stages, 
providing a truly representative picture of the chemical nature of exoplanets, and 
relating this directly to the type and chemical environment of the host star.
Hence, the accurate and precise determination of the host star fundamental properties is essential to Ariel for drawing a comprehensive picture of the underlying essence of these planetary systems.
We present here a structured approach for the characterisation of Ariel stars that accounts for the concepts of homogeneity and coherence among a large set of stellar parameters. We present here the studies and benchmark analyses we have been performing to determine robust stellar fundamental parameters, elemental abundances, activity indices, and stellar ages. 
In particular, we present results for the homogeneous estimation of the activity indices $S$ and \logRHK, and preliminary results for elemental abundances of Na, Al, Mg, Si, C, N.
In addition, we analyse the variation of a planetary spectrum, obtained with Ariel, as a function of the uncertainty on the stellar effective temperature. Finally, we present our observational campaign for precisely and homogeneously characterising all Ariel stars in order to perform a meaningful choice of final targets before the mission launch.

\keywords{Stars: abundances \and Stars: activity  \and Stars: fundamental parameters  \and Stars: general  \and planetary systems  \and Planets and satellites: atmospheres}

\end{abstract}

\section{Introduction}
\label{sec:intro}
The Ariel M4 space mission \cite{Tinetti2018}, recently adopted by ESA whose launch is foreseen for 2029, has the objective to observe the atmospheres of $\sim$ 1000 transiting exoplanets. By studying the chemical composition and thermal properties of a large and diversified sample of planets, Ariel will allow us to answer key questions such as {\it ``What are exoplanets made of?''}, {\it``How do planets and planetary systems form?''}, and {\it ``How do planets and their atmospheres evolve over time?''}. Most importantly, it will 
finally allow us to place the Solar System into a broader Galactic context. 
The diversity in atmospheric compositions observed today is expected to be linked to different planetary formation and evolutionary scenarios, and our Solar System is only one example among a vast range of possible outcomes. 
To achieve its goals, Ariel will therefore need to cover a relevant range of both planetary and stellar parameters in a statistically significant way.
The relation between a planet and its star is perpetually interweaved, so one cannot be dissected without accounting for the other.
It is therefore important to acquire an accurate knowledge of the stellar properties well before the Ariel launch, in order to explore the range of host-star parameters in its Reference Sample \cite{Edwards2019}, prioritise the observations, and finally perform a meaningful choice of Ariel targets.
In fact, the Reference Sample will evolve with time based on both the analysis of its targets' properties, and new planetary discoveries, to always comply the mission requirements and maximise its science outcome.

The characterisation of planet-hosting star(s) is an important step for understanding the nature of their planetary companion(s). On a general level, in relation to the transiting planetary population, it is salient for the following reasons: 

    \paragraph{The determination of the planetary radius, mass, and bulk composition:} the precise determination of stellar radii is crucial to retrieve an absolute value of the radius of a transiting planet; likewise, through synergies with the radial velocity method, the precision of the planetary mass would depend on the stellar one. In turn, the inference of both stellar radius and mass usually hinges on the stellar atmospheric parameters
    through the use of stellar models. 
    It is consequently important to adopt a coherent modelling approach with well calibrated models for determining the stellar mass and radius.
    The atmospheric parameters i.e. effective temperature (\teff), surface gravity (\logg), and metallicity (\feh) can be accurately determined through high-resolution spectroscopy. 
    
    \paragraph{The exact identification of atmospheric features:} stellar variability, caused by the interaction between magnetic fields and turbulent plasma, is responsible for intrinsic variations of the stellar spectrum that can be confused with planetary features (e.g. \cite{Robertson2015,Ballerini2012}) when occurring on transit (or phase-curve) timescales. It is therefore essential to acquire an exact knowledge of the stellar spectrum to at least the same level of the planetary signal, i.e. 10 -- 50 ppm relative to the star, for accurately determining the chemical composition and molecular abundances in a planetary atmosphere. 
    Furthermore, information on long-term stellar activity is important for studying the effect of the stellar irradiation onto the planet atmosphere and for constraining the evolution of the atmosphere itself e.g. \cite{Locci2019,Johnstone2020}
    
    \paragraph{The investigation of planetary formation and migration processes:} 
    determining the precise age of the host star and, by extension, of the planetary system is critical as it tells us how much time was available for the system to dynamically evolve. This, in turn, allows us to constrain the pathways that produced its current architecture. Different dynamical pathways (e.g. early or late migration) act on different timescales and have different implications for the composition of planets (e.g. \cite{turrini+2015,Turrini2020} and references therein). Knowing the age of the system provides the temporal dimension of the problem.
    In parallel, the atmospheric composition of the host star is itself a cipher key to decode the compositional signatures on planets left by their formation and migration histories (e.g. \cite{oberg+2011,johnson+2012,carter-bond+2012,thiabaud+2014,marboeuf+2014a,marboeuf+2014b,mordasini+2016,cridland+2019} and \cite{Turrini2018,Turrini2020} for a discussion in the context of Ariel). 
    Stellar abundances, in fact, provide the meter on which to measure whether planets are enriched or deprived of a given element. Beginning with the early results of \cite{oberg+2011}, various works point out that the comparison between the stellar metallicity and C/O ratio, with the analogous planetary quantities (e.g. the planetary C/O ratio being super-stellar or sub-stellar) provides information on the original formation region of the planet with respect to the H$_2$O, CO$_2$ and CO snow lines, and on the time when the planet migrated to its present orbit (e.g. \cite{Madhusudhan2016}, and references therein, and \cite{mordasini+2016,cridland+2019,Turrini2020,Turriniinprep} for recent examples). 
    Further work shows that similar considerations apply also to elemental ratios involving other elements (e.g. N, S, Ti, Al), and that their comparison with the stellar abundance ratios provides additional constraints on the planetary formation process \cite{Turrini2018,oberg+2019,bosman+2019,Turrini2020,Turriniinprep}. 
    Therefore, the precise knowledge of the host-star age, abundances, and elemental ratios, is the founding stone for interpreting the planetary composition, and the origin and evolution of planets. 

    \paragraph{The identification of correlations between planetary and stellar properties:}
    more and more observational evidence is being reported presenting correlations between the host-star properties and their companion properties. To cite some, we today know  correlations between planet radius versus host-star metallicity (e.g., \cite{Buchhave2014,Schlaufman2015}); eccentricity versus metallicity (e.g., \cite{Adibekyan2013,Wilson2018}); occurrence rate of gas-giant planets versus stellar metallicity (e.g. \cite{Santos2004,ValentiFischer2005,Sousa2008}); correlation between hot-Jupiters surface gravity and stellar chromospheric fluxes \cite{Hartman2010,Figueira2014,Lanza2014,Fossati2017}; low-mass planets occurrence rate versus stellar Mg/Si ratio \cite{Adibekyan2015}; or the fact that iron-poor planet-hosting stars are enhanced in $\alpha$-elements \cite{Adibekyan2012,Adibekyan2012b}.
    Such correlations are of great value in constraining both the physics of the planet-star lifetime relation, and formation/evolution processes.  To perform this kind of statistical studies it is though essential to work with samples of objects whose parameters have been derived in a uniform and  precise way (e.g. \cite{SWEET-Cat,Biazzo2015,Andreasen2017,Sousa2018}). \\
    
For Ariel the concept of {\it homogeneity} is crucial: fundamental parameters of host stars are usually found in the literature as the result of a case-by-case analysis performed by different teams, resulting in an inhomogeneous census of stars with planets. In order to achieve a comprehensive characterisation of a statistically significant number of planetary systems, as in Ariel goals, the accurate, precise and uniform determination of the fundamental properties of host stars is pivotal. 
We present here the characterisation study we have been performing in preparation of the Ariel mission to determine the properties of the known host stars of the Ariel Tier 1 Reference Sample \cite{Edwards2019} with a complete homogeneous approach.
We planned and tested different types of techniques (both model and empirical based) with the final goal to have a catalogue of stellar properties whose measurements are homogeneous, robust and precise.\\
The core of our analysis is based on atmospheric parameters retrieved from the \sweet\footnote{\url{https://www.astro.up.pt/resources/sweet-cat/}} catalogue (Stars With ExoplanETs Catalogue, \cite{SWEET-Cat}). At the time of writing \sweet\ is a catalogue of parameters for 2872 stars with planets, 642 of which have been analysed using the same uniform methodology. We note that the analyses we performed throughout this manuscript have been focusing on a total of 155 Ariel stars that are in common with the homogeneously determined sample of \sweet.

The manuscript is structured as follows: 
Section \ref{sec:parameters} presents the analysis performed to robustly determine the stellar atmospheric parameters such as \teff, \logg\ and \feh\ for the whole Ariel parameter space. 
Section \ref{sec:precisionparams} presents a study on the accuracy of the transit depth value as a function of the stellar fundamental parameters' uncertainties and wavelength range.
Section \ref{sec:abundances} describes the techniques used to estimate the elemental abundances of aluminium, carbon, nitrogen, magnesium, sodium, and silicon. 
Section \ref{sec:indexes} describes various techniques used to estimate useful activity indices (i.e. $S$ and \logRHK) from high-resolution spectra.
Section \ref{sec:ages} presents the inconsistency of the age values presented on a case-by-case study in the literature and our approach for the determination of the age of the Ariel stars. 
Finally, Section \ref{sec:campaign} presents the ground-based monitoring campaign to obtain high-resolution spectra for those stars in the Ariel Tier 1 that have not yet been observed with the spectral quality needed to perform a precise characterisation.
We present our conclusions in Sec. \ref{sec:conclusions}.

\section{Homogeneous stellar parameters}
\label{sec:parameters}

The atmospheric parameters \teff, \logg, and \feh\ are the usual input data for the determination of other stellar properties such as age, mass, activity indices and elemental abundances. Consequently, in order to provide a final set of homogeneous parameters which are precise, consistent and robust, we performed a benchmark study between three different methods for measuring such parameters.  The goal of this study is to develop an overall robust and coherent method that covers the whole Ariel stellar parameter space, and which will be used to homogeneously determine atmospheric parameters of all Ariel host stars.

\subsection{Data}
We analysed a total of $\sim$150 high-resolution, high S/N spectra of stars common to both the Ariel Tier 1 and the homogeneous sample of \sweet. The range of V magnitude spans values between $5 < V < 16$ mag,  with spectra covering a signal-to-noise ratio (S/N) between ~50 $<$ S/N $< \sim$ 800.

\subsection{Methods and results}
We used three different spectroscopic techniques to determine the atmospheric stellar parameters for the list of targets mentioned above.
The three methods employed for our analysis are the ``Fast Automatic MOOG Analysis'' (FAMA, \cite{Magrini13}),  the ``Fast Analysis of Spectra Made Automatically'' (FASMA, \cite{Tsantaki2018}), and the method used for determining \sweet\ parameters \cite{SWEET-Cat}. 
We performed a comparison between the values of \teff, \logg, and \feh~ obtained from each method, seeking possible outliers and trends among the parameters, and we evaluated the robustness of the results by comparing it to external benchmarks,
e.g. isochrones grids, gravities derived from both trigonometric distances (Gaia DR2), and photometric lightcurves.

Our results show a good agreement for a bulk population around the solar values (\teff= 5000 - 6000 K, \logg=4.2 - 4.6 dex), yet at low and high temperatures some methods tend to under/overestimate the surface gravity \logg. 
specifically, 
within the 5000 K - 6000 K temperature range we find no trends, however an offset is present between FAMA/FASMA and \sweet~ (+70 K and +34 K, respectively). For \teff$<$ 5000 K, the dispersion between the three methods increases, while for \teff$>$ 6000 K both FAMA and FASMA yield  temperatures cooler than \sweet~ ones.\\
Concerning the surface gravity determination, we detect trends in the results provided by both FAMA and FASMA. Specifically, the difference between their \logg~ and the \sweet~ estimate increases  with \sweet~ \logg. In the range between 4.2 - 4.6 dex the trend is almost negligible, however for \logg $>$ 4.6 the differences become particularly high. \\
In relation to the metallicity \feh, and considering its typical errors $\sim$ 0.10 dex, the three methods overall converge to similar final metallicities.
We note that the three methods rely on different line lists with different set of
atomic data, which cause them to converge on slightly different sets of \teff~ and
\logg~ values, which should satisfy both conditions of excitation equilibrium, and
ionization balance. In fact, while the value of \feh~ is usually more constant, \teff~ and \logg~ tend to vary along a local minimum.
Details specific to each method and complete set of comparison results (including all individual stellar parameters derived by FAMA and FASMA) are reported in a dedicated study by \cite{Brucalassi2020}. 

Further studies will be carried out within the parameters' range where the three spectroscopic methods do not provide consistent results. Finally, through the help of external benchmarks, we will select the method(s) that yield(s) the most robust result, depending upon the parameters space (\teff, \logg~ and \feh) each individual Ariel star falls in. The combination of the selected method(s) will be used to coherently build a homogeneous catalogue of stellar atmospheric parameters for the whole Ariel Reference Sample.

\section{Study of the effect of stellar parameters' uncertainties through ExoSim }
\label{sec:precisionparams}

In parallel to the estimation of homogeneous atmospheric parameters, we started analysing how their uncertainties can affect the determination of the retrieved planetary spectrum. 
More specifically, the uncertainties in \teff, \logg\ and \feh\ directly reflect onto the determination of the stellar limb-darkening, a phenomenon that has to be taken into account when analysing data of planets observed during their primary transit i.e. when they pass in front of the star.

\subsection{Stellar limb-darkening}
Stellar limb-darkening is the wavelength-dependent radial decrease in specific intensity as observed from the centre to the edge of the stellar disc. Intensity is traditionally represented as a function $I_{\lambda}(\mu)$, where $\lambda$ denotes the observing wavelength or passband, $\mu$ is defined as $\mu = \cos{\theta}$, and $\theta$ is the angle between the surface normal and the line of sight. Numerous functional forms, so-called limb-darkening laws,  have been proposed in the literature to approximate $I_{\lambda}(\mu)$. In this study we adopt the following four-coefficient law proposed by \cite{claret00}, (hereinafter ``claret-4''):
\begin{equation}
\label{eqn:ld_law_claret4}
\frac{I_{\lambda}(\mu)}{I_{\lambda}(1)} = 1 - \sum_{k=1}^{4} a_k(1-\mu^{k/2}) .
\end{equation}
Previous studies indicate that the claret-4 law should be used to guarantee the precision level required for the Ariel spectra \cite{hayek2012,morello17,changeat2020}.

\begin{figure*}[b]
\centering
\includegraphics[trim = 1.2cm 0cm 2cm 0.8cm, clip, width=.49\linewidth]{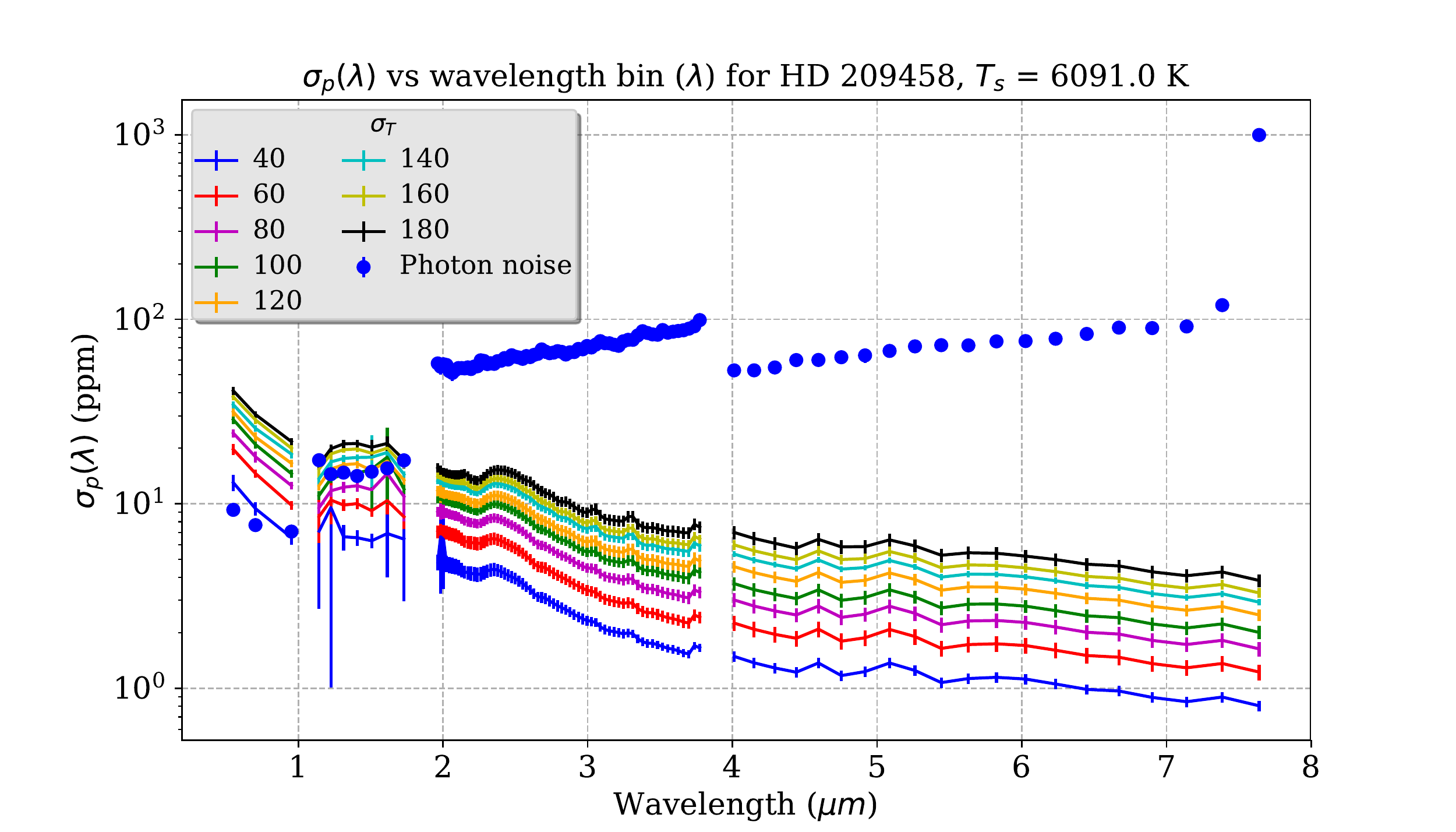}
\includegraphics[trim = 1.2cm 0cm 2cm 0.8cm, clip,width=.49\linewidth]{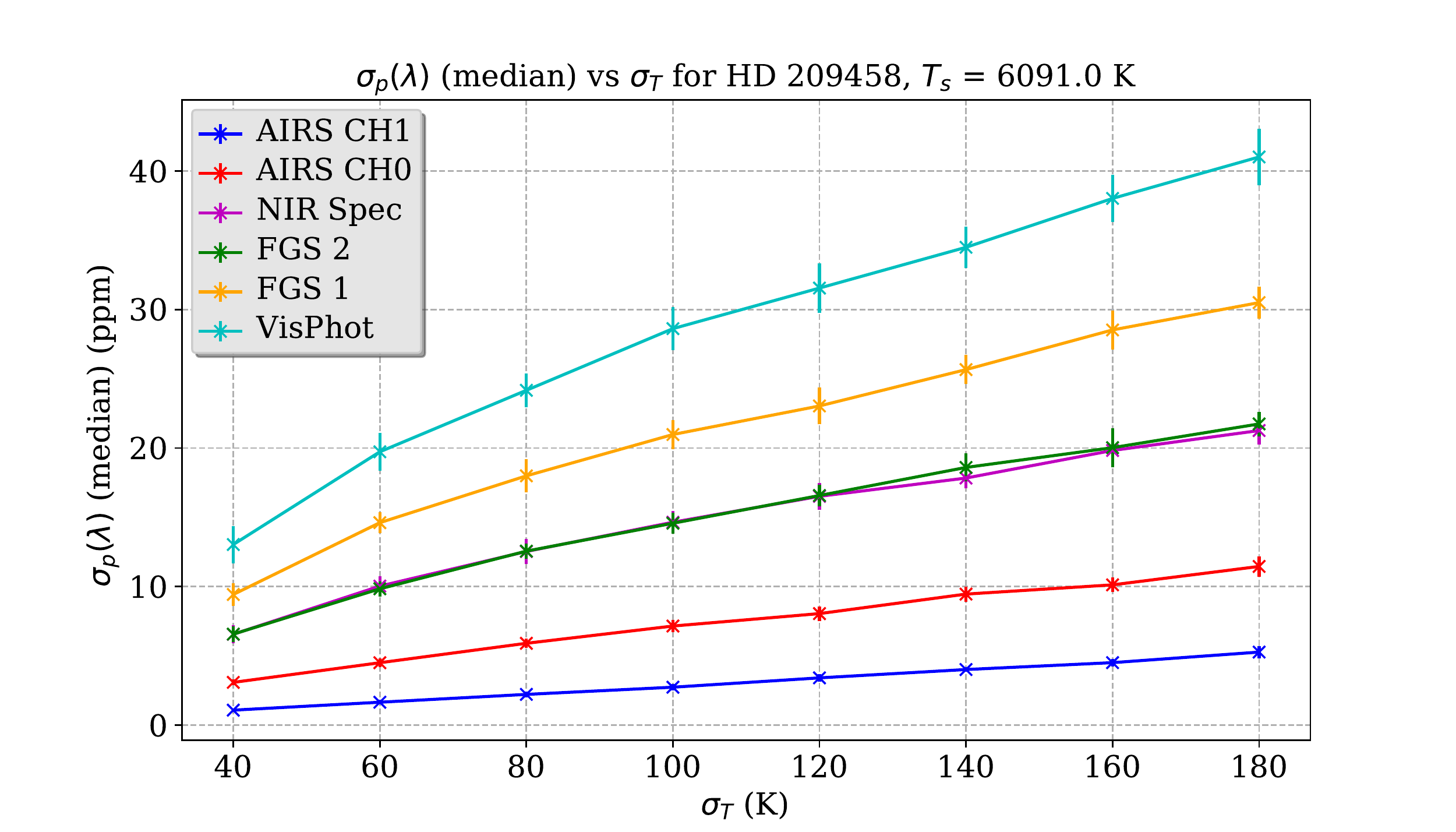}
\caption{{\it Left panel}: Simulations of HD-209458  (V = 7.63 mag, K = 6.31 mag) showing the standard deviation of the recovered transit depths, $\sigma_p(\lambda)$, versus wavelength, for different $\sigma_T$ values. The crosses give the mean value of $\sigma_p(\lambda)$ over 20 results (each result is obtained from a distribution of 100 transit depths), and the error bars give the standard deviation of these values.  Also shown for comparison is the fractional noise on the transit depth (for 1 transit observation) from stellar photon noise for HD-209458 b (assuming no temperature variation): dots give the mean and error bars give the standard deviation of 20 results, each consisting of 100 realisations.
 {\it Right panel}: 
 $\sigma_p(\lambda)$ versus $\sigma_T$ for the median wavelength bin in each of the three Ariel spectroscopic channels, and the three photometric channels.}
\label{fig:limb_darkening}
\end{figure*}

To investigate the uncertainty in the transit depth that results from the precision of stellar parameters, we used ExoSim \cite{Sarkar2020} to simulate the exoplanet transit observations of HD-209458 b with all the Ariel instrument channels.  The simulations generate wavelength-dependent light curves which applied in the generation of synthetic data. The input transit depth ($p =(R_p/R_*)^2$) was assumed to be constant at all wavelengths.  

The limb-darkening coefficients were calculated through the \texttt{ExoTETHyS}\footnote{\url{https://github.com/ucl-exoplanets/ExoTETHyS/}} package \cite{morello2020joss,morello2020}, based on the \texttt{PHOENIX}\_2012\_13 grid of stellar atmosphere models (\cite{claret12}, \cite{claret13}). With this configuration the limb-darkening coefficients depend on the stellar parameters \teff, \logg, and \feh.
The limb-darkening coefficients used for the synthetic data were fixed to those obtained for the HD-209458 model with \teff$ = 6091 \, K$, \logg$= 4.45$ dex, and \feh$=0$.

This data is then processed to extract spectral light curves in wavelength bins with spectral resolution of $R=100$ (AIRS CH0), $R = 30$ (AIRS CH1), and $R=15$ (NIRSpec).  No additional noise was implemented in these simulations.  The final step of this process involves the fitting of model light curves to the simulated data curves, to extract the transit depths, $p(\lambda)$.  

The light curves used for the fit however, are generated for stars with random temperature values, obtained from a Gaussian distribution centered at \teff$ = 6091 \, K$, and of standard deviation, $\sigma_T$, which represents the uncertainty in \teff. All other stellar parameters are kept identical to the simulated star.  The curve fitting process used a downhill simplex algorithm to find the best-fitting curve with the transit depth being the only free parameter.  The curve fit mismatch occurs between the simulated data and the model light curves, due to the use of the slightly wrong limb-darkening coefficients, causing a bias in the recovered fractional transit depth. This procedure is repeated 100 times in a Monte Carlo simulation, to obtain a distribution of transit depths, with a standard deviation, $\sigma_p(\lambda)$, an uncertainty associated with the precision of \teff, $\sigma_T$.  We note that dependence of the limb-darkening profile on the \logg\ and \feh\ is weaker than that on the \teff.  The whole process is repeated 20 times to obtain a mean and standard deviation for $\sigma_p(\lambda)$.

Figure \ref{fig:limb_darkening} (left) shows $\sigma_p(\lambda)$ as a function of wavelength for different values of $\sigma_T$. This term should be added in quadrature with the other sources of error in order to obtain the final error bars of the transit depth.  Figure \ref{fig:limb_darkening} (right) shows the median spectral bin results per spectroscopic channel, and photometric channel results plotted against $\sigma_T$.
We note that by determining the temperature of a solar-like star with a precision of 80 K we can have a potential uncertainty of $\sim$ 25 ppm in the visible channel (VisPhot).
The mid-infrared channels are not significantly affected by this source of error, given that the standard deviation is below 10 ppm at wavelengths longer than 3 $\mu$m even if $\sigma_T =$180 K.

To compare these results to the photon noise limit, we performed an ExoSim Monte Carlo simulation where HD-209458 b was simulated using 100 realisations with just stellar photon noise. The visible magnitude of the host star is V = 7.63 mag. Light curve fits were performed as previously, but no temperature variations were simulated.  A distribution of transit depths is again obtained for each spectral bin. The standard deviation of this distribution gives the fractional photon noise on the transit depth (for 1 transit) per bin.  As before we repeat this process 20 times, and obtain a mean and standard deviation of the photon noise result.  These results are shown in Figure \ref{fig:limb_darkening} (left).  Again we find that the mid-infrared channels are not significantly affected by the stellar \teff\ precision, with $\sigma_p(\lambda)$ falling $\sim$ an order of magnitude below the photon noise. For the FGS channels however the fractional photon noise is much smaller and falls below the uncertainty from stellar precision.  The absolute uncertainties from stellar precision remain overall low in these channels, but this result indicates that the precision on \teff~ must be considered in the overall transit depth uncertainty at these shorter wavelengths. We note that fainter stars have larger photon noise error, so that the added contribution due to the uncertainty on \teff~ could become negligible over the whole Ariel spectral range.

 An analogous study for other stellar types, which includes noise, uncertainties in \logg, \feh, and the possible dependence on the model of stellar atmosphere (PHOENIX, ATLAS, MARCS, STAGGER), will be presented in a dedicated upcoming paper.

\section{Homogeneous stellar abundances}
\label{sec:abundances}

When it comes to transiting exoplanets, what we are observing is a close-in population that 
did not form {\it in situ} 
but migrated inwards to their present position from different formation regions and at different times in the life of their respective systems. Constraining the starting position of the planetary seeds that will either form giant planets or super-Earths is a challenging task,
but it can be confronted by comparing the chemical composition of the planet with that of its host star \cite{oberg+2011,Madhusudhan2014,Madhusudhan2016}.
Stellar abundances reflect the chemical composition of the initial circumstellar disc of gas and dust, hence of the birth environment from which transiting planets and their atmospheres formed as the results of a combination of formation, migration, disc/atmospheric evolution and possible enrichment processes (e.g., \cite{turrini+2015,Madhusudhan2016}, and references therein, and \cite{mordasini+2016,cridland+2019,ShibataIkoma2019,Turrini2020,Turriniinprep}).
Therefore, knowing the stellar composition of planet hosts and comparing it to stars without detected planets is essential to understand planet formation mechanisms \cite{Adibekyan2012b,DelgadoMena2010,petigura2011,santos2015,Jofre2015,daSilva2015,suarezandres2017,DelgadoMena2018}

In the context of Ariel science we have selected here those elements which are visible in the Ariel spectral range, and of interests for constraining both planetary formation and stellar interior: the refractory  Na, Mg, Al, Si (Sec. \ref{sec:salMgsi}),  and the volatiles C, N (Sec. \ref{sec:CN}). We present here different methods to determine abundances from homogeneous atmospheric parameters, our preliminary results and the need to improve such results.

\subsection{Data}

We selected a sample of stars with \teff\,$>$\,4500 K in common between Ariel Tier 1 and the homogeneous sample of \sweet~ \cite{SWEET-Cat}, for which we already have good quality high resolution spectra obtained from different spectrographs: HARPS, UVES, SOPHIE, FEROS, EsPaDOns, FIES, NARVAL, and SARG. 
The minimum S/N of the spectra is 60 with a median value of 165. The sample is composed of a total of 155 stars whose spectroscopic parameters (\teff, \logg, \feh, and  microturbulent velocity $v_{micro}$) were derived using the same spectra we utilise here for deriving the abundances. We note here that the results presented in this work are preliminary and obtained with published stellar parameters in \sweet. Once the final set of parameters is defined (see Sec. \ref{sec:parameters}) we will present a more detailed work on chemical abundances in an upcoming paper.

\subsection{Na, Mg, Al, Si abundances determination}
\label{sec:salMgsi}

In order to obtain robust results we estimated and compared the abundances of the elements Al, Mg, Na and Si retrieved through two different approaches.
The first approach employed ARES to measure the equivalent widths (EWs) of the selected lines for each element. The number and choice of lines with their atomic parameters can be found in \cite{Adibekyan2012}. The radiative transfer code to derive the abundances is MOOG 2014 \cite{sneden2014}, assuming local thermodynamic equilibrium (LTE) and ATLAS stellar model atmospheres. 
The second approach used the DOOp code, based on DAOSPEC \cite{stetson},  to measure the EWs \cite{CantatGaudin2014} and the FAMA code \cite{Magrini13}, based on MOOG 2014, to derive the stellar abundances (which can determine also the atmospheric parameters, as reported in Sec. \ref{sec:parameters} and fully described in  \cite{Brucalassi2020}). It adopted MARCS stellar model atmospheres and the Gaia-ESO clean line-list \cite{heiter15}, from which we selected only lines with the best oscillator strengths ($\log~gf$).

We performed a differential analysis with respect to the Sun by measuring stellar abundances also in a Vesta solar spectrum obtained with HARPS at very high S/N ratio (retrieved from the ESO archive). Therefore, we provide the [X/H] ratios, defined as the difference of absolute logarithmic abundances between a star and the Sun:\\

$[\mathrm{X}/\mathrm{H}] = A (\mathrm{X}) - A(\mathrm{X})_{\odot} $

~\\
where $A(\mathrm{X})$ is defined as : \\

$A(\mathrm{X}) = \log[N({\rm X})/N({\rm H})]+12$. \\

\vspace{0.1cm}

For the purpose of evaluating the general behaviour of elements in the context of the Galactic chemical evolution we make use of the [X/Fe] ratios defined as:\\

$[\mathrm{X}/\mathrm{Fe}] = [\mathrm{X}/\mathrm{H}] - [\mathrm{Fe}/\mathrm{H}] $

\subsubsection*{Results:}
Among 155 targets some of the spectra had a good enough quality (S/N $\lesssim$ 70) for the derivation of stellar parameters but not good enough to derive chemical abundances so neither of the methods could analyse such stars. Moreover some stars were excluded by one or the other technique as no reliable abundance could be derived for a given element. We compared the results obtained for the sample of stars whose abundances were derived by both methods.
After comparing the [X/H] we considered as preliminary abundances the average value of both methods for those elements where the differences in [X/H] ratios between both methods was below two times the median absolute deviation (MAD). Except for the case of Si (for which we applied a cut of 3 times the MAD) the differences are small and within the individual abundance errors. Moreover there are no systematic differences caused by stellar parameters. We note that for the range of stellar parameters in our sample the Non-local thermodynamic equilibrium (NLTE) corrections are quite low and below the error level for Si and Mg (e.g. \cite{Osorio2015}). For example, \cite{Adibekyan2017} found NLTE corrections below 0.03 dex for Mg and below 0.01 dex for Si by using the database from MPIA (Max Planck Institute for Astronomy) (e.g. \cite{Bergemann2013}) to correct the abundances in the HARPS-GTO sample. In a similar way the abundances of Al for solar type stars are not significantly affected by 3D-NLTE effects (below 0.03 dex, \cite{Nordlander2017}. On the other hand, the abundances of Na are more severely affected by NLTE effects although they can be balanced out by considering differential abundances with respect to the Sun \cite{Amarsi2020}.

\begin{figure*}
\centering
\includegraphics[width=1\linewidth]{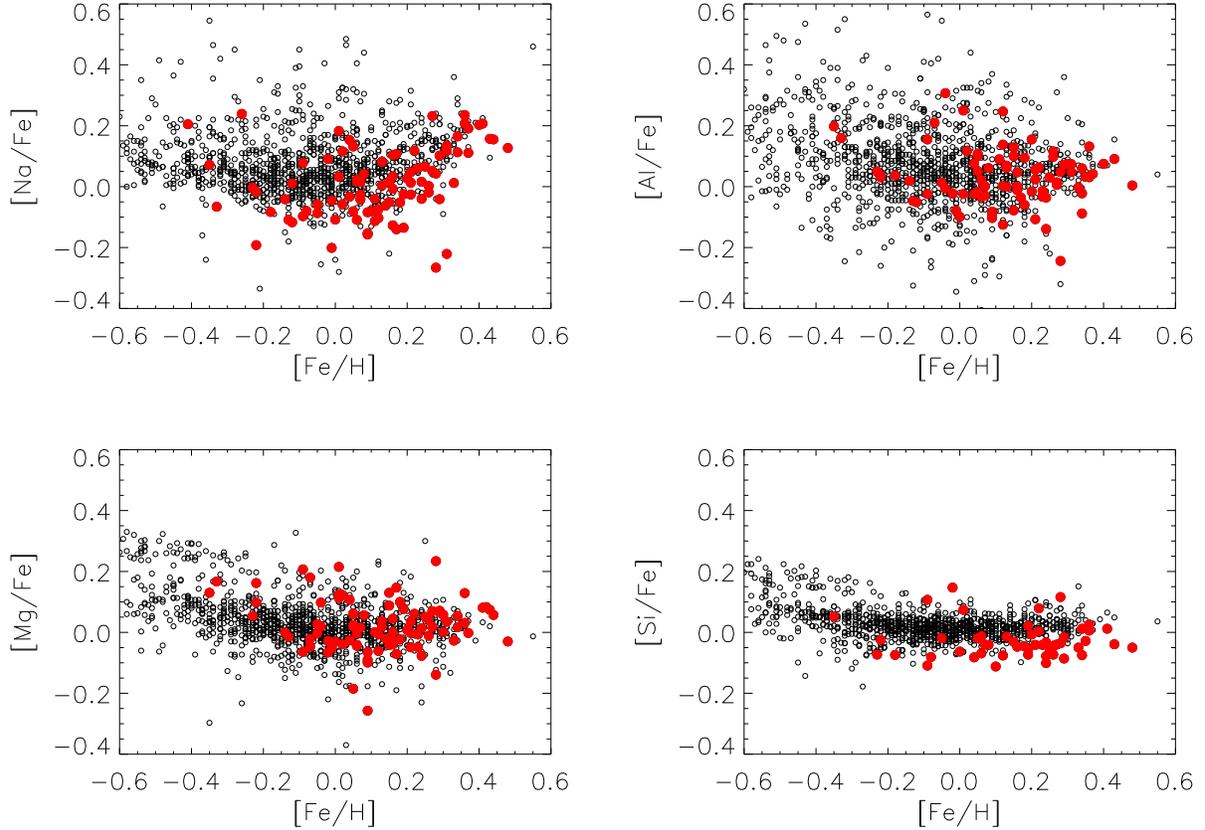}
\caption{Final [X/Fe] ratios as a function of [Fe/H] (red circles). The black empty circles represent the HARPS GTO 1111 FGK stars sample taken from \cite{Adibekyan2012}.}
\label{xfe_harps}
\end{figure*}

The uncertainties are calculated by considering the error on the measurement of the EW and the sensitivity of the abundances to the change in stellar parameters due to their errors. Since the errors on stellar parameters are larger as the values depart from solar, especially for cooler stars, the errors on abundances are also larger for cooler stars. The average errors for [X/H] ratios is 0.06 dex for Na and Si, and 0.07 dex for Mg and Al. We refer to \cite{DelgadoMena2017} for a detailed description on the error determination.
We also examined the precision errors as a function of S/N. Meanwhile for S/N values higher than 150 we can achieve precision errors below 0.05 dex in elemental abundance for most of the stars, for spectra with S/N below 100 the precision errors can go above 0.1 dex. We find small differences depending on the \teff\, of the stars (where cooler stars tend to have higher errors) but the main factor affecting the precision is the S/N of the spectra. Therefore, for the faint stars in the Ariel target list, for which it will be difficult to obtain spectra with S/N\,$>$\,100, we can expect average precision values around 0.1 dex or higher. 
Fig. \ref{xfe_harps} shows the preliminary [X/Fe] vs [Fe/H] of our sample of stars (red symbols) compared to the 1111 FGK stars within the HARPS GTO sample. In general the stars follow the expected Galactic chemical evolution trends for these elements \cite{Adibekyan2012,Bensby2014} and we only can see a small offset in the Si abundances that will be further explored in the future.

\subsection{C and N abundances determination}
\label{sec:CN}

Carbon abundance is estimated with two different methods:  the measurements of the EWs of absorption lines of atomic carbon C~I and the spectral synthesis of C$_{2}$ and CH molecular bands. 
In the wavelength range covered by our spectra, there are two C~I lines at 5052.144\,\AA\,and   
6587.610\,\AA\,in the Gaia-ESO line-list (\cite{heiter15}) classified with high-quality flags, and one line at 5380.325 \AA, with a lower quality flag, which can be used for C abundance determination. All three lines are discussed in \cite{amarsi19} and the effects of 3D model atmosphere and of NLTE are negligible, as also discussed in \cite{Franchini2020}.
We used them to derive carbon abundance in about 120 spectra of our sample with FAMA, fixing the stellar parameters to the \sweet\ ones, as done for the other elements.

We tested an additional procedure to derive the abundance for C and N, based on both the computation of a grid of synthetic spectra by varying the abundance of the element under study, and on the comparison of the synthetic grid to the observed spectrum. The synthesis of molecular bands might allow us to obtain the measurement of N abundances, whose atomic lines are not accessible in the wavelength range of our spectra.  
The synthesis is performed with the radiative transfer code {\it turbospectrum} \cite{AlvarezPlez1998} using the grid of MARCS model atmospheres \cite{Gustafsson2008}. We used an extended version of the Gaia-ESO linelist (U. Heiter, private communication) to cover the wavelength regions below the Gaia-ESO lowest wavelength, i.e. below \SI{4200}{\angstrom}. We include the molecular linelists for CN, C$_{2}$, CH (Masseron, private communication; see also B. Plez linelist repository and reference therein). The synthesis are performed assuming LTE, in 1D plane-parallel geometry (since the sample stars have \logg\ $>$ 3.5). The best-matching synthesis is chosen by a $\chi^2$ minimisation. 

For the carbon estimation through spectral synthesis method we select three molecular bands $[\SI{4208}{\angstrom}, \SI{4233}{\angstrom}]$, $[\SI{4280}{\angstrom}, \SI{4330}{\angstrom}]$ and $[\SI{4350}{\angstrom}, \SI{4400}{\angstrom}]$, from which the derived abundances do not show any correlation with \teff\ or \logg. For nitrogen estimation we used one band: $[\SI{4170}{\angstrom}, \SI{4220}{\angstrom}]$. For this preliminary study, C was measured assuming $\abratio{N}{Fe} = 0$, while N was derived assuming $\abratio{C}{Fe} = \abratio{\alpha}{Fe}$. For both C and N determinations, the synthesis was performed assuming $\abratio{O}{Fe} = \abratio{\alpha}{Fe}$.

\begin{figure}[b]
  \begin{center}
    \includegraphics[trim=0.cm .0cm 0.0cm 0.cm, clip,width=.75\textwidth]{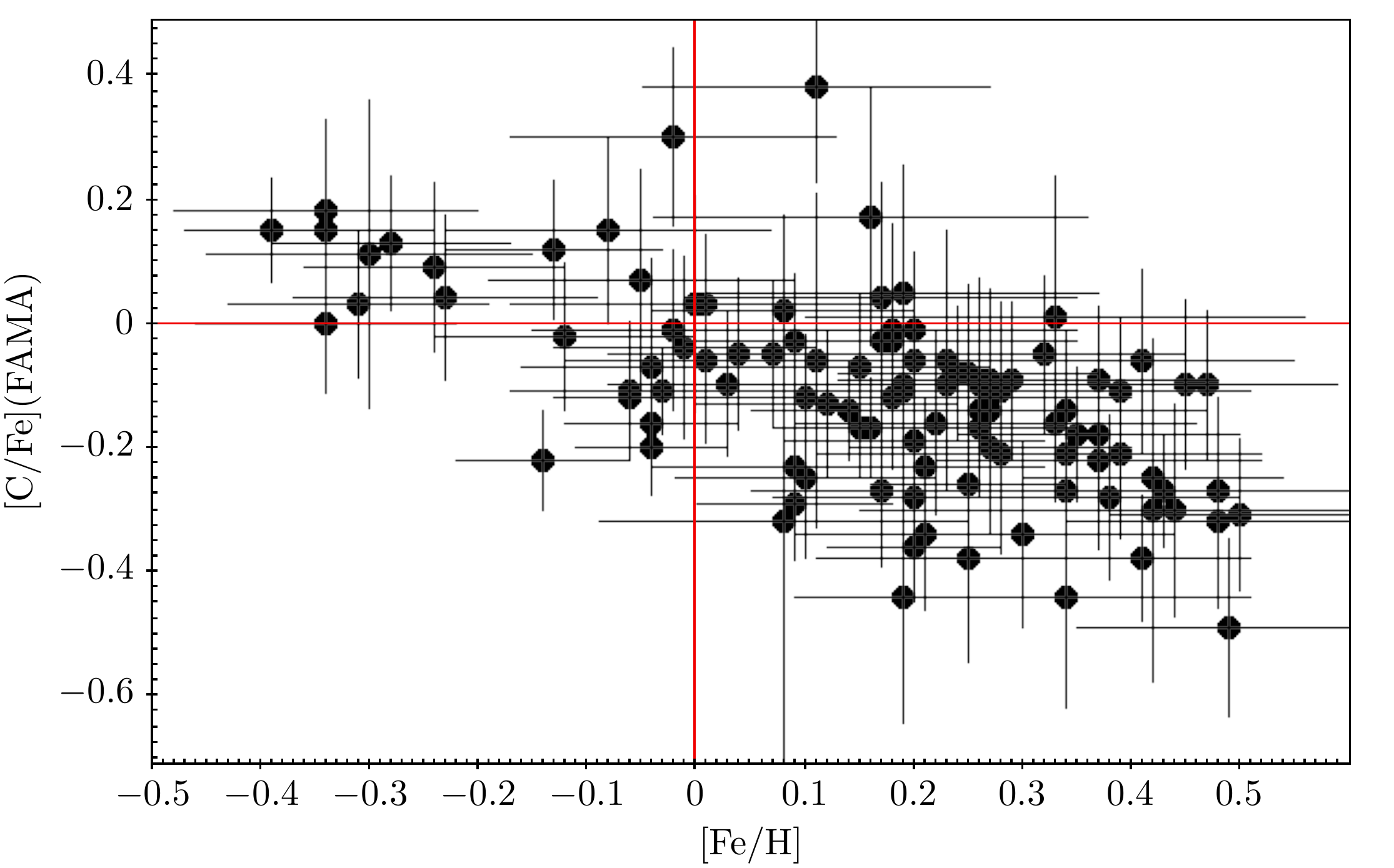}
    \caption{\label{cfe_fe} Carbon abundance over Fe vs [Fe/H] from atomic C~I lines.  Only stars with at least two C~I lines measured in their spectra are shown. The red lines indicate the solar abundances.}
\end{center}
\end{figure}

\begin{figure}[!h]
\centering
  \includegraphics[trim=0.1cm 0cm 0.3cm 1.cm, clip,width=.8\textwidth]{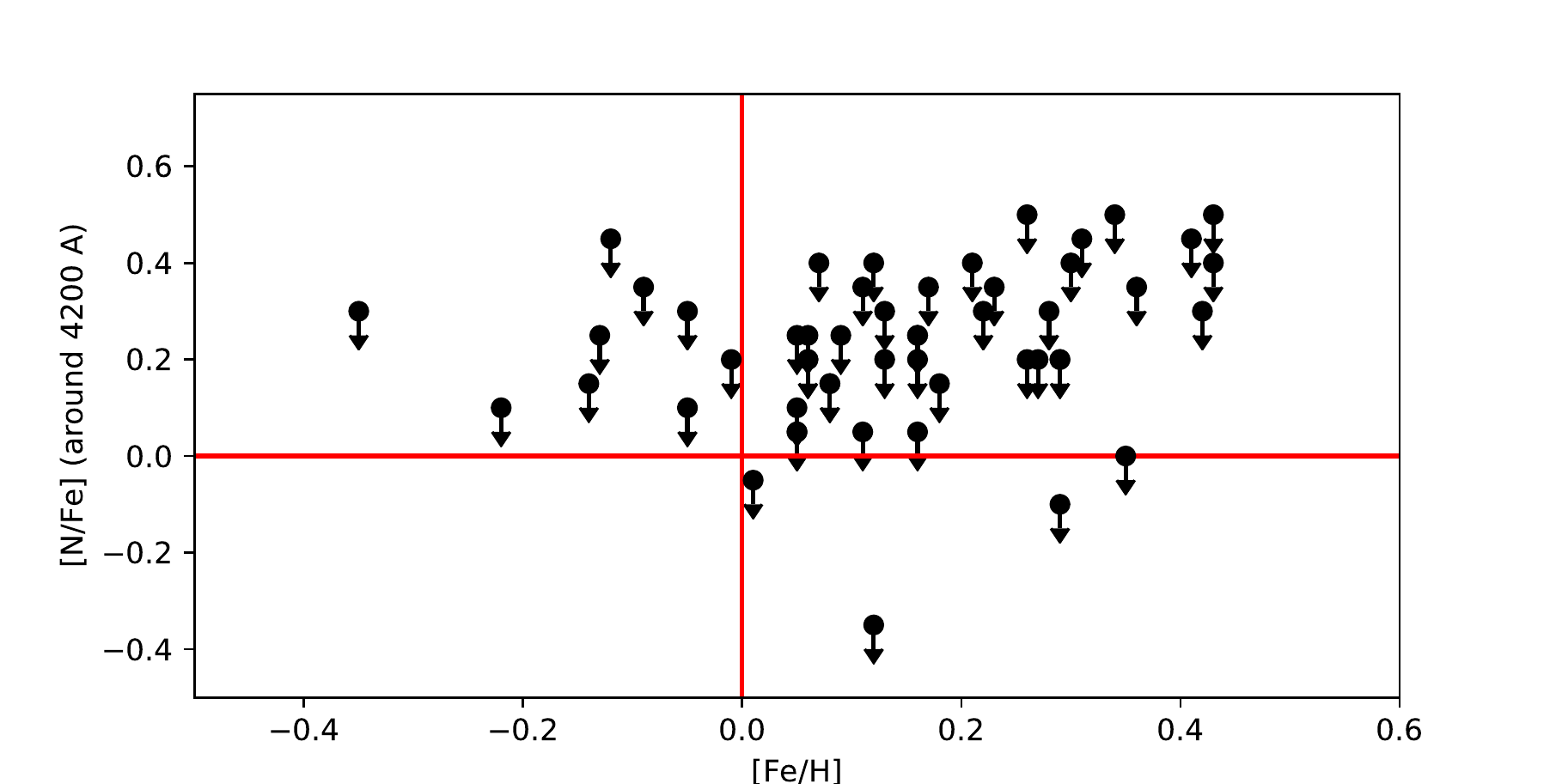}
  \caption{[N/Fe] vs [Fe/H] from molecular bands around \SI{4200}{\angstrom}. The red lines indicate the solar ratios. Only upper limit on the N abundances could be measured.}\label{AbDiag2}
\end{figure}

\subsubsection*{Results:}

The [C/Fe] ratios obtained with atomic lines are shown in Fig.~\ref{cfe_fe}. Only stars with at least two  measured lines in their spectra are plotted. The errors are the 1-$\sigma$ of the mean C abundance. The trend shown in Figure~\ref{cfe_fe} is the expected one for C (see e.g. \cite{Nissen2014,suarezandres2017,Franchini2020}), with higher [C/Fe] at lower [Fe/H], and a decreasing trend with increasing metallicity, although we find slightly lower abundances at higher metallicities than the cited works. This trend is typical of elements (partially) produced by core collapse supernovae, as C (which is also produced by low and intermediate mass stars). On the other hand, we could determine carbon from molecular bands only for a fraction of the stars, whose [C/Fe] ratios are mostly distributed around the solar values. However, for some of the stars at high metallicities the abundances seem to be enhanced as compared to the ones obtained by atomic lines.

Figure~\ref{AbDiag2} shows the trend for the nitrogen from molecular bands vs [Fe/H]. Nitrogen has both a primary and secondary production, and it is expected to increase with metallicity at high [Fe/H] (see. e.g., \cite{Ecuvillon2004a,suarezandres2016,magrini18}). Thus, its abundance ratio [N/Fe] vs [Fe/H] is expected to increase at high metallicity, where the secondary production dominates. Our results indicate an almost flat behaviour, up to [Fe/H]$\sim$0.2, and then a slightly increasing trend, but we acknowledge that this trend might not be real since we could only measure upper limits. There is also an offset at solar metallicity of about 0.2~dex, which can likely be corrected with an estimate of Solar nitrogen from the same molecular band. However, we recall that nitrogen from the molecular band [4170\,\AA, 4330\,\AA] in warm dwarf stars is very challenging as it is very weak and blended with other strong lines (see, e.g., \cite{sneden2014}).

The comparison with literature results shows that the two methods, atomic and molecular C, are in  quite good agreement at sub-solar metallicity, while at solar and super-solar metallicity the measurements of C abundance from atomic lines have to be preferred, since C from molecular bands might be overestimated. 

Concerning nitrogen abundance, due to the difficulties encountered, it should be preferable to analyse transitions from neutral N atoms or molecular bands in the near-infrared or near-ultraviolet regions (\cite{sneden2014,Ecuvillon2004a,suarezandres2016}, and references therein). Therefore we will apply for observing time (see Sec. \ref{sec:campaign}) to obtain new spectra covering the needed wavelength region.

Finally, we also plan to study oxygen abundances, which unfortunately are not measurable for many of our stars, since the optical lines covered by most of our spectra (6158\,\AA, 6300\,\AA) are very weak (especially for cooler objects) and a very high S/N spectrum is needed to obtain reliable abundances. For the stars for which we collect spectra covering redder wavelengths we will explore the use of the oxygen triplet in the NIR ($\sim$7775\,\AA). Nevertheless, in order to obtain homogeneous abundances ratios (e.g. C/O) we will try to use the same oxygen indicators for a great majority of the stars. In this regard, the choice of the weak oxygen lines is preferred since they are not severely affected by 3D effects or departures from LTE as does happen for the oxygen triplet \cite{Caffau2008}.

\section{Homogeneous stellar activity indices}
\label{sec:indexes}
Stellar activity is notably the major source of astrophysical noise in the search and characterisation of exoplanets. Photometric modulations or the distortion of spectral line profiles, induced by inhomogeneities on the stellar surface (e.g. starspots, faculae, granulation), are able to hamper the clear detection of a planet with the transit or the radial velocity method, respectively (e.g. \cite{1997ApJ...485..319S,2015PhDT.......193H,2018A&A...613A..50C,2004A&A...414.1139A,2013A&A...556A..19O}).
Stellar activity can also affect the planetary atmosphere characterisation, which is the main goal of Ariel (e.g. \cite{2013MNRAS.432.2917P,micela2015,2017ApJ...844...27Z,sarkar2018}, see also \cite{Cracchiolo2020} for the specific case of Ariel).
Knowing the activity level of the hosts can help quantify its impact on Ariel observations and importantly contribute to the target prioritisation process.

In the context of the characterisation of Ariel host stars, we homogeneously analysed high-resolution spectra available for part of the Ariel Reference Sample, aiming to derive spectroscopic activity indices. The CaII H\&K lines at 393.4 and 396.8 nm are particularly sensitive to chromospheric emission (e.g. \cite{wilson1968,noyes1984}), and were used to derive the Mount Wilson $S$ and the \logRHK indices as proxies of the stellar activity level.
From \logRHK we also obtained an independent estimate of the stellar ages, 
and an indication of the stellar rotation rates. The latter was finally compared with the rotation periods obtained from time series of the same high-resolution spectra and photometric archival data.

    \subsection{Data}
    
    We selected 92 planet-host stars by crossmatching the Ariel Reference Sample and the homogeneous sample of the \sweet~catalogue, for which high-resolution spectra from TNG/HARPS-N and ESO/HARPS are available.
    We analysed 1671 public archival spectra acquired with HARPS between 2004 and 2019 through various programs, and 891 archival spectra obtained with HARPS-N between 2012 and 2019 in the framework of the GAPS project \cite{covino2013}. 
    Table \ref{tab:sample} (currently placed at the end of the manuscript for better readability) summarises the stellar sample, the number of available spectra for each target, the average signal-to-noise ratio (S/N) of the corresponding spectra, and the results of our analysis. 
    For the study of activity through CaII H\&K lines presented here, we used the one-dimensional spectra corrected for the instrumental response and stretched to the solar barycentric reference frame (s1d files) produced by the HARPS and the HARPS-N pipelines.
    
    \subsection{Activity level from Calcium lines}
    \label{sec:logRhk}
    
    For all available spectra, we determined the HARPS-related $S$ index from the Ca line cores, and converted it to the Mount Wilson $S$ following the method \cite{lovis2011} developed specifically for HARPS spectra. In order to perform the colour-dependent $S$-to-$R^{'}_\mathrm{HK}$ conversion, we then extracted the $B-V$ colours from the SIMBAD \cite{wenger2000}, the Two-Micron All-Sky Survey (2MASS \cite{skrutskie2006}), and NASA Exoplanet archives, which we corrected for the effect of interstellar extinction. 
    For all our targets we obtained estimates of the A$_V$ extinctions (or its upper limits) using a 3D map of Galactic interstellar dust \cite{Vergely2021}, which provides in a large number of regions a better angular resolution than its previous version presented by \cite{Lallement2019}. Following, we computed the $E(B-V)$ and $E(V-K)$ colour excess assuming a standard extinction law A$_\lambda$/A$_V$ with R=3.1 \cite{Cardelli1989}.
    From there, we measured the intrinsic $(B-V)_0$ and $(V-K)_0$ colours. 
    When the $(B-V)_0$ 
    value of our objects was within the validity range of \cite{noyes1984}'s relationship, we used their calibration to convert the weighted-averaged $S$ value to \logRHK. Otherwise, we referred to \cite{astudillo-defru}'s relations, derived for broader ranges of $(B-V)_0$ or $(V-K)_0$. The error bars on \logRHK were obtained through error propagation from the uncertainty in the measure of $S$ and the published uncertainty of the $S$-\logRHK conversion coefficients. We removed from our sample 29 objects, for which the S/N of the co-added spectrum in the echelle order containing the CaII doublet was $\leq 5$. Also, we could not calculate the \logRHK~index for 16 of our targets, whose colours, or low ($\lesssim 0.1$) $S$-index fall off the reported calibration ranges. This left us with 47 \logRHK~index measurements, reported in Table \ref{tab:sample}.

    Thanks to the derived \logRHK~indices, we were then able to estimate rotation rates (Section \ref{sec:rotation})  and stellar ages for our sample. These latter were calculated following the empirical relation by \cite{mamajek2008}, which is valid for \textit{(i)} $0.5 \, \mathrm{mag} < B-V < 0.9 \, \mathrm{mag}$; \textit{(ii)} -5.1 $\lesssim$ \logRHK $\lesssim$ -4.0; and \textit{(iii)} 6.7 $\lesssim \log ({\rm age_{Gyr}}) \lesssim$ 9.9. 
    
    Given the large number of spectra available for some of our targets, we also inspected for variations in the $S$-index which could indicate activity cycles (e.g. \cite{Maldonado2019}). For this analysis, we also included the stars for which no $S$-to-\logRHK conversion was possible. We tentatively identified 3 objects showing $>3\sigma$ $S$-index variations around their weighted average value. However, all such cases are impacted by the low-S/N in the spectral order containing the CaII lines, or by possible instrumental issues whose investigation is beyond the scope of this study. As a result, such detected variations are likely not robust.

    \begin{figure*}[t!]
    \centering
    \includegraphics[width=0.8\textwidth]{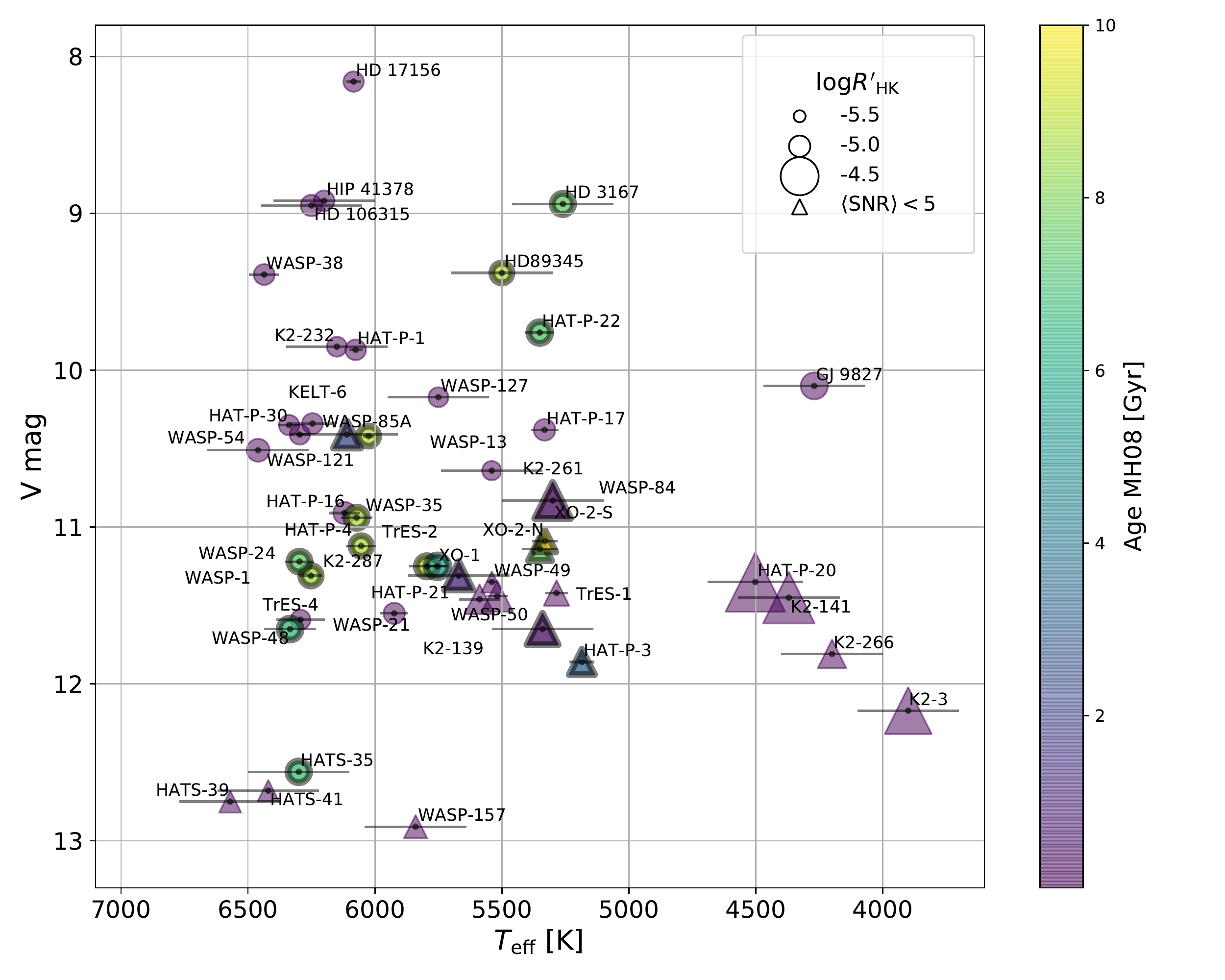}
    \caption{$V$ magnitude versus effective temperature for the subset of 47 Ariel stars with a valid \logRHK~index. The average \logRHK is indicated by the circle marker size, while the targets whose spectra have an average $\mathrm{S/N}<5$ are drawn as triangles.
    Colours represent the stellar age in Gyr derived from the average activity index. Symbols with a black edge correspond to targets whose \logRHK~and colour fall within \cite{mamajek2008}'s age calibration range, the other targets being assigned an age of $0$ Gyr (off the colour bar range) for representation purposes.}
    \label{teff_mag_rhk}
    \end{figure*}
    Figure \ref{teff_mag_rhk} presents the $V$ magnitude of the targets with a derived \logRHK~index versus their effective temperature. The mean activity level (obtained with a weighted average of the activity indices measured on individual spectra) is shown by the marker size: 
    circles indicate targets whose average S/N $>5$, while triangles represent an average S/N $<5$, which implies larger error bars for the \logRHK measure.
    
\begin{figure}
        \centering
         \includegraphics[width=0.8\textwidth]{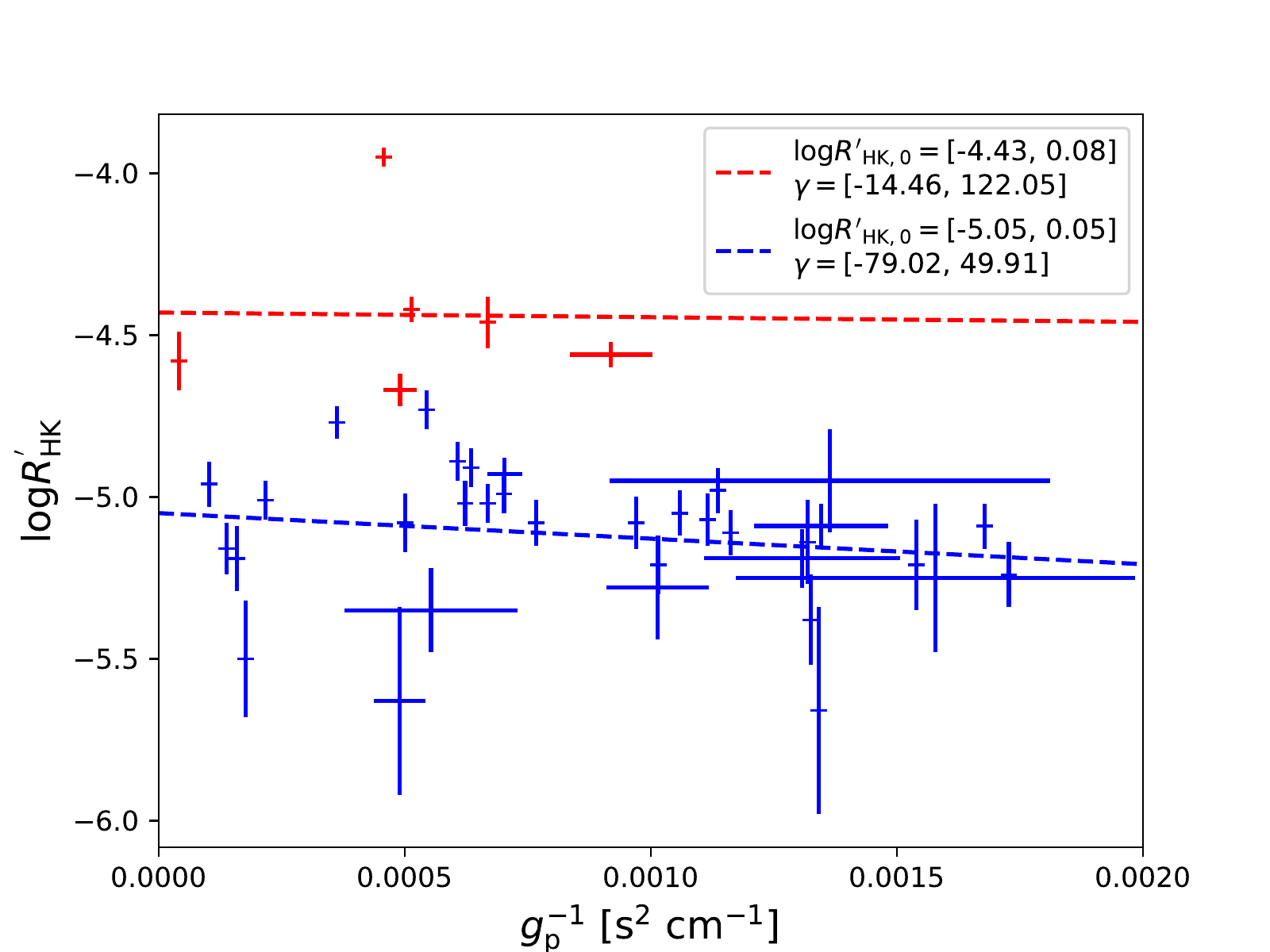}
        \caption{Measured stellar \logRHK versus the inverse of the planetary surface gravity. Two populations with high and low observed stellar activity level are tentatively identified in red and blue, respectively. Their linear fits are shown with dashed lines. For each population, we report in the legend (i) the $\log R_\mathrm{HK}^{'(0)}$ intercept value with its $1\sigma$ uncertainty, (ii) the slope with $1\sigma$ uncertainty [cm s$^{-2}$]. More details are provided in the text.}
        \label{fig:logRHKvslogg}
    \end{figure}

    We note that \cite{Hartman2010,Fossati2015,Fossati2017} reported a correlation between \logRHK and planetary surface gravity $g_\mathrm{p}$, possibly explained by evaporated planetary material which absorbs at the core of chromospheric resonance lines \cite{Lanza2014}. \cite{Fossati2015} identified two sub-populations, classified by their \logRHK versus $g_\mathrm{p}^{-1}$ trend, among active stars hosting close-in giant planets, and they attributed their separation to the Vaughan-Preston gap \cite{vaughan1980}. The Vaughan-Preston Gap is in fact a  low populated region with intermediate levels of activity, which divides the high level chromospheric-activity population, and the low level one. We looked for such correlations in our sample: in Figure \ref{fig:logRHKvslogg}, the target subset with available \logRHK is shown as a function of the inverse of the respective close-in planet's surface gravity.\footnote{The planetary parameters were downloaded from the Extrasolar Planets Encyclopaedia \cite{schneider2011}, the NASA Exoplanet Archive and the Exoplanet Orbit Database \cite{han2014}.} In order to compare our analysis with \cite{Fossati2015}'s result, we considered only those targets with $g_\mathrm{p}^{-1} < 0.03$ s$^2$ cm$^{-1}$ and \logRHK$> -6$. This allowed us to identify a higher-activity and a lower-activity population, using the scikit-learn package \cite{scikit-learn} implementation of a Gaussian mixture model. 
    Following the results of \cite{Fossati2015}, we adopted two components for our models. We fitted linear relations to the two populations and determined their intrinsic activity index $\log R_\mathrm{HK}^{'(0)}$ and slope $\gamma$, and the respective uncertainties with bootstrapping. We obtained $\log R_\mathrm{HK}^{'(0)} = -5.05 \pm0.05, \, \gamma=-79.02\pm49.91$ cm s$^{-2}$ and $\log R_\mathrm{HK}^{'(0)} = -4.43 \pm 0.08, \, \gamma = -14.46 \pm 122.05$ cm s$^{-2}$ for the two subsamples, respectively.
    
  We tested our model on the \cite{Fossati2015}'s dataset, and we obtained results in agreement at the $3 \sigma$ level\footnote{In our implementation, we left each component to have its own covariance matrix, and initialised the components through K-Means clustering.}.
 When applying the method on our dataset, 
    we found results consistent with \cite{Fossati2015} within $2\sigma$ level for the intercept of the active population, and the slope of the inactive population. The slope of the active population is consistent within 3$\sigma$, and the intercept of the inactive one is consistent within 5$\sigma$ with their results.  These differences can be caused by a selection effect and/or to the homoskedasticity of the \cite{Fossati2015} data, and/or to the diversity of the clustering methods. The investigation of such discrepancies is beyond the scope of this manuscript, however a more thorough analysis will be presented in \cite{Claudi2020}.

\subsection{Rotation periods}
\label{sec:rotation}
As a further characterisation, we used the available photometric and spectroscopic data for a subsample of the stars we analysed in the previous section, to search for their rotation periods. In the following, we show our method to extract photometric periods from transit surveys data and the time series analysis of a set of spectroscopic activity indices, aiming to detect the activity signature. We finally compare the results of these two approaches.

\subsubsection{Photometry}
To derive the photometric periods, we collected the available light curves of the stars in our sample (31 targets). We downloaded the data obtained from the transiting exoplanet surveys SuperWASP (from the NASA Exoplanets Archive\footnote{\url{https://exoplanetarchive.ipac.caltech.edu/docs/data.html}}), HATNet\footnote{\url{https://hatnet.org}} and HATS\footnote{\url{https://hatsouth.org/}}.  
Parameters such as the orbital periods, the transit midpoint times, and the transit duration with corresponding uncertainties are considered as well, since we used this information to identify all the portions in the time series where the transits are expected to occur. Once we removed these portions from the light curves, the resulting residuals should be only affected by the stellar contribution.

\begin{figure*}[h]
    \centering
    \includegraphics[trim=1.5cm 2cm 1.5cm 2cm, clip, width=0.7\textwidth,angle=270]{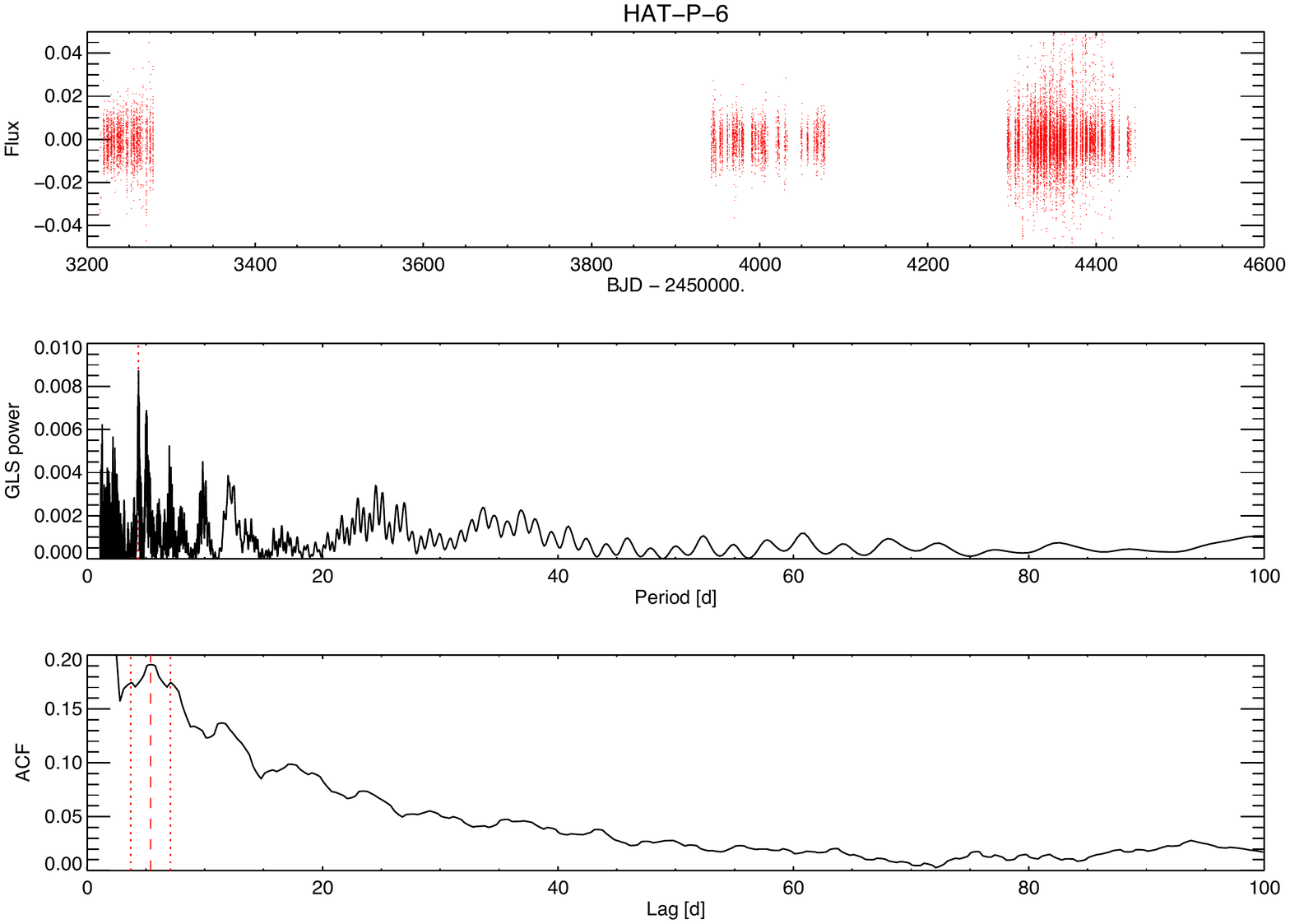}
    \caption{{\it Upper panel}: light curve of HAT-P-6, composed by more than 15.000 data points with the planetary transits removed. 
    Middle panel: GLS of the light curve. The highest periodicity is marked with a red dotted vertical line and is equal to $4.400 \pm 10^{-3}$ days. The FAP corresponding to this peak is lower than 0.01\%.} {\it Lower panel}: ACF of the same light curve, showing a similar periodicity at $5.4 \pm 1.7$ days. 
    \label{fig:p_rot}
    \end{figure*}

We used two different methods to measure the periodic content of the ``clean'' photometric time series. The first one is the Generalized Lomb-Scargle periodogram (GLS, \cite{Zechmeister2009}). In general, we applied the GLS to the whole time series, but when the gaps within the data are larger than one observing season, we also checked the single bulk of data. This is justified by the fact that the photometric signal described by period, amplitude and phase, can change during the activity cycle, depending on the different distribution of magnetic spots on the stellar disc. To evaluate the significance of the detected peaks, we calculate the False Alarm Probability (FAP) through a bootstrap random permutation technique.
The second method is the Auto-Correlation Function (ACF, \cite{McQuillan2013}).
As reported by \cite{McQuillan2013}, the ACF method seems to be more robust to find the photometric periods, as demonstrated with \textit{Kepler} data. When this method is applied to the light curve, a number of maxima can be observed in the ACF as a function of the lag, and the position of the highest peak and the half-width at half maximum (HWHM) of a Gaussian fit to that peak are assumed as the photometric period with the corresponding uncertainty. An example of the application of the two methods is shown in Figure \ref{fig:p_rot}, in the case of the HAT-P-6 photometric time series.
We notice that even if the main periodicity (indicated with a red vertical dashed line) is very close to the one recovered by the GLS, the peak height is not very different from the mean value of the adjacent areas of the ACF. Similar behaviour is found for other targets of the sample, possibly suggesting that the ACF method is not as suitable as in the case of space-based or high duty-cycle observations.
We compared the results of these analyses with the estimation of the stellar rotation period as obtained from the empirical relation by \cite{mamajek2008} between this parameter and the \logRHK index obtained in Sect. \ref{sec:logRhk} and reported in Table \ref{tab:sample}. 
In a few cases (HAT-P-6 and WASP-14), we suspect that the derived photometric period rather indicates the first harmonics of the real rotation period, since its value is half the one obtained with the spectroscopic indices (see the next section).

We reported the derived photometric periods in Table \ref{tab:gaps} (eighth column). Note that it was not possible to derive such a parameter for all the targets in our sample. The main reason is related to the spectral type of those stars (F or early G). Early type stars are in fact characterised by a lack of dark starspots, preventing us to observe and measure photometric modulations.

    \begin{table*}[b!]
    \centering
        \caption{Stellar sample for which we inspected spectroscopic and photometric data in order to evaluate the rotation period. Spectroscopic data are collected with HARPS-N at TNG within the GAPS collaboration, while photometric light curves are mainly provided by SuperWASP\protect\footnote{\url{https://exoplanetarchive.ipac.caltech.edu/docs/SuperWASPMission.html}}, HATNet\protect\footnote{\url{https://hatnet.org}} and HATS\protect\footnote{\url{https://hatsouth.org/}} surveys. For each star, we listed: the number of the available high-resolution spectra with the corresponding time span, the rotation period from the empirical relations, the main spectroscopic periodicity obtained with the GLS (after the removal of the transiting planet signature), the number of photometric data and the corresponding timespan, and the derived photometric periodicity.}
        \label{tab:gaps}
        \begin{threeparttable}[t]
        \centering
        \begin{tabular}{lccccccc}
           Star     & Nr.& Timespan   & Empirical  &  Spectr.  &  Nr. & Timespan  & Photom.  \\ 
           Name & spectra & [days]    & P$_{\rm rot}$ [d] & Period [d] &  data & [d] & Period [d]\\
         \hline
            HAT-P-1	& 16 & 1875       &	56.8$\pm$6.4 & - & 4787& 1029 & - \\
            HAT-P-3	& 15 & 792        & 18.0$\pm$4.1 & 29.7$\pm$0.2 & 8131 & 1138 & - \\
            HAT-P-4	& 29 & 2355       & 36.6$\pm$4.5 & 15.90$\pm$0.02 & 5717 & 141 & - \\
            HAT-P-6	& 18 & 1860       &	- & 9.07$\pm$0.01  & 19181 & 1232 & 4.4$\pm$0.1  \\
            HAT-P-8	& 14 & 1423       &	- & -  & 11162& 489 & 20.8$\pm$0.1\\
            HAT-P-14	& - & - &	- & - & 1993 & 37 & - \\
            HAT-P-16	& 21 & 1850   & 6.9$\pm$2.3 & 4.60$\pm$0.01 & 12552 & 223& - \\
            HAT-P-17	& 30 & 2513   & 55.4$\pm$6.5 & 44.0$\pm$0.1 & 14552 & 505 & -\\
            HAT-P-20	& 19 & 1137   &	- & 16.18$\pm$0.03 & 2655 & 193 & 14.6$\pm$0.1\\
            HAT-P-21	& 24 & 1495   &	- & 54.4$\pm$0.2 & 27866 & 584& - \\
            HAT-P-22	& 36 & 2595   &	47.7$\pm$5.9 & - & 25971 & 3724& - \\
            HAT-P-23	& - & - &	- & - & 4780 & 103 & -\\
            HAT-P-24	& 17 & 1260   &	- & 10.81$\pm$0.01&12323 & 554&  11.1$\pm$1.7\\
            HAT-P-29	& 32 & 2684   &	- & 23.0$\pm$0.1 & 3603 & 140& - \\
            HAT-P-30	& 15 & 1386   &	24.4$\pm$3.5 & - & 12612 & 898& - \\
            HAT-P-31	& 15 & 1143   &	- & 16.76$\pm$0.07 & 9173 & 117& - \\
            HATS-22     &  - &   -    &     -             &     -           & 13168 & 823 &  52.8$\pm$0.9 \\
            HATS-25       &   - &   - & -        &        -        & 6044 & 147 &  70$\pm$4 \\
            HATS-27       &   - &  -  & -     &         -       & 10618 & 444&  9.04$\pm$0.01 \\
            HATS-60       &  -  &  -  & -  &         -       & 16651 & 202&  22.9$\pm$0.5 \\
            HD-17156	& 18 & 1860   &	27.7$\pm$3.7 & 19.00$\pm$0.04 & 142 & 1171& -\\
            WASP-1	& 13 & 1434       &	36.1 $\pm$4.4 & 11.54$\pm$0.01 & 12308& 1294&  11.1$\pm$1.8\\
            WASP-2	& - & - &	- & - & 12856 & 1600& 27.8$\pm$1.3\\
            WASP-3	& 2 & 26          & - & - & 5799 & 1543& 7.6$\pm$0.5\\
            WASP-12	& 23 & 1845 &	- & 50.8$\pm$0.2 & 5435 & 150 & 67.1$\pm$5.0\\
            WASP-13	& 19 & 918        &	46.5$\pm$5.1& 57.0$\pm$0.1 & 12192 & 866& -\\
            WASP-14	& 15 & 1230 &	- & 23.55$\pm$0.04 & 6214 & 1151& 10.6$\pm$0.1\\
            XO-1	& 15 & 1244       &	19.0$\pm$3.4 & - & 8665 & 150& -\\
            XO-2N	& 60 & 2637       &	42.2$\pm$5.8 & - & 7936 & 501& -\\
            XO-2S	& 135 & 2486      & 46.0$\pm$5.6 & 39.0$\pm$0.1 & - & -& -\\
            XO-3	& 21 & 1482       &	- & 8.28$\pm$0.01 & 2969 & 733& -\\
            XO-4	& 17 & 1165       &	- & - & 1830 & 63& 6.8$\pm$1.8\\    
    \hline
        \end{tabular}
        \end{threeparttable}%
    \end{table*}
    
\subsubsection{Spectroscopy}
\begin{figure*}[!t]
\centering
  \includegraphics[trim=1.5cm 3cm 1.5cm 3.5cm, clip,width=.38\textwidth,angle=270]{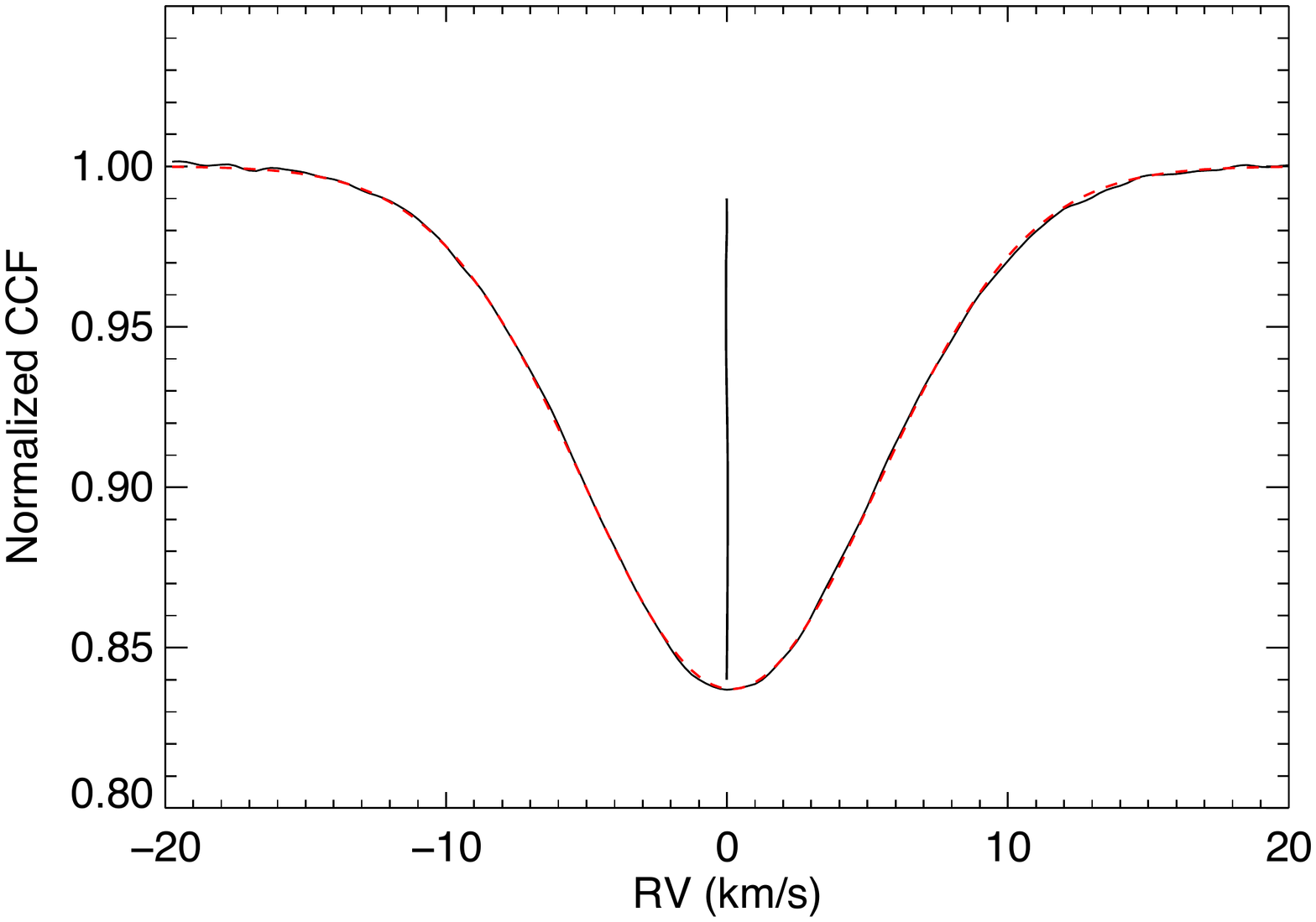}
  \hfill
  \includegraphics[trim=1.5cm 3cm 1.5cm 3.5cm, clip,width=.38\textwidth,angle=270]{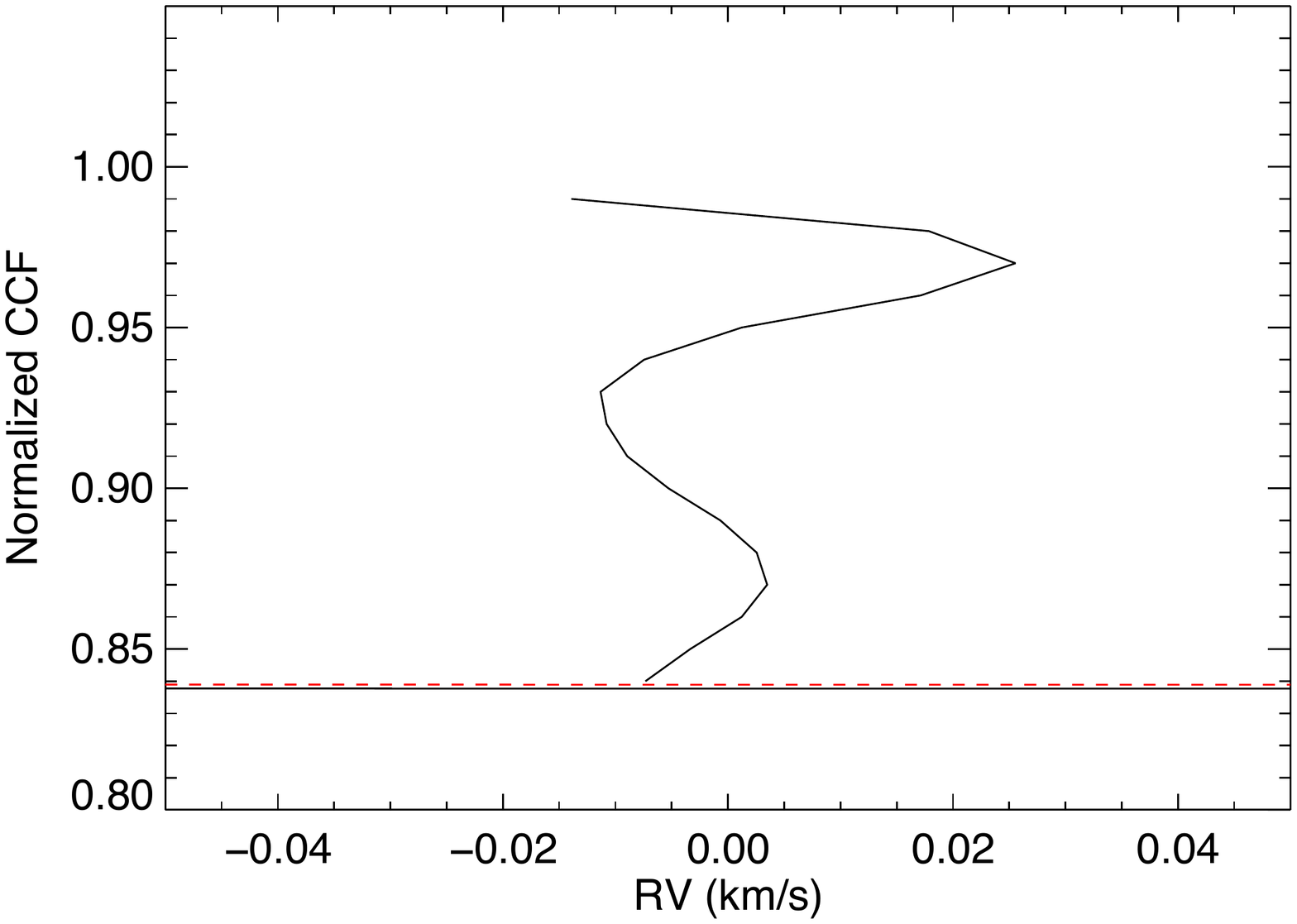}
  \caption{{\it Left panel}: plot of the Cross-Correlation Function for one of the HARPS-N spectra we analysed for the target HAT-P-6. The CCF is depicted with a black line while a Gaussian fit is represented with a red dashed line. Since the CCF is not perfectly symmetric, a line bisector helps to verify the potential distortion of the spectral lines (black vertical line). {\it Right panel}: The actual shape of the bisector can be observed by zooming the central part of the CCF. }\label{fig:bis}
\end{figure*}

Time series of high-resolution spectra from HARPS-N \cite{cosentino14} at TNG are available for 25 targets analysed here, collected within the GAPS collaboration \cite{covino2013}. Spectroscopic time series are useful to derive information on the stellar activity, including the stellar rotation periods, thanks to a number of activity indicators that can be obtained from the line profiles. We exploited the HARPS-N standard reduction, performed through the Cross Correlation Function (CCF) of the spectrum with a numerical mask representing the spectral features of a specific spectral type (see \cite{pepe2002}). The resulting CCF is the weighted and scaled average of the lines in the correlation mask (see Figure \ref{fig:bis}, left panel), and for this reason it can be used as a proxy for variations of the line profiles, like the ones induced by the activity-related phenomena.
We considered the time series of the radial velocity (RV), the bisector inverse slope (BIS, \cite{2001A&A...379..279Q}) that is the difference between the bisector value in the upper and lower part of the CCF (see Figure \ref{fig:bis}, right panel), and the CCF Full Width at Half Maximum, that are directly provided by the HARPS-N pipeline. We also analysed the asymmetry indices of the CCF, $\Delta V$ and V$_{\rm asy (mod)}$. 
The asymmetry indices can be used to define the effects of the activity on the stellar spectral lines due to processes occurring in the photosphere (e.g., magnetoconvection, see \cite{cegla2019}). The $\Delta V$ \cite{2006A&A...453..309N} is the difference between the RV obtained with a Gaussian fit of the CCF versus the one obtained with a bi-Gaussian fit that accounts for the asymmetry of the profile, and the $V_{\rm asy(mod)}$ (\cite{lanza2018} and the references therein) is defined as the difference between the RV information from the red and blue part of the CCF. The BIS shows the impact of the passage of dark spots on the stellar disc as the star rotates, with respect to the noise floor caused by the convective blueshift. The information from the BIS can be different from the one provided by the $V_{\rm asy(mod)}$ since the former is evaluated on limited ranges of the CCF, the latter on the whole profile and thus in some cases, it can be more sensitive to the asymmetry. 
To obtain these indices, we applied the publicly available IDL procedure described in \cite{lanza2018}.

Our RV periodograms show very clearly the periodicities of the transiting planets, so we first removed the corresponding signal before searching for signatures related to the stellar rotation/activity.
The main drawback of this analysis is the sparse data sampling of our time series (typically 20 RV data points spread over a few years), that prevents our estimates to be robust. This is confirmed by the large values of FAP, obtained with bootstrap permutations. For this reason, for each target we only considered those periodicities simultaneously detected in the GLS' of more than one time series (RV, BIS,...) and, just like we did with the photometry, we make use of the empirical relations in \cite{noyes1984} and \cite{mamajek2008} as a further reference for our investigations. In general, the highest peaks of the GLS periodograms of the spectroscopic time series show values quite similar with respect to the expected ones, but we stress that more data would improve their significance. Table \ref{tab:gaps} shows a summary of the periodicities found with photometric (obtained with ACF or GLS) and spectroscopic data (GLS). The values we reported are the result of the Levenberg-Marquardt fit used by the GLS routine with the formal errors or the result of a gaussian fit when ACF is used. If no value is provided in the table, then our methods did not returned any reliable periodicity. 
As a final check, we applied the pooled variance method to the $S$ index and other spectroscopic activity indicators trying to characterise the time-scales of the stellar activity, including the rotation periods (see e.g. \cite{Maldonado2019}).
Unfortunately, such a technique suffers from the sparse sampling of the time series, so the information we obtained was mainly used to verify the results from the GLS analysis and not to make robust predictions.

In conclusion, the recovery of the stellar rotation periods of potential targets of Ariel could be challenging. Photometry suffers when early spectral type stars (F and G) are considered, since they show no or few star spots that are responsible for modulations on the stellar brightness related to the rotation. Some advantages can be obtained if we consider spectroscopic time series of those targets, but in this case the availability of a dense monitoring with high-resolution data is lesser than the one of the photometric light curves. For this reason we are working to collect more data in order to confirm such findings, as presented in section \ref{sec:campaign}.
A robust derivation of the stellar rotation periods from archive data, or data obtained on purpose, would also allow to estimate the stellar age in an independent way as derived, for instance, from isochrones. With the homogeneous estimate of the stellar mass, we plan to apply gyrochronology to all the stars with a derived rotation period.

\section{The need for homogeneous stellar ages}
\label{sec:ages}

\begin{figure}[b!]
    \centering
    \includegraphics[trim = 0 0 0 0, clip,width=.6\textwidth]{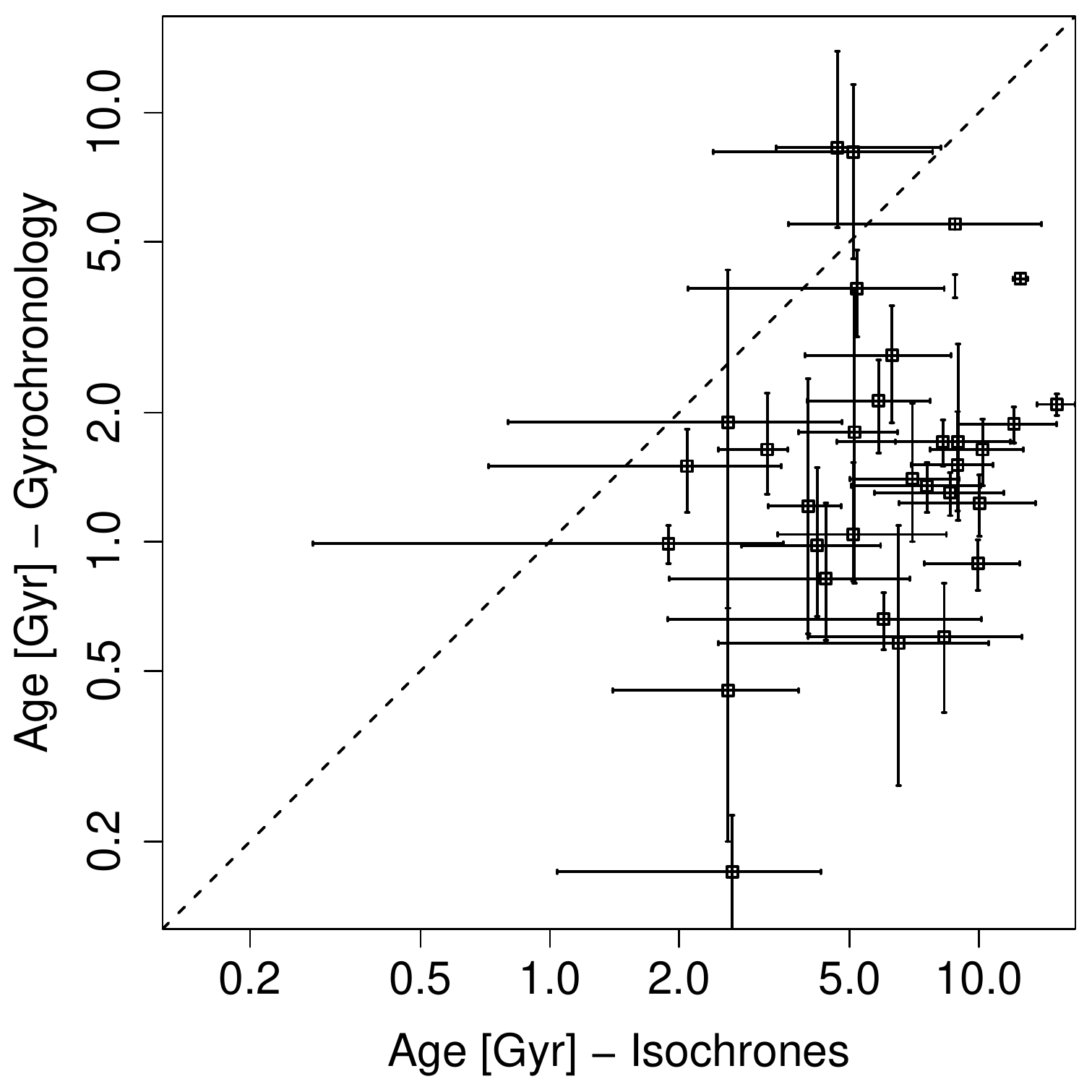}
    \caption{Stellar ages for a sample of 32 Ariel stars whose age has been determined both by gyrochronology ($y$ axis) and by isochrone fitting ($x$ axis), as found in the literature. The dotted line represents the identity line. The stellar mass ranges from 0.69 \Msun\ to 1.49 \Msun. Data is reported in Table \ref{tab:ageliterature}.}
    \label{fig:GyroVSiso}
\end{figure}

Stellar age is extremely important as it provides the broader context necessary to interpret the formation and evolution of the planet(s) in a system, both in terms of composition and architecture.
Age is also correlated to stellar rotation and chromospheric activity (e.g., \cite{Skumanich1972,Wilson1963,mamajek2008}), important aspects to be taken into account when analysing observations of a transiting planet (see Sec.~\ref{sec:indexes}).

By inspecting the literature on Ariel stars, we found large inconsistencies between ages estimated by different teams and/or using different techniques. To illustrate this, Fig.~\ref{fig:GyroVSiso} shows literature ages estimated via gyrochronology versus those obtained from isochrone fitting (see also Table \ref{tab:ageliterature}). For a sample of 32 stars hosting transiting planets and covering a mass range from 0.69 \Msun -- 1.49 \Msun, gyrochronology values are systematically lower than those provided by stellar models. 
Works by \cite{Brown2014,Maxted2015} tried to interpret such a discrepancy as an indication of tidal interaction between the planets and their host stars, but neither study could draw strong conclusions. \cite{Maxted2015} concluded that on a case-by-case basis it is not always clear that there is good evidence for the tidal spin-up of the host star by the planet.

\begin{table}[t!]
\begin{threeparttable}
\caption{ List of targets whose age has been plotted in Fig. \ref{fig:GyroVSiso}. We report their age estimated through isochrones fitting and gyrochronology. References from which data was collected are reported as notes}
\begin{tabular}{l | c | c }
\hline\noalign{\smallskip}
Star Name & Isochrones Age [Gyr] &  Gyrochronology Age [Gyr]  \\
\noalign{\smallskip}\hline\noalign{\smallskip}
55Cnc       & 5.10 $\pm$ 2.70 \tnote{a}  &               8.10 $\pm$ 3.54 \tnote{b} \\
CoRoT-2     & 2.66 $\pm$ 1.62 \tnote{b}  &               0.17 $\pm$ 0.06 \tnote{b}\\ 
HAT-P-11    & 5.20 $\pm$  3.10 \tnote{a}     &               3.89 $\pm$ 0.89\tnote{b}\\
HAT-P-19    & 8.80 $\pm$  5.20 \tnote{c}     &               5.50 $\substack{+1.30 \\-1.80}$ \tnote{d}\\
HAT-P-21    &  10.20 $\pm$ 2.50 \tnote{e}    &               1.64 $\pm$ 0.29\tnote{b}\\
HD-219134    &  12.50 $\pm$ 0.50 \tnote{f}    &               4.10 $\pm$  na \tnote{f}\\
  K2-29     &  2.60 $\pm$ 1.20 \tnote{g}    &               0.45 $\pm$ 0.25 \tnote{g}\\
WASP-10     &  6.00 $\pm$ 4.12 \tnote{b}     &               0.66 $\pm$ 0.10\tnote{b}\\
WASP-106  &  7.00 $\pm$ 2.00 \tnote{h}    &               1.40 $\substack{+0.70 \\-0.40}$ \tnote{h}\\
WASP-107  &  8.30 $\pm$ 4.30 \tnote{i}     &               0.60 $\pm$ 0.20 \tnote{i}\\
WASP-113  &  5.10 $\substack{+3.30 \\-1.70}$ \tnote{j} &       1.04 $\substack{+ 0.49\\-0.23}$   \tnote{j}\\
WASP-114  &  4.20 $\substack{+1.70 \\-1.40}$ \tnote{j} &       0.98 $\substack{+0.51\\-0.31}$    \tnote{j}\\
WASP-135  &  4.40 $\pm$ 2.50 \tnote{k} &                  0.82 $\substack{+0.41 \\-0.23}$\tnote{k}\\
WASP-151  &  5.13 $\pm$ 1.33 \tnote{l}    &               1.80 $\substack{+2.03 \\-1.00}$ \tnote{l}\\
WASP-153  &  4.00 $\pm$ 0.77 \tnote{l}    &               1.21 $\substack{+1.19 \\-0.60}$ \tnote{l}\\
WASP-156  &  6.50 $\pm$ 4.03 \tnote{l}    &               0.58 $\substack{+0.51 \\-0.31}$ \tnote{l}\\
WASP-19   &  9.95 $\pm$ 2.49 \tnote{b}      &               0.89 $\pm$ 0.12\tnote{b}\\
WASP-4    &  6.27 $\pm$ 2.34 \tnote{b}      &               2.72 $\pm$ 0.83\tnote{b}\\
WASP-41   &  8.25 $\pm$ 3.59 \tnote{b}      &               1.71 $\pm$ 0.21\tnote{b}\\ 
WASP-46   &  10.03 $\pm$ 3.51\tnote{b}      &               1.23 $\pm$ 0.20\tnote{b}\\
WASP-5    &  5.84 $\pm$ 1.86 \tnote{b}      &               2.13 $\pm$ 0.52\tnote{b}\\
WASP-50   &  8.57 $\pm$ 2.86 \tnote{b}     &                1.30 $\pm$ 0.15\tnote{b}\\
WASP-57   &  2.60 $\substack{+2.20 \\-1.80}$ \tnote{m}     &  1.90 $\substack{+2.40 \\-1.20 }$ \tnote{m}\\ 
WASP-64   &  8.94 $\substack{+3.15 \\-2.55}$ \tnote{n}     &  1.71 $\substack{+1.18 \\-0.59 }$ \tnote{n}\\
WASP-65   &  8.92 $\substack{+1.87 \\-1.97}$ \tnote{n}     &  1.51  $\substack{+0.50 \\- 0.33}$ \tnote{n}\\
WASP-69   &  15.20 $\pm$ 1.55 \tnote{b}               &  2.09 $\pm$ 0.12 \tnote{b}\\
WASP-70A  &  4.68 $\substack{+3.47 \\-1.31}$ \tnote{n}     &  8.29 $\substack{+5.62 \\-2.89}$ \tnote{n}\\
WASP-71   &  3.21 $\substack{+0.38 \\-0.74}$ \tnote{n}     &  1.64 $\substack{+0.58 \\-0.35}$ \tnote{n}\\
WASP-77A  &  7.57 $\pm$ 2.53 \tnote{b}                &    1.35 $\pm$ 0.18\tnote{b}\\
WASP-84   &  1.89 $\pm$ 1.61 \tnote{b}                &    0.99 $\pm$ 0.10\tnote{b}\\
WASP-85A  &  2.09 $\pm$ 1.37 \tnote{b}                &    1.50 $\pm$ 0.33\tnote{b}\\
WASP-89   &  12.07 $\pm$ 3.11 \tnote{b}               &    1.88 $\pm$ 0.18 \tnote{b}\\
\noalign{\smallskip}\hline
\end{tabular}
\label{tab:ageliterature}
 \begin{tablenotes}
      \small
      \item References: a. \cite{Bonfanti2016}; b. \cite{Maxted2015};
      c. \cite{Bonomo2017}; d.\cite{Mallonn2015}; e.\cite{Bakos2011}; f. \cite{Johnson2016}; g. \cite{Santerne2016}; h. \cite{Smith2014}; i. \cite{Mocnik2017}; j. \cite{Barros2016}; k. \cite{Spake2016}; l. \cite{Demangeon2018}; m. \cite{Faedi2013}; n. \cite{Brown2014}
    \end{tablenotes}
\end{threeparttable}
\end{table}

For Ariel to achieve one of its main goal of shedding light on the formation and evolution of planetary systems, a robust determination of the stellar age 
is crucial, as it provides a temporal dimension of the problem. Furthermore, this estimate needs to be homogeneous over the whole Ariel Reference sample, to allow for both statistical studies, and chemical and orbital architecture comparison.

Even if they applied a uniform technique for their analysis, the two studies reported above did not use a homogeneous set of stellar parameters as input.
In order to perform a statistical analysis, we are in the process of estimating the ages of Ariel stars via different methods, both model-based and empirical, by using our homogeneous set of parameters as input (see Sec.~\ref{sec:parameters}).
Precisely, we will be using stellar models \cite{Bossini2020}, spectral analysis, and gyrochronology (see Sec.~\ref{sec:indexes}). This specific study is expected to be incremental, and unfold over the time  
with the continuous collection of new data (see Sec. \ref{sec:campaign}) and with the revision of the existing one, in light of an homogeneous framework.
The comparison on a relatively large sample of stars will enable us to unveil the possible underlying causes of the methods discrepancies, and to identify a technique that allows for consistent results. The continuation of this work will be presented in upcoming papers.

\section{Ground based observational campaign}
\label{sec:campaign}

Ariel Tier 1 Reference Sample consists today of 487 known planetary systems, and it will be incremented in the upcoming years before the launch \cite{Edwards2019}.  The range of Gaia $G$ magnitude, as retrieved from NASA's Exoplanet Archive\footnote{https://exoplanetarchive.ipac.caltech.edu/}, covers 5.2 $< G <$ 16.6, with median $G$ = 12.\\
In such a sample not all the stars have yet been observed and/or characterised with the level of precision needed to provide both precise stellar parameters (\teff, \logg, and \feh) and elemental abundances. Consequently, we have begun an observational campaign in both northern and southern hemisphere to obtain spectra with S/N $>$ 100 in the visible spectral band, for those Ariel candidates lacking such a precise and homogeneous characterisation, to measure intrinsic uncertainties of the order of $\sim$ 80 K in \teff,  $\sim$ 0.10 dex in \logg, and $\sim$ 0.05 dex in elemental abundances.
A higher S/N ($>$ 150) would allow us to decrease even more the uncertainties: a spectrum with S/N $\sim$ 300 would for instance allow us to measure the \teff with intrinsic precision of $\sigma_T$ $\sim$ 40 K.

The combination between high resolution, large telescope collecting area, and accessibility made UVES@VLT and HARPS-N@TNG our best candidates for a spectroscopic campaign. Four proposals have been accepted for this year 2020 (two proposals of 30 and 22.20 hours with UVES, in the southern hemisphere, and two proposals of 9 hours each with HARPS-N, in the northern hemisphere). Unfortunately, due to the world-wide closure of the observatories caused by the Covid-19 pandemic in the first half of the year, observations of one of our UVES proposal have been delayed to the first semester of 2021.
Ariel stars observational campaign will further continue in the future and the outcome will be made available to the community.

\section{Conclusions}
\label{sec:conclusions}
We presented here a structured approach to the study of the transiting host-stars belonging to the Tier 1 of the Ariel Reference Sample \cite{Edwards2019}. Unveiling the true nature of planetary systems, their formation and evolution is only possible if both planets and their host stars are characterised in an homogeneous and self-consistent way. 

In this manuscript we used publicly available high-resolution spectra and photometric light curves to determine elemental abundances (C, N, Na, Mg, Al, Si), and stellar activity indices ($S$ and \logRHK) for a sub sample of Ariel stars, by using for input uniform stellar parameters reported in the \sweet\ catalogue \cite{SWEET-Cat}. 

Elemental abundances were determined via equivalent widths by two different approaches: 
 using ARES and MOOG or with the Fast Automatic MOOG Analysis.
Our preliminary results show average errors of 0.06 dex for Na and Si, and 0.07 dex for Al and Mg. Such abundances are found to follow the expected Galactic chemical evolution trends. 
The abundances for C and N were estimated through spectral synthesis. For carbon in particular, we compared abundances obtained from both molecular bands and atomic lines (i.e. carbon C~I measured through EWs), to the literature, finding a good agreement at sub-solar metallicity.
A specific work on homogeneous stellar abundances for the Ariel stellar candidates is in preparation.

We quantified CaII H\&K line core emission by measuring the Mount Wilson $S$ activity index on HARPS and HARPS-N spectra, and homogeneously converted this latter to \logRHK whenever a suitable colour index was available. On a sample of 47 stars for which the \logRHK was measured, we report an activity index {between $-6.1$ and $-4.0$,} 
with a median value of -5.1.  
This was expected, as most of our sample was selected from radial velocity campaigns, which privilege inactive stellar hosts. We confirmed the presence of a low- and a high-activity population, as highlighted by \cite{Fossati2015}, and recovered the correlation between \logRHK and planetary surface gravity reported by \cite{Hartman2010,Figueira2014,Fossati2015,Fossati2017}. 

The analysis of the available photometric and spectroscopic time series for a subset of targets in the \sweet\ catalogue, allowed us to derive a first estimate of their rotation periods. We performed a periodogram analysis of both the light curves and the time series of spectroscopic activity indices (in particular, CCF asymmetry indicators, besides radial velocities after the removal of the Keplerian signal of the known transiting planet) and compared the resulting periods with the empirical values obtained from the \logRHK and the relation in \cite{mamajek2008}.  Spectroscopic indices resulted to be more suitable to derive the stellar rotation period, being the line profiles more sensitive to the stellar activity with respect to the available photometric data. 
Larger spectroscopic datasets will be useful to improve the significance of our findings and to extend our analysis to a larger number of targets in the Ariel stellar sample.

In parallel, in \cite{Brucalassi2020}, we performed a benchmark study between three different spectroscopic techniques to determine stellar parameters in order to identify the most robust method(s) to uniformly determine \teff, \feh\ and \logg\ all along the Ariel stellar parameter space. The uncertainty on these parameters directly reflects on the limb-darkening estimation, phenomenon that is important to take into account to precisely derive a planetary spectrum.
We consequently investigated the uncertainty in the planetary transit depth as a function of the stellar temperature precision and Ariel wavelength range, from visible to infrared.
We found that a 80 K (40 K) error can produce a maximum bias of $\sim$ 25 ppm (12 ppm) on the spectrum of a planet like HD-209458 b transiting its star and observed by Ariel.

Finally, we highlighted the large discrepancies between literature ages obtained with different methods and heterogeneous sets of input parameters. We showed that, for a sub-sample of Ariel stars, the isochrone age estimate is larger than the gyrochronology estimate, an effect already discussed in the literature and that could be caused by tidal interaction between the planet and the host star \cite{Brown2014,Maxted2015}. In upcoming works we plan on revising such a relation with the use of ages determined through both stellar models \cite{Bossini2020} and gyrochronology, whose input will be the homogeneous set of Ariel atmospheric parameters.

Currently the Ariel reference sample amounts to 487 known planetary systems, but such number is destined to increase with new planetary discoveries and detailed observations by missions like Gaia \cite{Perryman2014}, TESS \cite{Ricker2014}, CHEOPS \cite{Broeg2013} and PLATO \cite{PLATO}, along with ground-based surveys. 
The work presented here is hence part of a long-term incremental analysis that will be carried out over the years till the launch of the Ariel mission, which will be of benefit for Ariel and for the whole exoplanet community.

\begin{ThreePartTable}
\begin{longtable}{lcccccccc}
    \caption{Stars for which the $S$ index was evaluated (typical uncertainties are $< 0.01$). The $\log R'_\mathrm{HK}$ is
    indicated for targets with magnitudes in the suitable  range for the conversion (see Section \ref{sec:logRhk}). Reported colours are corrected for extinction as described in the text. We also present the stellar ages derived with \cite{mamajek2008}'s empirical relation, when the \logRHK values are within the appropriate range of the calibration. S/N is the average signal-to-noise ratio measured among the set of spectra for each target. We then list the number of available spectra per object, and whether the $S$ index shows a $>3\sigma$ time variation.}
    \label{tab:sample} \\
    \hline 
    Star Name &  $(B-V)_0$ & $(V-K)_0$& $S$ & \logRHK & Age [MH08] (Gyr) & $\langle \mathrm{S/N} \rangle$ & N. spectra & $\Delta>3\sigma$\\
    \hline
    \endfirsthead
    Star Name &  $(B-V)_0$ & $(V-K)_0$& $S$ & \logRHK & Age [MH08] (Gyr) & $\langle \mathrm{S/N} \rangle$ & N. spectra & $\Delta>3\sigma$\\
    \hline
    ... \\
    \endhead
    ... \\ \hline
    \endfoot
    \hline \hline
    \endlastfoot
      GJ 9827&1.47&2.90 & 0.25   $\pm$   0.01&	      -5.63$\pm$	   0.29 &$^{(\ast)}$  &  5.47 &    7	&	    no \\
    HAT-P-1&0.92&0.93 & 0.15   $\pm$   0.00&	      -5.11$\pm$	   0.07 &$^{(\ast)}$  & 12.82 &   16	&	    no \\
   HAT-P-14&0.40&1.09 & 0.16   $\pm$   0.00& $^{(\dagger)}$ &	  --  & 20.45 &   14	&	    no \\
   HAT-P-16&0.45&1.29 & 0.17   $\pm$   0.00&	      -5.19$\pm$	    0.1 &$^{(\ast)}$   &  9.86 &   21	&	    no \\
   HAT-P-17&0.94&1.82 & 0.16   $\pm$   0.00&	      -5.08$\pm$	   0.07 &$^{(\ast)}$   &  7.93 &   30	&	    no \\
   HAT-P-20&1.23&2.74 & 1.22   $\pm$   0.01&	      -4.58$\pm$	   0.09 &$^{(\ast)}$   &  1.60 &   19	&	    no \\
   HAT-P-21&1.09&1.32 & 0.31   $\pm$   0.00&	      -4.96$\pm$	   0.07 &$^{(\ast)}$   &  4.77 &   24	&	    no \\
   HAT-P-22&0.86&1.91 & 0.16   $\pm$   0.00&	      -5.01$\pm$	   0.06 &	6.79 $\pm$	 1.19  & 13.34 &   33	&	    no \\
   HAT-P-24&0.43&1.15 & 0.15   $\pm$   0.00& $^{(\dagger)}$ &	  -- &  6.79 &   17   & 	  no \\
   HAT-P-29&0.40&1.34 & 0.16   $\pm$   0.00& $^{(\dagger)}$ &	  -- &  3.73 &   32   & 	  no \\
    HAT-P-3&0.66&2.40 & 0.22   $\pm$   0.00&	      -4.73$\pm$	   0.06 &	2.31 $\pm$	  0.4  &  2.97 &   13	&	    no \\
   HAT-P-30&0.59&1.18 & 0.14   $\pm$   0.00&	      -5.21$\pm$	   0.09 &$^{(\ast)}$  & 14.90 &   15	&	    no \\
   HAT-P-31&0.29&1.42 & 0.13   $\pm$   0.00& $^{(\dagger)}$ &	  --  &  5.66 &   14	&	    no \\
    HAT-P-4&0.70&1.31 & 0.14   $\pm$   0.00&	      -5.08$\pm$	   0.07 &	8.09 $\pm$	 1.42  &  6.41 &   29	&	    no \\
    HAT-P-6&0.32&0.91 & 0.19   $\pm$   0.00& $^{(\dagger)}$ &	      --   & 16.96 &   18   &	       yes \\
    HAT-P-8&0.37&1.29 & 0.15   $\pm$   0.00& $^{(\dagger)}$ &	      --   & 19.64 &   14   &		no \\
    HATS-27&0.41&1.06 & 0.17   $\pm$   0.01& $^{(\dagger)}$ &	      --   &  2.21 &   14   &		no \\
    HATS-35&0.65&1.30 & 0.16   $\pm$   0.00&	      -4.99$\pm$	   0.06 &	6.41 $\pm$	 1.12  & 16.90 &  128	&	    no \\
    HATS-39&0.45&1.15 & 0.17   $\pm$   0.00&	      -5.25$\pm$	   0.23 &$^{(\ast)}$ &  2.86 &   16   & 	  no \\
    HATS-41&0.47&1.09 & 0.16   $\pm$   0.01&	      -5.35$\pm$	    0.5 &$^{(\ast)}$ &  2.40 &    5   & 	  no \\
    HATS-64&0.38&0.92 & 0.07   $\pm$   0.00& $^{(\dagger)}$ &	  --   &  1.82 &   16	&	   yes \\
  HD 106315&0.45&1.08 & 0.16   $\pm$   0.00&	      -5.35$\pm$	   0.13 &$^{(\ast)}$   & 42.37 &   94	&	    no \\
   HD 17156&0.62& -- & 0.14   $\pm$   0.00&	      -5.16$\pm$	   0.08 &$^{(\ast)}$   & 35.28 &   26	&	    no \\
    HD 3167&0.82&1.86 & 0.16   $\pm$   0.00&	      -5.02$\pm$	   0.06 &	6.98 $\pm$	 1.22  & 17.82 &   50	&	    no \\
    HD89345&0.77&1.65 & 0.14   $\pm$   0.00&	      -5.09$\pm$	   0.07 &	8.27 $\pm$	 1.45  & 21.23 &   13	&	    no \\
  HIP 41378&0.49&1.18 & 0.15   $\pm$   0.00&	      -5.39$\pm$	   0.14 &$^{(\ast)}$   & 23.13 &  205	&	    no \\
     K2-139&0.69&1.75 & 0.34   $\pm$   0.01&	      -4.46$\pm$	   0.08 &	0.46 $\pm$	 0.08  &  3.18 &    6	&	    no \\
     K2-141&0.69&3.04 & 0.92   $\pm$   0.02&	      -3.95$\pm$	   0.03 &$^{(\ast)}$   &  3.34 &   29	&	    no \\
     K2-232&0.59&1.37 & 0.14   $\pm$   0.00&	      -5.28$\pm$	   0.16 &$^{(\ast)}$   &  5.26 &    8	&	    no \\
     K2-237&0.40&0.85 & 0.20   $\pm$   0.00& $^{(\dagger)}$ &	  --   &  6.24 &    7	&	    no \\
     K2-261&0.91&1.73 & 0.13   $\pm$   0.00&	      -5.19$\pm$	   0.09 &$^{(\ast)}$   &  8.37 &    7	&	    no \\
     K2-266&1.19&2.90 & 0.28   $\pm$   0.01&	      -5.16$\pm$	   0.11 &$^{(\ast)}$   &  3.57 &   15	&	    no \\
     K2-280 & 0.66 & 1.59 &  0.12  $\pm$ 0.00&  $^{(\bullet)}$ &           --        &  3.66 &    20 &               no\\
     K2-287&0.62&1.91 & 0.28   $\pm$   0.00&	      -4.56$\pm$	   0.04 &	 0.9 $\pm$	 0.16  &  1.16 &  421	&	    no \\
       K2-3&1.35&3.60 & 0.76   $\pm$   0.02&	      -4.95$\pm$	   0.16 &$^{(\ast)}$   &  1.06 &  138	&	    no \\
     KELT-6&0.49&1.23 & 0.15   $\pm$   0.00&	      -5.38$\pm$	   0.14 &$^{(\ast)}$   & 14.92 &   42	&	    no \\
     TrES-1&0.92&1.56 & 0.22   $\pm$   0.00&	      -4.89$\pm$	   0.06 &$^{(\ast)}$   &  3.31 &   19	&	    no \\
     TrES-2&0.59&1.37 & 0.16   $\pm$   0.00&	      -5.08$\pm$	   0.09 &	8.09 $\pm$	 1.42  &  6.67 &   15	&	    no \\
     TrES-4&0.50&1.19 & 0.14   $\pm$   0.00&	      -5.66$\pm$	   0.32 &$^{(\ast)}$   &  6.52 &   19	&	    no \\
     WASP-1&0.70&0.90 & 0.15   $\pm$   0.00&	      -5.07$\pm$	   0.08 &	7.91 $\pm$	 1.38  &  6.12 &   13	&	    no \\
    WASP-12 & 0.51 & 1.21 &  0.10  $\pm$     0.00 &  $^{(\bullet)}$       &     --        &  5.77   &  23 &               no\\
   WASP-121&0.47&1.08 & 0.18   $\pm$   0.00&	      -5.05$\pm$	   0.07 &$^{(\ast)}$  &  9.23 &  140   &	   no \\
   WASP-127&0.63&1.52 & 0.14   $\pm$   0.00&	      -5.21$\pm$	   0.09 &$^{(\ast)}$  &  7.78 &  107   &	   no \\
    WASP-13&0.78&1.28 & 0.14   $\pm$   0.00&	      -5.09$\pm$	   0.07 &	8.27 $\pm$	 1.45  & 11.35 &   19	&	    no \\
    WASP-14&0.43&1.10 & 0.15   $\pm$   0.00& $^{(\dagger)}$ &	  --  & 20.43 &   15	&	    no \\
   WASP-157&0.03&2.09 & 0.19   $\pm$   0.01&	      -6.11$\pm$	   1.39 &$^{(\ast)}$   &  2.92 &    4	&	    no \\
    WASP-21&0.51&1.39 & 0.16   $\pm$   0.00&	      -5.21$\pm$	   0.14 &$^{(\ast)}$   &  6.25 &   10	&	    no \\
    WASP-24&0.75&0.99 & 0.15   $\pm$   0.00&	      -5.02$\pm$	   0.07 &	6.98 $\pm$	 1.22  &  5.64 &   14	&	    no \\
     WASP-3&0.42&1.21 & 0.14   $\pm$   0.00& $^{(\dagger)}$ &	      --   & 21.35 &	2   &	       no \\
    WASP-31&0.32&0.92 & 0.15   $\pm$   0.00& $^{(\dagger)}$ &	      --   &  4.26 &   17   &		no \\
    WASP-35&0.60&1.37 & 0.16   $\pm$   0.00&	      -5.08$\pm$	   0.08 &	8.09 $\pm$	 1.42  &  7.98 &   13	&	    no \\
    WASP-38&0.46&1.34 & 0.15   $\pm$   0.00&	       -5.5$\pm$	   0.18 &$^{(\ast)}$   & 25.66 &   18	&	    no \\
    WASP-48&0.63&1.19 & 0.17   $\pm$   0.00&	      -4.98$\pm$	   0.07 &	6.22 $\pm$	 1.09  &  6.20 &   12	&	    no \\
    WASP-49&0.57&1.58 & 0.15   $\pm$   0.00&	      -5.14$\pm$	   0.13 &$^{(\ast)}$  &  4.92 &    9   &	   no \\
    WASP-50&0.98&1.44 & 0.33   $\pm$   0.00&	      -4.77$\pm$	   0.05 &$^{(\ast)}$  &  3.67 &   12   &	   no \\
    WASP-54&0.59&1.34 & 0.14   $\pm$   0.00&	      -5.24$\pm$	    0.1 &$^{(\ast)}$  & 12.90 &   16   &	   no \\
    WASP-84&0.81&1.96 & 0.45   $\pm$   0.01&	      -4.42$\pm$	   0.04 &	0.35 $\pm$	 0.06  &  3.33 &   30	&	    no \\
   WASP-85A&0.74&1.66 & 0.25   $\pm$   0.01&	      -4.67$\pm$	   0.05 &	 1.7 $\pm$	  0.3  &  3.74 &   13	&	    no \\
       XO-1&0.61&1.68 & 0.18   $\pm$   0.00&	      -4.91$\pm$	   0.06 &	4.95 $\pm$	 0.87  &  5.33 &   15	&	    no \\
     XO-2-N&0.85&1.81 & 0.18   $\pm$   0.00&	      -4.93$\pm$	   0.05 &	5.31 $\pm$	 0.93  &  4.38 &   57	&	    no \\
     XO-2-S&0.83&1.80 & 0.16   $\pm$   0.00&	      -5.01$\pm$	   0.06 &	6.79 $\pm$	 1.19  &  4.80 &  130	&	    no \\
       XO-3&0.35&0.88 & 0.23   $\pm$   0.00& $^{(\dagger)}$ &	      -- & 19.74 &   21   &	     yes \\
       XO-4&0.31&1.33 & 0.15   $\pm$   0.00& $^{(\dagger)}$ &	      -- & 13.76 &   17   &	      no \\
    \hline 
\end{longtable}
    \begin{tablenotes}
\item [$^{(\dagger)}$] {\footnotesize  Out-of-range magnitudes for the $S$-to-\logRHK\ conversion.} 
\item [$^{(\bullet)}$] {\footnotesize While the $(B-V)_0$ value is within the range for the $S$-to-\logRHK\ conversion, the resulting non-logarithmic $R^{'}_\mathrm{HK}$ is negative; following \cite{noyes1984}'s calibration, this happens for  $S \lesssim 0.1$ and $(B-V)_0 \lesssim 0.7$ mag.}
\item [$^{(\ast)}$] {\footnotesize{\logRHK~out of \cite{mamajek2008}'s calibration range for age calculation.}} 
\end{tablenotes}
\end{ThreePartTable}

\begin{acknowledgements}
We thank R. Lallement for extinction estimates based on the 3D dust map of Vergely et al., 2021, in advance of publication. We thank I. Pagano for the useful discussions. C.D. acknowledges financial support from the State Agency for Research of the Spanish MCIU through the
\textquotedblleft Center of Excellence Severo Ochoa\textquotedblright\ award to the Instituto de Astrof\'isica de Andaluc\'ia (SEV-2017-0709), and the Group project Ref. PID2019-110689RB-I00/AEI/10.13039/501100011033; S.B.,  G.M. and D.T. acknowledge the support of the ARIEL ASI-INAF agreement n. 2018-22-HH.0; D.T. acknowledges the support of the Italian National Institute of Astrophysics (INAF) through the INAF Main Stream project ``Ariel and the astrochemical link between circumstellar discs and planets" (CUP: C54I19000700005); D.B. acknowledges support by FCT through the research grants UIDB/04434/2020, UIDP/04434/2020 and PTDC/FIS-AST/30389/2017, and by
FEDER - Fundo Europeu de Desenvolvimento Regional through COMPETE2020 -
Programa Operacional Competitividade e Internacionaliza\c{c}\~{a}o (grant:
POCI-01-0145-FEDER-030389).
G.B. acknowledges support from CHEOPS ASI-INAF agreement n. 2019-29-HH.0; S.S. was supported by United Kingdom Space Agency (UKSA) grant: ST/S002456/1.; 
E.D.M., V.A. and S.G.S. acknowledge the support from Funda\c{c}\~ao para a Ci\^encia e a Tecnologia (FCT) through national funds and from FEDER through COMPETE2020 by the following grants: UID/FIS/04434/2019, UIDB/04434/2020 and UIDP/04434/2020; PTDC/FIS-AST/32113/2017 and POCI-01-0145-FEDER-032113; PTDC/FIS-AST/28953/2017 and POCI-
01-0145-FEDER-028953; 
E.D.M., V.A., S.G.S. also acknowledge the support from FCT through Investigador FCT contracts IF/00849/2015/CP1273/CT0003, IF/00650/2015/CP1273/CT0001, IF/00028/2014/CP1215/CT0002; 
T.C. acknowledges support from the European Union's Horizon 2020 research and innovation programme under the Marie Sk\l{}odowska-Curie grant agreement No.~792848 (PULSATION). T.C. is supported by Funda\c c\~ao para a Ci\^encia e a Tecnologia (FCT) in the form of a work contract (CEECIND/00476/2018); 
G.M. was supported by the LabEx P2IO, the French ANR contract 05-BLANNT09-573739; 
P.K. acknowledges the support of INTER-TRANSFER grant LTT20015.
The work on the stellar activity indices was based on data from the following GAPS observation programs:
A26TAC\_70 and A27TAC\_52, and from observations collected with HARPS/ESO (Chile) facilities (ESO programs IDs 0100.C-0474, 0100.C-0487, 0100.C-0750, 0100.C-0750, 0100.C-0808, 072.C-0488, 074.C-0364, 082.C-0718, 087.C-0831, 089.C-0050, 089.C-0151, 090.C-0540, 091.C-0184, 091.C-0184, 093.C-0417, 094.C-0428, 095.C-0367, 095.C-0718, 096.C-0183, 096.C-0417, 096.C-0417, 096.C-0762, 097.C-0571, 097.C-0571, 097.C-0948, 098.C-0292, 098.C-0292, 098.C-0304, 098.C-0304, 098.C-0820, 098.C-0860, 099.C-0303, 099.C-0374, 099.C-0491, 099.C-0491, 183.C-0437, 183.C-0972, 191.C-0873, 198.C-0169, 198.C-0838, 60.A-9036, 60.A-9700).
This research has made use of the NASA Exoplanet Archive, which is operated by the California Institute of Technology, under contract with the National Aeronautics and Space Administration under the Exoplanet Exploration Program. This publication makes use of data products from the Two Micron All Sky Survey, which is a joint project of the University of Massachusetts and the Infrared Processing and Analysis Center/California Institute of Technology, funded by the National Aeronautics and Space Administration and the National Science Foundation. This research has made use of the SIMBAD database, operated at CDS, Strasbourg, France.
\end{acknowledgements}

\bibliographystyle{spmpsci}      
\bibliography{Danielski_SC_R2_arxiv}   

\begin{thebibliography}{100}
\providecommand{\url}[1]{{#1}}
\providecommand{\urlprefix}{URL }
\expandafter\ifx\csname urlstyle\endcsname\relax
  \providecommand{\doi}[1]{DOI~\discretionary{}{}{}#1}\else
  \providecommand{\doi}{DOI~\discretionary{}{}{}\begingroup
  \urlstyle{rm}\Url}\fi

\bibitem{Adibekyan2017}
{Adibekyan}, V., {Gon{\c{c}}alves da Silva}, H.M., {Sousa}, S.G., {Santos},
  N.C., {Delgado Mena}, E., {Hakobyan}, A.A.: {Mg/Si Mineralogical Ratio of
  Low-Mass Planet Hosts. Correction for the NLTE Effects}.
\newblock Astrophysics \textbf{60}(3), 325--332 (2017).
\newblock \doi{10.1007/s10511-017-9486-5}

\bibitem{Adibekyan2015}
{Adibekyan}, V., {Santos}, N.C., {Figueira}, P., {Dorn}, C., {Sousa}, S.G.,
  {Delgado-Mena}, E., {Israelian}, G., {Hakobyan}, A.A., {Mordasini}, C.: {From
  stellar to planetary composition: Galactic chemical evolution of Mg/Si
  mineralogical ratio}.
\newblock \aap \textbf{581}, L2 (2015).
\newblock \doi{10.1051/0004-6361/201527059}

\bibitem{Adibekyan2012b}
{Adibekyan}, V.Z., {Delgado Mena}, E., {Sousa}, S.G., {Santos}, N.C.,
  {Israelian}, G., {Gonz{\'a}lez Hern{\'a}ndez}, J.I., {Mayor}, M., {Hakobyan},
  A.A.: {Exploring the {\ensuremath{\alpha}}-enhancement of metal-poor
  planet-hosting stars. The Kepler and HARPS samples}.
\newblock \aap \textbf{547}, A36 (2012).
\newblock \doi{10.1051/0004-6361/201220167}

\bibitem{Adibekyan2013}
{Adibekyan}, V.Z., {Figueira}, P., {Santos}, N.C., {Mortier}, A., {Mordasini},
  C., {Delgado Mena}, E., {Sousa}, S.G., {Correia}, A.C.M., {Israelian}, G.,
  {Oshagh}, M.: {Orbital and physical properties of planets and their hosts:
  new insights on planet formation and evolution}.
\newblock \aap \textbf{560}, A51 (2013).
\newblock \doi{10.1051/0004-6361/201322551}

\bibitem{Adibekyan2012}
{Adibekyan}, V.Z., {Sousa}, S.G., {Santos}, N.C., {Delgado Mena}, E.,
  {Gonz{\'a}lez Hern{\'a}ndez}, J.I., {Israelian}, G., {Mayor}, M.,
  {Khachatryan}, G.: {Chemical abundances of 1111 FGK stars from the HARPS GTO
  planet search program. Galactic stellar populations and planets}.
\newblock \aap \textbf{545}, A32 (2012).
\newblock \doi{10.1051/0004-6361/201219401}

\bibitem{2004A&A...414.1139A}
{Aigrain}, S., {Favata}, F., {Gilmore}, G.: {Characterising stellar
  micro-variability for planetary transit searches}.
\newblock \aap \textbf{414}, 1139--1152 (2004).
\newblock \doi{10.1051/0004-6361:20034039}

\bibitem{AlvarezPlez1998}
{Alvarez}, R., {Plez}, B.: {Near-infrared narrow-band photometry of M-giant and
  Mira stars: models meet observations}.
\newblock \aap \textbf{330}, 1109--1119 (1998)

\bibitem{amarsi19}
{Amarsi}, A.M., {Barklem}, P.S., {Collet}, R., {Grevesse}, N., {Asplund}, M.:
  {3D non-LTE line formation of neutral carbon in the Sun}.
\newblock \aap \textbf{624}, A111 (2019).
\newblock \doi{10.1051/0004-6361/201833603}

\bibitem{Amarsi2020}
{Amarsi}, A.M., {Lind}, K., {Osorio}, Y., {Nordlander}, T., {Bergemann}, M.,
  {Reggiani}, H., {Wang}, E.X., {Buder}, S., {Asplund}, M., {Barklem}, P.S.,
  {Wehrhahn}, A., {Sk{\'u}lad{\'o}ttir}, {\'A}., {Kobayashi}, C., {Karakas},
  A.I., {Gao}, X.D., {Bland-Hawthorn}, J., {de Silva}, G.M., {Kos}, J.,
  {Lewis}, G.F., {Martell}, S.L., {Sharma}, S., {Simpson}, J.D., {Zucker},
  D.B., {{\v{C}}otar}, K., {Horner}, J., {Galah Collaboration}: {The GALAH
  Survey: non-LTE departure coefficients for large spectroscopic surveys}.
\newblock \aap \textbf{642}, A62 (2020).
\newblock \doi{10.1051/0004-6361/202038650}

\bibitem{Andreasen2017}
{Andreasen}, D.T., {Sousa}, S.G., {Tsantaki}, M., {Teixeira}, G.D.C.,
  {Mortier}, A., {Santos}, N.C., {Su{\'a}rez-Andr{\'e}s}, L., {Delgado-Mena},
  E., {Ferreira}, A.C.S.: {SWEET-Cat update and FASMA. A new minimization
  procedure for stellar parameters using high-quality spectra}.
\newblock \aap \textbf{600}, A69 (2017).
\newblock \doi{10.1051/0004-6361/201629967}

\bibitem{astudillo-defru}
{Astudillo-Defru}, N., {Delfosse}, X., {Bonfils}, X., {Forveille}, T., {Lovis},
  C., {Rameau}, J.: {Magnetic activity in the HARPS M dwarf sample. The
  rotation-activity relationship for very low-mass stars through R'(HK)}.
\newblock Astronomy \& Astrophysics \textbf{600}, A13 (2017).
\newblock \doi{10.1051/0004-6361/201527078}

\bibitem{Bakos2011}
{Bakos}, G.{\'A}., {Hartman}, J., {Torres}, G., {Latham}, D.W., {Kov{\'a}cs},
  G., {Noyes}, R.W., {Fischer}, D.A., {Johnson}, J.A., {Marcy}, G.W., {Howard},
  A.W., {Kipping}, D., {Esquerdo}, G.A., {Shporer}, A., {B{\'e}ky}, B.,
  {Buchhave}, L.A., {Perumpilly}, G., {Everett}, M., {Sasselov}, D.D.,
  {Stefanik}, R.P., {L{\'a}z{\'a}r}, J., {Papp}, I., {S{\'a}ri}, P.:
  {HAT-P-20b-HAT-P-23b: Four Massive Transiting Extrasolar Planets}.
\newblock \apj \textbf{742}(2), 116 (2011).
\newblock \doi{10.1088/0004-637X/742/2/116}

\bibitem{Ballerini2012}
{Ballerini}, P., {Micela}, G., {Lanza}, A.F., {Pagano}, I.: {Multiwavelength
  flux variations induced by stellar magnetic activity: effects on planetary
  transits}.
\newblock \aap \textbf{539}, A140 (2012).
\newblock \doi{10.1051/0004-6361/201117102}

\bibitem{Barros2016}
{Barros}, S.C.C., {Brown}, D.J.A., {H{\'e}brard}, G., {G{\'o}mez Maqueo Chew},
  Y., {Anderson}, D.R., {Boumis}, P., {Delrez}, L., {Hay}, K.L., {Lam}, K.W.F.,
  {Llama}, J., {Lendl}, M., {McCormac}, J., {Skiff}, B., {Smalley}, B.,
  {Turner}, O., {Vanhuysse}, M., {Armstrong}, D.J., {Boisse}, I., {Bouchy}, F.,
  {Collier Cameron}, A., {Faedi}, F., {Gillon}, M., {Hellier}, C., {Jehin}, E.,
  {Liakos}, A., {Meaburn}, J., {Osborn}, H.P., {Pepe}, F., {Plauchu-Frayn}, I.,
  {Pollacco}, D., {Queloz}, D., {Rey}, J., {Spake}, J., {S{\'e}gransan}, D.,
  {Triaud}, A.H.M., {Udry}, S., {Walker}, S.R., {Watson}, C.A., {West}, R.G.,
  {Wheatley}, P.J.: {WASP-113b and WASP-114b, two inflated hot Jupiters with
  contrasting densities}.
\newblock \aap \textbf{593}, A113 (2016).
\newblock \doi{10.1051/0004-6361/201526517}

\bibitem{Bensby2014}
{Bensby}, T., {Feltzing}, S., {Oey}, M.S.: {Exploring the Milky Way stellar
  disk. A detailed elemental abundance study of 714 F and G dwarf stars in the
  solar neighbourhood}.
\newblock \aap \textbf{562}, A71 (2014).
\newblock \doi{10.1051/0004-6361/201322631}

\bibitem{Bergemann2013}
{Bergemann}, M., {Kudritzki}, R.P., {W{\"u}rl}, M., {Plez}, B., {Davies}, B.,
  {Gazak}, Z.: {Red Supergiant Stars as Cosmic Abundance Probes. II. NLTE
  Effects in J-band Silicon Lines}.
\newblock \apj \textbf{764}(2), 115 (2013).
\newblock \doi{10.1088/0004-637X/764/2/115}

\bibitem{Biazzo2015}
{Biazzo}, K., {Gratton}, R., {Desidera}, S., {Lucatello}, S., {Sozzetti}, A.,
  {Bonomo}, A.S., {Damasso}, M., {Gand olfi}, D., {Affer}, L., {Boccato}, C.,
  {Borsa}, F., {Claudi}, R., {Cosentino}, R., {Covino}, E., {Knapic}, C.,
  {Lanza}, A.F., {Maldonado}, J., {Marzari}, F., {Micela}, G., {Molaro}, P.,
  {Pagano}, I., {Pedani}, M., {Pillitteri}, I., {Piotto}, G., {Poretti}, E.,
  {Rainer}, M., {Santos}, N.C., {Scandariato}, G., {Zanmar Sanchez}, R.: {The
  GAPS programme with HARPS-N at TNG. X. Differential abundances in the XO-2
  planet-hosting binary}.
\newblock \aap \textbf{583}, A135 (2015).
\newblock \doi{10.1051/0004-6361/201526375}

\bibitem{Bonfanti2016}
{Bonfanti}, A., {Ortolani}, S., {Nascimbeni}, V.: {Age consistency between
  exoplanet hosts and field stars}.
\newblock \aap \textbf{585}, A5 (2016).
\newblock \doi{10.1051/0004-6361/201527297}

\bibitem{Bonomo2017}
{Bonomo}, A.S., {Desidera}, S., {Benatti}, S., {Borsa}, F., {Crespi}, S.,
  {Damasso}, M., {Lanza}, A.F., {Sozzetti}, A., {Lodato}, G., {Marzari}, F.,
  {Boccato}, C., {Claudi}, R.U., {Cosentino}, R., {Covino}, E., {Gratton}, R.,
  {Maggio}, A., {Micela}, G., {Molinari}, E., {Pagano}, I., {Piotto}, G.,
  {Poretti}, E., {Smareglia}, R., {Affer}, L., {Biazzo}, K., {Bignamini}, A.,
  {Esposito}, M., {Giacobbe}, P., {H{\'e}brard}, G., {Malavolta}, L.,
  {Maldonado}, J., {Mancini}, L., {Martinez Fiorenzano}, A., {Masiero}, S.,
  {Nascimbeni}, V., {Pedani}, M., {Rainer}, M., {Scand ariato}, G.: {The GAPS
  Programme with HARPS-N at TNG . XIV. Investigating giant planet migration
  history via improved eccentricity and mass determination for 231 transiting
  planets}.
\newblock \aap \textbf{602}, A107 (2017).
\newblock \doi{10.1051/0004-6361/201629882}

\bibitem{bosman+2019}
{Bosman}, A.D., {Cridland}, A.J., {Miguel}, Y.: {Jupiter formed as a pebble
  pile around the N$_{2}$ ice line}.
\newblock \aap \textbf{632}, L11 (2019).
\newblock \doi{10.1051/0004-6361/201936827}

\bibitem{Bossini2020}
{Bossini}, D., {Rodrigues}, T., {Campante}, T., {Sacco}, G.G., {Danielski}, C.,
  {Khan}, S.e.a.: {Homogeneous determination of stellar age in the Ariel
  Reference Sample}.
\newblock in prep.

\bibitem{Broeg2013}
{Broeg}, C., {Fortier}, A., {Ehrenreich}, D., {Alibert}, Y., {Baumjohann}, W.,
  {Benz}, W., {Deleuil}, M., {Gillon}, M., {Ivanov}, A., {Liseau}, R., {Meyer},
  M., {Oloffson}, G., {Pagano}, I., {Piotto}, G., {Pollacco}, D., {Queloz}, D.,
  {Ragazzoni}, R., {Renotte}, E., {Steller}, M., {Thomas}, N.: {CHEOPS: A
  transit photometry mission for ESA's small mission programme}.
\newblock In: European Physical Journal Web of Conferences, \emph{European
  Physical Journal Web of Conferences}, vol.~47, p. 03005 (2013).
\newblock \doi{10.1051/epjconf/20134703005}

\bibitem{Brown2014}
{Brown}, D.J.A.: {Discrepancies between isochrone fitting and gyrochronology
  for exoplanet host stars?}
\newblock \mnras \textbf{442}(2), 1844--1862 (2014).
\newblock \doi{10.1093/mnras/stu950}

\bibitem{Brucalassi2020}
{Brucalassi}, A., {Tsantaki}, M., {Magrini}, L., {Sousa}, S., {Danielski}, C.,
  {Biazzo}, K., {Casali}, G., {Van der Swaelmen}, M., {Rainer}, M.,
  {Adibekyan}, V., {Delgado-Mena}, E., {Sanna}, N.: {Determination of stellar
  parameters for Ariel targets: a comparison analysis between different
  spectroscopic methods}.
\newblock Experimental Astronomy  (2021).
\newblock \doi{10.1007/s10686-020-09695-4}

\bibitem{Buchhave2014}
{Buchhave}, L.A., {Bizzarro}, M., {Latham}, D.W., {Sasselov}, D., {Cochran},
  W.D., {Endl}, M., {Isaacson}, H., {Juncher}, D., {Marcy}, G.W.: {Three
  regimes of extrasolar planet radius inferred from host star metallicities}.
\newblock \nat \textbf{509}(7502), 593--595 (2014).
\newblock \doi{10.1038/nature13254}

\bibitem{Caffau2008}
{Caffau}, E., {Ludwig}, H.G., {Steffen}, M., {Ayres}, T.R., {Bonifacio}, P.,
  {Cayrel}, R., {Freytag}, B., {Plez}, B.: {The photospheric solar oxygen
  project. I. Abundance analysis of atomic lines and influence of atmospheric
  models}.
\newblock \aap \textbf{488}(3), 1031--1046 (2008).
\newblock \doi{10.1051/0004-6361:200809885}

\bibitem{CantatGaudin2014}
{Cantat-Gaudin}, T., {Donati}, P., {Pancino}, E., {Bragaglia}, A., {Vallenari},
  A., {Friel}, E.D., {Sordo}, R., {Jacobson}, H.R., {Magrini}, L.: {DOOp, an
  automated wrapper for DAOSPEC}.
\newblock \aap \textbf{562}, A10 (2014).
\newblock \doi{10.1051/0004-6361/201322533}

\bibitem{Cardelli1989}
{Cardelli}, J.A., {Clayton}, G.C., {Mathis}, J.S.: {The relationship between
  infrared, optical, and ultraviolet extinction}.
\newblock \apj \textbf{345}, 245--256 (1989).
\newblock \doi{10.1086/167900}

\bibitem{2018A&A...613A..50C}
{Carleo}, I., {Benatti}, S., {Lanza}, A.F., {Gratton}, R., {Claudi}, R.,
  {Desidera}, S., {Mace}, G.N., {Messina}, S., {Sanna}, N., {Sissa}, E.,
  {Ghedina}, A., {Ghinassi}, F., {Guerra}, J., {Harutyunyan}, A., {Micela}, G.,
  {Molinari}, E., {Oliva}, E., {Tozzi}, A., {Baffa}, C., {Baruffolo}, A.,
  {Bignamini}, A., {Buchschacher}, N., {Cecconi}, M., {Cosentino}, R., {Endl},
  M., {Falcini}, G., {Fantinel}, D., {Fini}, L., {Fugazza}, D., {Galli}, A.,
  {Giani}, E., {Gonz{\'a}lez}, C., {Gonz{\'a}lez-{\'A}lvarez}, E.,
  {Gonz{\'a}lez}, M., {Hernandez}, N., {Hernandez Diaz}, M., {Iuzzolino}, M.,
  {Kaplan}, K.F., {Kidder}, B.T., {Lodi}, M., {Malavolta}, L., {Maldonado}, J.,
  {Origlia}, L., {Perez Ventura}, H., {Puglisi}, A., {Rainer}, M., {Riverol},
  L., {Riverol}, C., {San Juan}, J., {Scuderi}, S., {Seemann}, U., {Sokal},
  K.R., {Sozzetti}, A., {Sozzi}, M.: {Multi-band high resolution spectroscopy
  rules out the hot Jupiter BD+20 1790b. First data from the GIARPS
  Commissioning}.
\newblock \aap \textbf{613}, A50 (2018).
\newblock \doi{10.1051/0004-6361/201732350}

\bibitem{carter-bond+2012}
{Carter-Bond}, J.C., {O'Brien}, D.P., {Raymond}, S.N.: {The Compositional
  Diversity of Extrasolar Terrestrial Planets. II. Migration Simulations}.
\newblock \apj \textbf{760}(1), 44 (2012).
\newblock \doi{10.1088/0004-637X/760/1/44}

\bibitem{cegla2019}
{Cegla}, H.M., {Watson}, C.A., {Shelyag}, S., {Mathioudakis}, M., {Moutari},
  S.: {Stellar Surface Magnetoconvection as a Source of Astrophysical Noise.
  III. Sun-as-a-Star Simulations and Optimal Noise Diagnostics}.
\newblock \apj \textbf{879}(1), 55 (2019).
\newblock \doi{10.3847/1538-4357/ab16d3}

\bibitem{changeat2020}
{Changeat}, Q., {Al-Refaie}, A., {Mugnai}, L.V., {Edwards}, B., {Waldmann},
  I.P., {Pascale}, E., {Tinetti}, G.: {Alfnoor: A Retrieval Simulation of the
  Ariel Target List}.
\newblock \aj \textbf{160}(2), 80 (2020).
\newblock \doi{10.3847/1538-3881/ab9a53}

\bibitem{claret00}
{Claret}, A.: {A new non-linear limb-darkening law for LTE stellar atmosphere
  models. Calculations for -5.0 $<$= log[M/H] $<$= +1, 2000 K $<$= T$_{eff} <$=
  50000 K at several surface gravities}.
\newblock \aap \textbf{363}, 1081--1190 (2000)

\bibitem{claret12}
{Claret}, A., {Hauschildt}, P.H., {Witte}, S.: {New limb-darkening coefficients
  for PHOENIX/1D model atmospheres. I. Calculations for 1500 K $\le$ Teff $\le$
  4800 K Kepler, CoRot, Spitzer, uvby, UBVRIJHK, Sloan, and 2MASS photometric
  systems}.
\newblock A\&A \textbf{546}, A14 (2012).
\newblock \doi{10.1051/0004-6361/201219849}

\bibitem{claret13}
{Claret}, A., {Hauschildt}, P.H., {Witte}, S.: {New limb-darkening coefficients
  for Phoenix/1d model atmospheres. II. Calculations for 5000 K $\le$ Teff
  $\le$ 10 000 K Kepler, CoRot, Spitzer, uvby, UBVRIJHK, Sloan, and 2MASS
  photometric systems}.
\newblock \aap \textbf{552}, A16 (2013).
\newblock \doi{10.1051/0004-6361/201220942}

\bibitem{Claudi2020}
{Claudi} R.~{et al}, .: {the GAPS Programme with HARPS-N at TNG: Investigating
  the correlations between the host activity and the transiting system
  parameters}.
\newblock in prep.

\bibitem{cosentino14}
{Cosentino}, R., {Lovis}, C., {Pepe}, F., {Collier Cameron}, A., {Latham},
  D.W., {Molinari}, E., {Udry}, S., {Bezawada}, N., {Buchschacher}, N.,
  {Figueira}, P., {Fleury}, M., {Ghedina}, A., {Glenday}, A.G., {Gonzalez}, M.,
  {Guerra}, J., {Henry}, D., {Hughes}, I., {Maire}, C., {Motalebi}, F.,
  {Phillips}, D.F.: {HARPS-N @ TNG, two year harvesting data: performances and
  results}.
\newblock In: \procspie, \emph{Society of Photo-Optical Instrumentation
  Engineers (SPIE) Conference Series}, vol. 9147, p. 91478C (2014).
\newblock \doi{10.1117/12.2055813}

\bibitem{covino2013}
{Covino}, E., {Esposito}, M., {Barbieri}, M., {Mancini}, L., {Nascimbeni}, V.,
  {Claudi}, R., {Desidera}, S., {Gratton}, R., {Lanza}, A.F., {Sozzetti}, A.,
  {Biazzo}, K., {Affer}, L., {Gandolfi}, D., {Munari}, U., {Pagano}, I.,
  {Bonomo}, A.S., {Collier Cameron}, A., {H{\'e}brard}, G., {Maggio}, A.,
  {Messina}, S., {Micela}, G., {Molinari}, E., {Pepe}, F., {Piotto}, G.,
  {Ribas}, I., {Santos}, N.C., {Southworth}, J., {Shkolnik}, E., {Triaud},
  A.H.M.J., {Bedin}, L., {Benatti}, S., {Boccato}, C., {Bonavita}, M., {Borsa},
  F., {Borsato}, L., {Brown}, D., {Carolo}, E., {Ciceri}, S., {Cosentino}, R.,
  {Damasso}, M., {Faedi}, F., {Mart{\'\i}nez Fiorenzano}, A.F., {Latham}, D.W.,
  {Lovis}, C., {Mordasini}, C., {Nikolov}, N., {Poretti}, E., {Rainer}, M.,
  {Rebolo L{\'o}pez}, R., {Scandariato}, G., {Silvotti}, R., {Smareglia}, R.,
  {Alcal{\'a}}, J.M., {Cunial}, A., {Di Fabrizio}, L., {Di Mauro}, M.P.,
  {Giacobbe}, P., {Granata}, V., {Harutyunyan}, A., {Knapic}, C., {Lattanzi},
  M., {Leto}, G., {Lodato}, G., {Malavolta}, L., {Marzari}, F., {Molinaro}, M.,
  {Nardiello}, D., {Pedani}, M., {Prisinzano}, L., {Turrini}, D.: {The GAPS
  programme with HARPS-N at TNG. I. Observations of the Rossiter-McLaughlin
  effect and characterisation of the transiting system Qatar-1}.
\newblock \aap \textbf{554}, A28 (2013).
\newblock \doi{10.1051/0004-6361/201321298}

\bibitem{Cracchiolo2020}
{Cracchiolo}, G., {Micela}, G., {Peres}, G.: {Correcting the effect of stellar
  spots on ARIEL transmission spectra}.
\newblock \mnras  (2020).
\newblock \doi{10.1093/mnras/staa3621}

\bibitem{cridland+2019}
{Cridland}, A.J., {van Dishoeck}, E.F., {Alessi}, M., {Pudritz}, R.E.:
  {Connecting planet formation and astrochemistry. A main sequence for C/O in
  hot exoplanetary atmospheres}.
\newblock \aap \textbf{632}, A63 (2019).
\newblock \doi{10.1051/0004-6361/201936105}

\bibitem{daSilva2015}
{da Silva}, R., {Milone}, A.d.C., {Rocha-Pinto}, H.J.: {Homogeneous abundance
  analysis of FGK dwarf, subgiant, and giant stars with and without giant
  planets}.
\newblock \aap \textbf{580}, A24 (2015).
\newblock \doi{10.1051/0004-6361/201525770}

\bibitem{DelgadoMena2018}
{Delgado Mena}, E., {Adibekyan}, V.Z., {Figueira}, P., {Gonz{\'a}lez
  Hern{\'a}ndez}, J.I., {Santos}, N.C., {Tsantaki}, M., {Sousa}, S.G., {Faria},
  J.P., {Su{\'a}rez-Andr{\'e}s}, L., {Israelian}, G.: {Chemical Abundances of
  Neutron-capture Elements in Exoplanet-hosting Stars}.
\newblock \pasp \textbf{130}(991), 094202 (2018).
\newblock \doi{10.1088/1538-3873/aacc1f}

\bibitem{DelgadoMena2010}
{Delgado Mena}, E., {Israelian}, G., {Gonz{\'a}lez Hern{\'a}ndez}, J.I.,
  {Bond}, J.C., {Santos}, N.C., {Udry}, S., {Mayor}, M.: {Chemical Clues on the
  Formation of Planetary Systems: C/O Versus Mg/Si for HARPS GTO Sample}.
\newblock \apj \textbf{725}(2), 2349--2358 (2010).
\newblock \doi{10.1088/0004-637X/725/2/2349}

\bibitem{DelgadoMena2017}
{Delgado Mena}, E., {Tsantaki}, M., {Adibekyan}, V.Z., {Sousa}, S.G., {Santos},
  N.C., {Gonz{\'a}lez Hern{\'a}ndez}, J.I., {Israelian}, G.: {Chemical
  abundances of 1111 FGK stars from the HARPS GTO planet search program. II.
  Cu, Zn, Sr, Y, Zr, Ba, Ce, Nd, and Eu}.
\newblock \aap \textbf{606}, A94 (2017).
\newblock \doi{10.1051/0004-6361/201730535}

\bibitem{Demangeon2018}
{Demangeon}, O.D.S., {Faedi}, F., {H{\'e}brard}, G., {Brown}, D.J.A., {Barros},
  S.C.C., {Doyle}, A.P., {Maxted}, P.F.L., {Collier Cameron}, A., {Hay}, K.L.,
  {Alikakos}, J., {Anderson}, D.R., {Armstrong}, D.J., {Boumis}, P., {Bonomo},
  A.S., {Bouchy}, F., {Delrez}, L., {Gillon}, M., {Haswell}, C.A., {Hellier},
  C., {Jehin}, E., {Kiefer}, F., {Lam}, K.W.F., {Lendl}, M., {Mancini}, L.,
  {McCormac}, J., {Norton}, A.J., {Osborn}, H.P., {Palle}, E., {Pepe}, F.,
  {Pollacco}, D.L., {Prieto-Arranz}, J., {Queloz}, D., {S{\'e}gransan}, D.,
  {Smalley}, B., {Triaud}, A.H.M.J., {Udry}, S., {West}, R., {Wheatley}, P.J.:
  {The discovery of WASP-151b, WASP-153b, WASP-156b: Insights on giant planet
  migration and the upper boundary of the Neptunian desert}.
\newblock \aap \textbf{610}, A63 (2018).
\newblock \doi{10.1051/0004-6361/201731735}

\bibitem{Ecuvillon2004a}
{Ecuvillon}, A., {Israelian}, G., {Santos}, N.C., {Mayor}, M., {Garc{\'\i}a
  L{\'o}pez}, R.J., {Randich}, S.: {Nitrogen abundances in planet-harbouring
  stars}.
\newblock \aap \textbf{418}, 703--715 (2004).
\newblock \doi{10.1051/0004-6361:20035717}

\bibitem{Edwards2019}
{Edwards}, B., {Mugnai}, L., {Tinetti}, G., {Pascale}, E., {Sarkar}, S.: {An
  Updated Study of Potential Targets for Ariel}.
\newblock \aj \textbf{157}(6), 242 (2019).
\newblock \doi{10.3847/1538-3881/ab1cb9}

\bibitem{Faedi2013}
{Faedi}, F., {Pollacco}, D., {Barros}, S.C.C., {Brown}, D., {Collier Cameron},
  A., {Doyle}, A.P., {Enoch}, R., {Gillon}, M., {G{\'o}mez Maqueo Chew}, Y.,
  {H{\'e}brard}, G., {Lendl}, M., {Liebig}, C., {Smalley}, B., {Triaud},
  A.H.M.J., {West}, R.G., {Wheatley}, P.J., {Alsubai}, K.A., {Anderson}, D.R.,
  {Armstrong}, D., {Bento}, J., {Bochinski}, J., {Bouchy}, F., {Busuttil}, R.,
  {Fossati}, L., {Fumel}, A., {Haswell}, C.A., {Hellier}, C., {Holmes}, S.,
  {Jehin}, E., {Kolb}, U., {McCormac}, J., {Miller}, G.R.M., {Moutou}, C.,
  {Norton}, A.J., {Parley}, N., {Queloz}, D., {Santerne}, A., {Skillen}, I.,
  {Smith}, A.M.S., {Udry}, S., {Watson}, C.: {WASP-54b, WASP-56b, and WASP-57b:
  Three new sub-Jupiter mass planets from SuperWASP}.
\newblock \aap \textbf{551}, A73 (2013).
\newblock \doi{10.1051/0004-6361/201220520}

\bibitem{Figueira2014}
{Figueira}, P., {Oshagh}, M., {Adibekyan}, V.Z., {Santos}, N.C.: {Revisiting
  the correlation between stellar activity and planetary surface gravity}.
\newblock \aap \textbf{572}, A51 (2014).
\newblock \doi{10.1051/0004-6361/201424902}

\bibitem{Fossati2015}
{Fossati}, L., {Ingrassia}, S., {Lanza}, A.F.: {A Bimodal Correlation between
  Host Star Chromospheric Emission and the Surface Gravity of Hot Jupiters}.
\newblock \apjl \textbf{812}(2), L35 (2015).
\newblock \doi{10.1088/2041-8205/812/2/L35}

\bibitem{Fossati2017}
{Fossati}, L., {Marcelja}, S.E., {Staab}, D., {Cubillos}, P.E., {France}, K.,
  {Haswell}, C.A., {Ingrassia}, S., {Jenkins}, J.S., {Koskinen}, T., {Lanza},
  A.F., {Redfield}, S., {Youngblood}, A., {Pelzmann}, G.: {The effect of ISM
  absorption on stellar activity measurements and its relevance for exoplanet
  studies}.
\newblock \aap \textbf{601}, A104 (2017).
\newblock \doi{10.1051/0004-6361/201630339}

\bibitem{Franchini2020}
{Franchini}, M., {Morossi}, C., {Di Marcantonio}, P., {Chavez}, M.,
  {Adibekyan}, V.Z., {Bayo}, A., {Bensby}, T., {Bragaglia}, A., {Calura}, F.,
  {Duffau}, S., {Gonneau}, A., {Heiter}, U., {Kordopatis}, G., {Romano}, D.,
  {Sbordone}, L., {Smiljanic}, R., {Tautvai{\v{s}}ien{\.{e}}}, G., {Van der
  Swaelmen}, M., {Delgado Mena}, E., {Gilmore}, G., {Randich}, S., {Carraro},
  G., {Hourihane}, A., {Magrini}, L., {Morbidelli}, L., {Sousa}, S., {Worley},
  C.C.: {The Gaia-ESO Survey: Carbon Abundance in the Galactic Thin and Thick
  Disks}.
\newblock \apj \textbf{888}(2), 55 (2020).
\newblock \doi{10.3847/1538-4357/ab5dc4}

\bibitem{Gustafsson2008}
{Gustafsson}, B., {Edvardsson}, B., {Eriksson}, K., {J{\o}rgensen}, U.G.,
  {Nordlund}, {\r{A}}., {Plez}, B.: {A grid of MARCS model atmospheres for
  late-type stars. I. Methods and general properties}.
\newblock \aap \textbf{486}(3), 951--970 (2008).
\newblock \doi{10.1051/0004-6361:200809724}

\bibitem{han2014}
{Han}, E., {Wang}, S.X., {Wright}, J.T., {Feng}, Y.K., {Zhao}, M., {Fakhouri},
  O., {Brown}, J.I., {Hancock}, C.: {Exoplanet Orbit Database. II. Updates to
  Exoplanets.org}.
\newblock \pasp \textbf{126}(943), 827 (2014).
\newblock \doi{10.1086/678447}

\bibitem{Hartman2010}
{Hartman}, J.D.: {A Correlation Between Stellar Activity and the Surface
  Gravity of Hot Jupiters}.
\newblock \apjl \textbf{717}(2), L138--L142 (2010).
\newblock \doi{10.1088/2041-8205/717/2/L138}

\bibitem{hayek2012}
{Hayek}, W., {Sing}, D., {Pont}, F., {Asplund}, M.: {Limb darkening laws for
  two exoplanet host stars derived from 3D stellar model atmospheres.
  Comparison with 1D models and HST light curve observations}.
\newblock \aap \textbf{539}, A102 (2012).
\newblock \doi{10.1051/0004-6361/201117868}

\bibitem{2015PhDT.......193H}
{Haywood}, R.D.: {Hide and Seek: Radial-Velocity Searches for Planets around
  Active Stars}.
\newblock Ph.D. thesis, University of St Andrews (2015)

\bibitem{heiter15}
{Heiter}, U., {Lind}, K., {Asplund}, M., {Barklem}, P.S., {Bergemann}, M.,
  {Magrini}, L., {Masseron}, T., {Mikolaitis}, {\v{S}}., {Pickering}, J.C.,
  {Ruffoni}, M.P.: {Atomic and molecular data for optical stellar
  spectroscopy}.
\newblock \physscr \textbf{90}(5), 054010 (2015).
\newblock \doi{10.1088/0031-8949/90/5/054010}

\bibitem{Jofre2015}
{Jofr{\'e}}, E., {Petrucci}, R., {Saffe}, C., {Saker}, L., {Artur de la
  Villarmois}, E., {Chavero}, C., {G{\'o}mez}, M., {Mauas}, P.J.D.: {Stellar
  parameters and chemical abundances of 223 evolved stars with and without
  planets}.
\newblock \aap \textbf{574}, A50 (2015).
\newblock \doi{10.1051/0004-6361/201424474}

\bibitem{Johnson2016}
{Johnson}, M.C., {Endl}, M., {Cochran}, W.D., {Meschiari}, S., {Robertson}, P.,
  {MacQueen}, P.J., {Brugamyer}, E.J., {Caldwell}, C., {Hatzes}, A.P.,
  {Ram{\'\i}rez}, I., {Wittenmyer}, R.A.: {A 12-year Activity Cycle for the
  Nearby Planet Host Star HD 219134}.
\newblock \apj \textbf{821}(2), 74 (2016).
\newblock \doi{10.3847/0004-637X/821/2/74}

\bibitem{johnson+2012}
{Johnson}, T.V., {Mousis}, O., {Lunine}, J.I., {Madhusudhan}, N.: {Planetesimal
  Compositions in Exoplanet Systems}.
\newblock \apj \textbf{757}(2), 192 (2012).
\newblock \doi{10.1088/0004-637X/757/2/192}

\bibitem{Johnstone2020}
{Johnstone}, C.P.: {Hydrodynamic Escape of Water Vapor Atmospheres near Very
  Active Stars}.
\newblock \apj \textbf{890}(1), 79 (2020).
\newblock \doi{10.3847/1538-4357/ab6224}

\bibitem{Lallement2019}
{Lallement}, R., {Babusiaux}, C., {Vergely}, J.L., {Katz}, D., {Arenou}, F.,
  {Valette}, B., {Hottier}, C., {Capitanio}, L.: {Gaia-2MASS 3D maps of
  Galactic interstellar dust within 3 kpc}.
\newblock \aap \textbf{625}, A135 (2019).
\newblock \doi{10.1051/0004-6361/201834695}

\bibitem{Lanza2014}
{Lanza}, A.F.: {On the correlation between stellar chromospheric flux and the
  surface gravity of close-in planets}.
\newblock \aap \textbf{572}, L6 (2014).
\newblock \doi{10.1051/0004-6361/201425051}

\bibitem{lanza2018}
{Lanza}, A.F., {Malavolta}, L., {Benatti}, S., {Desidera}, S., {Bignamini}, A.,
  {Bonomo}, A.S., {Esposito}, M., {Figueira}, P., {Gratton}, R., {Scandariato},
  G., {Damasso}, M., {Sozzetti}, A., {Biazzo}, K., {Claudi}, R.U., {Cosentino},
  R., {Covino}, E., {Maggio}, A., {Masiero}, S., {Micela}, G., {Molinari}, E.,
  {Pagano}, I., {Piotto}, G., {Poretti}, E., {Smareglia}, R., {Affer}, L.,
  {Boccato}, C., {Borsa}, F., {Boschin}, W., {Giacobbe}, P., {Knapic}, C.,
  {Leto}, G., {Maldonado}, J., {Mancini}, L., {Martinez Fiorenzano}, A.,
  {Messina}, S., {Nascimbeni}, V., {Pedani}, M., {Rainer}, M.: {The GAPS
  Programme with HARPS-N at TNG. XVII. Line profile indicators and kernel
  regression as diagnostics of radial-velocity variations due to stellar
  activity in solar-like stars}.
\newblock \aap \textbf{616}, A155 (2018).
\newblock \doi{10.1051/0004-6361/201731010}

\bibitem{Locci2019}
{Locci}, D., {Cecchi-Pestellini}, C., {Micela}, G.: {Photo-evaporation of
  close-in gas giants orbiting around G and M stars}.
\newblock \aap \textbf{624}, A101 (2019).
\newblock \doi{10.1051/0004-6361/201834491}

\bibitem{lovis2011}
{Lovis}, C., {Dumusque}, X., {Santos}, N.C., {Bouchy}, F., {Mayor}, M., {Pepe},
  F., {Queloz}, D., {S{\'e}gransan}, D., {Udry}, S.: {The HARPS search for
  southern extra-solar planets. XXXI. Magnetic activity cycles in solar-type
  stars: statistics and impact on precise radial velocities}.
\newblock arXiv e-prints arXiv:1107.5325 (2011)

\bibitem{Madhusudhan2016}
{Madhusudhan}, N., {Ag{\'u}ndez}, M., {Moses}, J.I., {Hu}, Y.: {Exoplanetary
  Atmospheres{\textemdash}Chemistry, Formation Conditions, and Habitability}.
\newblock \ssr \textbf{205}(1-4), 285--348 (2016).
\newblock \doi{10.1007/s11214-016-0254-3}

\bibitem{Madhusudhan2014}
{Madhusudhan}, N., {Amin}, M.A., {Kennedy}, G.M.: {Toward Chemical Constraints
  on Hot Jupiter Migration}.
\newblock \apjl \textbf{794}(1), L12 (2014).
\newblock \doi{10.1088/2041-8205/794/1/L12}

\bibitem{Magrini13}
{Magrini}, L., {Randich}, S., {Friel}, E., {Spina}, L., {Jacobson}, H.,
  {Cantat-Gaudin}, T., {Donati}, P., {Baglioni}, R., {Maiorca}, E.,
  {Bragaglia}, A., {Sordo}, R., {Vallenari}, A.: {FAMA: An automatic code for
  stellar parameter and abundance determination}.
\newblock \aap \textbf{558}, A38 (2013).
\newblock \doi{10.1051/0004-6361/201321844}

\bibitem{magrini18}
{Magrini}, L., {Vincenzo}, F., {Randich}, S., {Pancino}, E., {Casali}, G.,
  {Tautvai{\v{s}}ien{\.{e}}}, G., {Drazdauskas}, A., {Mikolaitis}, {\v{S}}.,
  {Minkevi{\v{c}}i{\={u}}t{\.{e}}}, R., {Stonkut{\.{e}}}, E., {Chorniy}, Y.,
  {Bagdonas}, V., {Kordopatis}, G., {Friel}, E., {Roccatagliata}, V.,
  {Jim{\'e}nez-Esteban}, F.M., {Gilmore}, G., {Vallenari}, A., {Bensby}, T.,
  {Bragaglia}, A., {Korn}, A.J., {Lanzafame}, A.C., {Smiljanic}, R., {Bayo},
  A., {Casey}, A.R., {Costado}, M.T., {Franciosini}, E., {Hourihane}, A.,
  {Jofr{\'e}}, P., {Lewis}, J., {Monaco}, L., {Morbidelli}, L., {Sacco}, G.,
  {Worley}, C.: {The Gaia-ESO Survey: The N/O abundance ratio in the Milky
  Way}.
\newblock \aap \textbf{618}, A102 (2018).
\newblock \doi{10.1051/0004-6361/201833224}

\bibitem{Maldonado2019}
{Maldonado}, J., {Phillips}, D.F., {Dumusque}, X., {Collier Cameron}, A.,
  {Haywood}, R.D., {Lanza}, A.F., {Micela}, G., {Mortier}, A., {Saar}, S.H.,
  {Sozzetti}, A., {Rice}, K., {Milbourne}, T., {Cecconi}, M., {Cegla}, H.M.,
  {Cosentino}, R., {Costes}, J., {Ghedina}, A., {Gonzalez}, M., {Guerra}, J.,
  {Hern{\'a}ndez}, N., {Li}, C.H., {Lodi}, M., {Malavolta}, L., {Molinari}, E.,
  {Pepe}, F., {Piotto}, G., {Poretti}, E., {Sasselov}, D., {San Juan}, J.,
  {Thompson}, S., {Udry}, S., {Watson}, C.: {Temporal evolution and
  correlations of optical activity indicators measured in Sun-as-a-star
  observations}.
\newblock \aap \textbf{627}, A118 (2019).
\newblock \doi{10.1051/0004-6361/201935233}

\bibitem{Mallonn2015}
{Mallonn}, M., {von Essen}, C., {Weingrill}, J., {Strassmeier}, K.G., {Ribas},
  I., {Carroll}, T.A., {Herrero}, E., {Granzer}, T., {Claret}, A., {Schwope},
  A.: {Transmission spectroscopy of the inflated exo-Saturn HAT-P-19b}.
\newblock \aap \textbf{580}, A60 (2015).
\newblock \doi{10.1051/0004-6361/201423778}

\bibitem{mamajek2008}
{Mamajek}, E.E., {Hillenbrand}, L.A.: {Improved Age Estimation for Solar-Type
  Dwarfs Using Activity-Rotation Diagnostics}.
\newblock \apj \textbf{687}(2), 1264--1293 (2008).
\newblock \doi{10.1086/591785}

\bibitem{marboeuf+2014b}
{Marboeuf}, U., {Thiabaud}, A., {Alibert}, Y., {Cabral}, N., {Benz}, W.: {From
  planetesimals to planets: volatile molecules}.
\newblock \aap \textbf{570}, A36 (2014).
\newblock \doi{10.1051/0004-6361/201423431}

\bibitem{marboeuf+2014a}
{Marboeuf}, U., {Thiabaud}, A., {Alibert}, Y., {Cabral}, N., {Benz}, W.: {From
  stellar nebula to planetesimals}.
\newblock \aap \textbf{570}, A35 (2014).
\newblock \doi{10.1051/0004-6361/201322207}

\bibitem{Maxted2015}
{Maxted}, P.F.L., {Serenelli}, A.M., {Southworth}, J.: {Comparison of
  gyrochronological and isochronal age estimates for transiting exoplanet host
  stars}.
\newblock \aap \textbf{577}, A90 (2015).
\newblock \doi{10.1051/0004-6361/201525774}

\bibitem{McQuillan2013}
{McQuillan}, A., {Aigrain}, S., {Mazeh}, T.: {Measuring the rotation period
  distribution of field M dwarfs with Kepler}.
\newblock \mnras \textbf{432}(2), 1203--1216 (2013).
\newblock \doi{10.1093/mnras/stt536}

\bibitem{micela2015}
{Micela}, G.: {EChO spectra and stellar activity - I. Correcting the infrared
  signal using simultaneous optical spectroscopy}.
\newblock Experimental Astronomy \textbf{40}(2-3), 723--732 (2015).
\newblock \doi{10.1007/s10686-014-9430-1}

\bibitem{mordasini+2016}
{Mordasini}, C., {van Boekel}, R., {Molli{\`e}re}, P., {Henning}, T.,
  {Benneke}, B.: {The Imprint of Exoplanet Formation History on Observable
  Present-day Spectra of Hot Jupiters}.
\newblock \apj \textbf{832}(1), 41 (2016).
\newblock \doi{10.3847/0004-637X/832/1/41}

\bibitem{morello2020joss}
{Morello}, G., {Claret}, A., {Martin-Lagarde}, M., {Cossou}, C., {Tsiara}, A.,
  {Lagage}, P.O.: {ExoTETHyS: Tools for Exoplanetary Transits around host
  stars}.
\newblock The Journal of Open Source Software \textbf{5}(46), 1834 (2020).
\newblock \doi{10.21105/joss.01834}

\bibitem{morello2020}
{Morello}, G., {Claret}, A., {Martin-Lagarde}, M., {Cossou}, C., {Tsiaras}, A.,
  {Lagage}, P.O.: {The ExoTETHyS Package: Tools for Exoplanetary Transits
  around Host Stars}.
\newblock \aj \textbf{159}(2), 75 (2020).
\newblock \doi{10.3847/1538-3881/ab63dc}

\bibitem{morello17}
{Morello}, G., {Tsiaras}, A., {Howarth}, I.D., {Homeier}, D.: {High-precision
  Stellar Limb-darkening in Exoplanetary Transits}.
\newblock AJ \textbf{154}, 111 (2017).
\newblock \doi{10.3847/1538-3881/aa8405}

\bibitem{Mocnik2017}
{Mo{\v{c}}nik}, T., {Hellier}, C., {Anderson}, D.R., {Clark}, B.J.M.,
  {Southworth}, J.: {Starspots on WASP-107 and pulsations of WASP-118}.
\newblock \mnras \textbf{469}(2), 1622--1629 (2017).
\newblock \doi{10.1093/mnras/stx972}

\bibitem{2006A&A...453..309N}
{Nardetto}, N., {Mourard}, D., {Kervella}, P., {Mathias}, P., {M{\'e}rand}, A.,
  {Bersier}, D.: {High resolution spectroscopy for Cepheids distance
  determination. I. Line asymmetry}.
\newblock \aap \textbf{453}(1), 309--319 (2006).
\newblock \doi{10.1051/0004-6361:20054333}

\bibitem{Nissen2014}
{Nissen}, P.E., {Chen}, Y.Q., {Carigi}, L., {Schuster}, W.J., {Zhao}, G.:
  {Carbon and oxygen abundances in stellar populations}.
\newblock \aap \textbf{568}, A25 (2014).
\newblock \doi{10.1051/0004-6361/201424184}

\bibitem{Nordlander2017}
{Nordlander}, T., {Lind}, K.: {Non-LTE aluminium abundances in late-type
  stars}.
\newblock \aap \textbf{607}, A75 (2017).
\newblock \doi{10.1051/0004-6361/201730427}

\bibitem{noyes1984}
{Noyes}, R.W., {Hartmann}, L.W., {Baliunas}, S.L., {Duncan}, D.K., {Vaughan},
  A.H.: {Rotation, convection, and magnetic activity in lower main-sequence
  stars.}
\newblock ApJ \textbf{279}, 763--777 (1984).
\newblock \doi{10.1086/161945}

\bibitem{oberg+2011}
{{\"O}berg}, K.I., {Qi}, C., {Fogel}, J.K.J., {Bergin}, E.A., {Andrews}, S.M.,
  {Espaillat}, C., {Wilner}, D.J., {Pascucci}, I., {Kastner}, J.H.: {Disk
  Imaging Survey of Chemistry with SMA. II. Southern Sky Protoplanetary Disk
  Data and Full Sample Statistics}.
\newblock \apj \textbf{734}(2), 98 (2011).
\newblock \doi{10.1088/0004-637X/734/2/98}

\bibitem{oberg+2019}
{{\"O}berg}, K.I., {Wordsworth}, R.: {Jupiter's Composition Suggests its Core
  Assembled Exterior to the N$_{2}$ Snowline}.
\newblock \aj \textbf{158}(5), 194 (2019).
\newblock \doi{10.3847/1538-3881/ab46a8}

\bibitem{2013A&A...556A..19O}
{Oshagh}, M., {Santos}, N.C., {Boisse}, I., {Bou{\'e}}, G., {Montalto}, M.,
  {Dumusque}, X., {Haghighipour}, N.: {Effect of stellar spots on
  high-precision transit light-curve}.
\newblock \aap \textbf{556}, A19 (2013).
\newblock \doi{10.1051/0004-6361/201321309}

\bibitem{Osorio2015}
{Osorio}, Y., {Barklem}, P.S., {Lind}, K., {Belyaev}, A.K., {Spielfiedel}, A.,
  {Guitou}, M., {Feautrier}, N.: {Mg line formation in late-type stellar
  atmospheres. I. The model atom}.
\newblock \aap \textbf{579}, A53 (2015).
\newblock \doi{10.1051/0004-6361/201525846}

\bibitem{scikit-learn}
Pedregosa, F., Varoquaux, G., Gramfort, A., Michel, V., Thirion, B., Grisel,
  O., Blondel, M., Prettenhofer, P., Weiss, R., Dubourg, V., Vanderplas, J.,
  Passos, A., Cournapeau, D., Brucher, M., Perrot, M., Duchesnay, E.:
  Scikit-learn: Machine learning in {P}ython.
\newblock Journal of Machine Learning Research \textbf{12}, 2825--2830 (2011)

\bibitem{pepe2002}
{Pepe}, F., {Mayor}, M., {Galland}, F., {Naef}, D., {Queloz}, D., {Santos},
  N.C., {Udry}, S., {Burnet}, M.: {The CORALIE survey for southern extra-solar
  planets VII. Two short-period Saturnian companions to <ASTROBJ>HD
  108147</ASTROBJ> and <ASTROBJ>HD 168746</ASTROBJ>}.
\newblock \aap \textbf{388}, 632--638 (2002).
\newblock \doi{10.1051/0004-6361:20020433}

\bibitem{Perryman2014}
{Perryman}, M., {Hartman}, J., {Bakos}, G.{\'A}., {Lindegren}, L.: {Astrometric
  Exoplanet Detection with Gaia}.
\newblock ApJ \textbf{797}(1), 14 (2014).
\newblock \doi{10.1088/0004-637X/797/1/14}

\bibitem{petigura2011}
{Petigura}, E.A., {Marcy}, G.W.: {Carbon and Oxygen in Nearby Stars: Keys to
  Protoplanetary Disk Chemistry}.
\newblock \apj \textbf{735}(1), 41 (2011).
\newblock \doi{10.1088/0004-637X/735/1/41}

\bibitem{2013MNRAS.432.2917P}
{Pont}, F., {Sing}, D.K., {Gibson}, N.P., {Aigrain}, S., {Henry}, G., {Husnoo},
  N.: {The prevalence of dust on the exoplanet HD 189733b from Hubble and
  Spitzer observations}.
\newblock \mnras \textbf{432}(4), 2917--2944 (2013).
\newblock \doi{10.1093/mnras/stt651}

\bibitem{2001A&A...379..279Q}
{Queloz}, D., {Henry}, G.W., {Sivan}, J.P., {Baliunas}, S.L., {Beuzit}, J.L.,
  {Donahue}, R.A., {Mayor}, M., {Naef}, D., {Perrier}, C., {Udry}, S.: {No
  planet for HD 166435}.
\newblock \aap \textbf{379}, 279--287 (2001).
\newblock \doi{10.1051/0004-6361:20011308}

\bibitem{PLATO}
{Rauer}, H., {Catala}, C., {Aerts}, C., {Appourchaux}, T., {Benz}, W.,
  {Brandeker}, A., {Christensen-Dalsgaard}, J., {Deleuil}, M., {Gizon}, L.,
  {Goupil}, M.J., {G{\"u}del}, M., {Janot-Pacheco}, E., {Mas-Hesse}, M.,
  {Pagano}, I., {Piotto}, G., {Pollacco}, D., {Santos}, {\.{C}}., {Smith}, A.,
  {Su{\'a}rez}, J.C., {Szab{\'o}}, R., {Udry}, S., {Adibekyan}, V., {Alibert},
  Y., {Almenara}, J.M., {Amaro-Seoane}, P., {Eiff}, M.A.v., {Asplund}, M.,
  {Antonello}, E., {Barnes}, S., {Baudin}, F., {Belkacem}, K., {Bergemann}, M.,
  {Bihain}, G., {Birch}, A.C., {Bonfils}, X., {Boisse}, I., {Bonomo}, A.S.,
  {Borsa}, F., {Brand {\~a}o}, I.M., {Brocato}, E., {Brun}, S., {Burleigh}, M.,
  {Burston}, R., {Cabrera}, J., {Cassisi}, S., {Chaplin}, W., {Charpinet}, S.,
  {Chiappini}, C., {Church}, R.P., {Csizmadia}, S., {Cunha}, M., {Damasso}, M.,
  {Davies}, M.B., {Deeg}, H.J., {D{\'\i}az}, R.F., {Dreizler}, S., {Dreyer},
  C., {Eggenberger}, P., {Ehrenreich}, D., {Eigm{\"u}ller}, P., {Erikson}, A.,
  {Farmer}, R., {Feltzing}, S., {de Oliveira Fialho}, F., {Figueira}, P.,
  {Forveille}, T., {Fridlund}, M., {Garc{\'\i}a}, R.A., {Giommi}, P.,
  {Giuffrida}, G., {Godolt}, M., {Gomes da Silva}, J., {Granzer}, T.,
  {Grenfell}, J.L., {Grotsch-Noels}, A., {G{\"u}nther}, E., {Haswell}, C.A.,
  {Hatzes}, A.P., {H{\'e}brard}, G., {Hekker}, S., {Helled}, R., {Heng}, K.,
  {Jenkins}, J.M., {Johansen}, A., {Khodachenko}, M.L., {Kislyakova}, K.G.,
  {Kley}, W., {Kolb}, U., {Krivova}, N., {Kupka}, F., {Lammer}, H., {Lanza},
  A.F., {Lebreton}, Y., {Magrin}, D., {Marcos-Arenal}, P., {Marrese}, P.M.,
  {Marques}, J.P., {Martins}, J., {Mathis}, S., {Mathur}, S., {Messina}, S.,
  {Miglio}, A., {Montalban}, J., {Montalto}, M., {Monteiro}, M.J.P.F.G.,
  {Moradi}, H., {Moravveji}, E., {Mordasini}, C., {Morel}, T., {Mortier}, A.,
  {Nascimbeni}, V., {Nelson}, R.P., {Nielsen}, M.B., {Noack}, L., {Norton},
  A.J., {Ofir}, A., {Oshagh}, M., {Ouazzani}, R.M., {P{\'a}pics}, P., {Parro},
  V.C., {Petit}, P., {Plez}, B., {Poretti}, E., {Quirrenbach}, A., {Ragazzoni},
  R., {Raimondo}, G., {Rainer}, M., {Reese}, D.R., {Redmer}, R., {Reffert}, S.,
  {Rojas-Ayala}, B., {Roxburgh}, I.W., {Salmon}, S., {Santerne}, A.,
  {Schneider}, J., {Schou}, J., {Schuh}, S., {Schunker}, H., {Silva-Valio}, A.,
  {Silvotti}, R., {Skillen}, I., {Snellen}, I., {Sohl}, F., {Sousa}, S.G.,
  {Sozzetti}, A., {Stello}, D., {Strassmeier}, K.G., {{\v{S}}vanda}, M.,
  {Szab{\'o}}, G.M., {Tkachenko}, A., {Valencia}, D., {Van Grootel}, V.,
  {Vauclair}, S.D., {Ventura}, P., {Wagner}, F.W., {Walton}, N.A., {Weingrill},
  J., {Werner}, S.C., {Wheatley}, P.J., {Zwintz}, K.: {The PLATO 2.0 mission}.
\newblock Experimental Astronomy \textbf{38}(1-2), 249--330 (2014).
\newblock \doi{10.1007/s10686-014-9383-4}

\bibitem{Ricker2014}
{Ricker}, G.R., {Winn}, J.N., {Vanderspek}, R., {Latham}, D.W., {Bakos},
  G.{\'A}., {Bean}, J.L., {Berta-Thompson}, Z.K., {Brown}, T.M., {Buchhave},
  L., {Butler}, N.R., {Butler}, R.P., {Chaplin}, W.J., {Charbonneau}, D.,
  {Christensen-Dalsgaard}, J., {Clampin}, M., {Deming}, D., {Doty}, J., {De
  Lee}, N., {Dressing}, C., {Dunham}, E.W., {Endl}, M., {Fressin}, F., {Ge},
  J., {Henning}, T., {Holman}, M.J., {Howard}, A.W., {Ida}, S., {Jenkins}, J.,
  {Jernigan}, G., {Johnson}, J.A., {Kaltenegger}, L., {Kawai}, N., {Kjeldsen},
  H., {Laughlin}, G., {Levine}, A.M., {Lin}, D., {Lissauer}, J.J., {MacQueen},
  P., {Marcy}, G., {McCullough}, P.R., {Morton}, T.D., {Narita}, N., {Paegert},
  M., {Palle}, E., {Pepe}, F., {Pepper}, J., {Quirrenbach}, A., {Rinehart},
  S.A., {Sasselov}, D., {Sato}, B., {Seager}, S., {Sozzetti}, A., {Stassun},
  K.G., {Sullivan}, P., {Szentgyorgyi}, A., {Torres}, G., {Udry}, S.,
  {Villasenor}, J.: {Transiting Exoplanet Survey Satellite (TESS)}.
\newblock In: SPIE, \emph{Society of Photo-Optical Instrumentation Engineers
  (SPIE) Conference Series}, vol. 9143, p. 914320 (2014).
\newblock \doi{10.1117/12.2063489}

\bibitem{Robertson2015}
{Robertson}, P., {Roy}, A., {Mahadevan}, S.: {Stellar Activity Mimics a
  Habitable-zone Planet around Kapteyn's Star}.
\newblock \apjl \textbf{805}(2), L22 (2015).
\newblock \doi{10.1088/2041-8205/805/2/L22}

\bibitem{1997ApJ...485..319S}
{Saar}, S.H., {Donahue}, R.A.: {Activity-Related Radial Velocity Variation in
  Cool Stars}.
\newblock \apj \textbf{485}(1), 319--327 (1997).
\newblock \doi{10.1086/304392}

\bibitem{Santerne2016}
{Santerne}, A., {H{\'e}brard}, G., {Lillo-Box}, J., {Armstrong}, D.J.,
  {Barros}, S.C.C., {Demangeon}, O., {Barrado}, D., {Debackere}, A., {Deleuil},
  M., {Delgado Mena}, E., {Montalto}, M., {Pollacco}, D., {Osborn}, H.P.,
  {Sousa}, S.G., {Abe}, L., {Adibekyan}, V., {Almenara}, J.M., {Andr{\'e}}, P.,
  {Arlic}, G., {Barthe}, G., {Bendjoya}, P., {Behrend}, R., {Boisse}, I.,
  {Bouchy}, F., {Boussier}, H., {Bretton}, M., {Brown}, D.J.A., {Carry}, B.,
  {Cailleau}, A., {Conseil}, E., {Coulon}, G., {Courcol}, B., {Dauchet}, B.,
  {Dalouzy}, J.C., {Deldem}, M., {Desormi{\`e}res}, O., {Dubreuil}, P.,
  {Fehrenbach}, J.M., {Ferratfiat}, S., {Girelli}, R., {Gregorio}, J.,
  {Jaecques}, S., {Kugel}, F., {Kirk}, J., {Labrevoir}, O., {Lachuri{\'e}},
  J.C., {Lam}, K.W.F., {Le Guen}, P., {Martinez}, P., {Maurin}, L.P.A.,
  {McCormac}, J., {Pioppa}, J.B., {Quadri}, U., {Rajpurohit}, A., {Rey}, J.,
  {Rivet}, J.P., {Roy}, R., {Santos}, N.C., {Signoret}, F., {Strabla}, L.,
  {Suarez}, O., {Toublanc}, D., {Tsantaki}, M., {Vienney}, J.M., {Wilson},
  P.A., {Bachschmidt}, M., {Colas}, F., {Gerteis}, O., {Louis}, P., {Mario},
  J.C., {Marlot}, C., {Montier}, J., {Perroud}, V., {Pic}, V., {Romeuf}, D.,
  {Ubaud}, S., {Verilhac}, D.: {K2-29 b/WASP-152 b: An Aligned and Inflated Hot
  Jupiter in a Young Visual Binary}.
\newblock \apj \textbf{824}(1), 55 (2016).
\newblock \doi{10.3847/0004-637X/824/1/55}

\bibitem{santos2015}
{Santos}, N.C., {Adibekyan}, V., {Mordasini}, C., {Benz}, W., {Delgado-Mena},
  E., {Dorn}, C., {Buchhave}, L., {Figueira}, P., {Mortier}, A., {Pepe}, F.,
  {Santerne}, A., {Sousa}, S.G., {Udry}, S.: {Constraining planet structure
  from stellar chemistry: the cases of CoRoT-7, Kepler-10, and Kepler-93}.
\newblock \aap \textbf{580}, L13 (2015).
\newblock \doi{10.1051/0004-6361/201526850}

\bibitem{Santos2004}
{Santos}, N.C., {Israelian}, G., {Mayor}, M.: {Spectroscopic [Fe/H] for 98
  extra-solar planet-host stars. Exploring the probability of planet
  formation}.
\newblock \aap \textbf{415}, 1153--1166 (2004).
\newblock \doi{10.1051/0004-6361:20034469}

\bibitem{SWEET-Cat}
{Santos}, N.C., {Sousa}, S.G., {Mortier}, A., {Neves}, V., {Adibekyan}, V.,
  {Tsantaki}, M., {Delgado Mena}, E., {Bonfils}, X., {Israelian}, G., {Mayor},
  M., {Udry}, S.: {SWEET-Cat: A catalogue of parameters for Stars With
  ExoplanETs. I. New atmospheric parameters and masses for 48 stars with
  planets}.
\newblock A\&A \textbf{556}, A150 (2013).
\newblock \doi{10.1051/0004-6361/201321286}

\bibitem{sarkar2018}
{Sarkar}, S., {Argyriou}, I., {Vandenbussche}, B., {Papageorgiou}, A.,
  {Pascale}, E.: {Stellar pulsation and granulation as noise sources in
  exoplanet transit spectroscopy in the ARIEL space mission}.
\newblock \mnras \textbf{481}(3), 2871--2877 (2018).
\newblock \doi{10.1093/mnras/sty2453}

\bibitem{Sarkar2020}
{Sarkar}, S., {Pascale}, E., {Papageorgiou}, A., {Johnson}, L.J., {Waldmann},
  I.: {ExoSim: the Exoplanet Observation Simulator}.
\newblock arXiv e-prints arXiv:2002.03739 (2020)

\bibitem{Schlaufman2015}
{Schlaufman}, K.C.: {A Continuum of Planet Formation between 1 and 4 Earth
  Radii}.
\newblock \apjl \textbf{799}(2), L26 (2015).
\newblock \doi{10.1088/2041-8205/799/2/L26}

\bibitem{schneider2011}
{Schneider}, J., {Dedieu}, C., {Le Sidaner}, P., {Savalle}, R., {Zolotukhin},
  I.: {Defining and cataloging exoplanets: the exoplanet.eu database}.
\newblock \aap \textbf{532}, A79 (2011).
\newblock \doi{10.1051/0004-6361/201116713}

\bibitem{ShibataIkoma2019}
{Shibata}, S., {Ikoma}, M.: {Capture of solids by growing proto-gas giants:
  effects of gap formation and supply limited growth}.
\newblock \mnras \textbf{487}(4), 4510--4524 (2019).
\newblock \doi{10.1093/mnras/stz1629}

\bibitem{skrutskie2006}
{Skrutskie}, M.F., {Cutri}, R.M., {Stiening}, R., {Weinberg}, M.D.,
  {Schneider}, S., {Carpenter}, J.M., {Beichman}, C., {Capps}, R., {Chester},
  T., {Elias}, J., {Huchra}, J., {Liebert}, J., {Lonsdale}, C., {Monet}, D.G.,
  {Price}, S., {Seitzer}, P., {Jarrett}, T., {Kirkpatrick}, J.D., {Gizis},
  J.E., {Howard}, E., {Evans}, T., {Fowler}, J., {Fullmer}, L., {Hurt}, R.,
  {Light}, R., {Kopan}, E.L., {Marsh}, K.A., {McCallon}, H.L., {Tam}, R., {Van
  Dyk}, S., {Wheelock}, S.: {The Two Micron All Sky Survey (2MASS)}.
\newblock \aj \textbf{131}(2), 1163--1183 (2006).
\newblock \doi{10.1086/498708}

\bibitem{Skumanich1972}
{Skumanich}, A.: {Time Scales for CA II Emission Decay, Rotational Braking, and
  Lithium Depletion}.
\newblock \apj \textbf{171}, 565 (1972).
\newblock \doi{10.1086/151310}

\bibitem{Smith2014}
{Smith}, A.M.S., {Anderson}, D.R., {Armstrong}, D.J., {Barros}, S.C.C.,
  {Bonomo}, A.S., {Bouchy}, F., {Brown}, D.J.A., {Collier Cameron}, A.,
  {Delrez}, L., {Faedi}, F., {Gillon}, M., {G{\'o}mez Maqueo Chew}, Y.,
  {H{\'e}brard}, G., {Jehin}, E., {Lendl}, M., {Louden}, T.M., {Maxted},
  P.F.L., {Montagnier}, G., {Neveu-VanMalle}, M., {Osborn}, H.P., {Pepe}, F.,
  {Pollacco}, D., {Queloz}, D., {Rostron}, J.W., {Segransan}, D., {Smalley},
  B., {Triaud}, A.H.M.J., {Turner}, O.D., {Udry}, S., {Walker}, S.R., {West},
  R.G., {Wheatley}, P.J.: {WASP-104b and WASP-106b: two transiting hot Jupiters
  in 1.75-day and 9.3-day orbits}.
\newblock \aap \textbf{570}, A64 (2014).
\newblock \doi{10.1051/0004-6361/201424752}

\bibitem{sneden2014}
{Sneden}, C., {Lucatello}, S., {Ram}, S.R., {Brooke}, J.S.A., {Bernath}, P.:
  {Line Lists for the A 
  CN, 13C14N, and 12C15N, and Application to Astronomical Spectra}.
\newblock \apjs \textbf{214}, A24 (2014).
\newblock \doi{10.1088/0067-0049/214/2/26}

\bibitem{Sousa2018}
{Sousa}, S.G., {Adibekyan}, V., {Delgado-Mena}, E., {Santos}, N.C.,
  {Andreasen}, D.T., {Ferreira}, A.C.S., {Tsantaki}, M., {Barros}, S.C.C.,
  {Demangeon}, O., {Israelian}, G., {Faria}, J.P., {Figueira}, P., {Mortier},
  A., {Brand{\~a}o}, I., {Montalto}, M., {Rojas-Ayala}, B., {Santerne}, A.:
  {SWEET-Cat updated. New homogenous spectroscopic parameters}.
\newblock \aap \textbf{620}, A58 (2018).
\newblock \doi{10.1051/0004-6361/201833350}

\bibitem{Sousa2008}
{Sousa}, S.G., {Santos}, N.C., {Mayor}, M., {Udry}, S., {Casagrande}, L.,
  {Israelian}, G., {Pepe}, F., {Queloz}, D., {Monteiro}, M.J.P.F.G.:
  {Spectroscopic parameters for 451 stars in the HARPS GTO planet search
  program. Stellar [Fe/H] and the frequency of exo-Neptunes}.
\newblock \aap \textbf{487}(1), 373--381 (2008).
\newblock \doi{10.1051/0004-6361:200809698}

\bibitem{Spake2016}
{Spake}, J.J., {Brown}, D.J.A., {Doyle}, A.P., {H{\'e}brard}, G., {McCormac},
  J., {Armstrong}, D.J., {Pollacco}, D., {G{\'o}mez Maqueo Chew}, Y.,
  {Anderson}, D.R., {Barros}, S.C.C., {Bouchy}, F., {Boumis}, P., {Bruno}, G.,
  {Collier Cameron}, A., {Courcol}, B., {Davies}, G.R., {Faedi}, F., {Hellier},
  C., {Kirk}, J., {Lam}, K.W.F., {Liakos}, A., {Louden}, T., {Maxted}, P.F.L.,
  {Osborn}, H.P., {Palle}, E., {Prieto Arranz}, J., {Udry}, S., {Walker}, S.R.,
  {West}, R.G., {Wheatley}, P.J.: {WASP-135b: A Highly Irradiated, Inflated Hot
  Jupiter Orbiting a G5V Star}.
\newblock \pasp \textbf{128}(960), 024401 (2016).
\newblock \doi{10.1088/1538-3873/128/960/024401}

\bibitem{stetson}
{Stetson}, P.B., {Pancino}, E.: {DAOSPEC: An Automatic Code for Measuring
  Equivalent Widths in High-Resolution Stellar Spectra}.
\newblock \pasp \textbf{120}(874), 1332 (2008).
\newblock \doi{10.1086/596126}

\bibitem{suarezandres2016}
{Su{\'a}rez-Andr{\'e}s}, L., {Israelian}, G., {Gonz{\'a}lez Hern{\'a}ndez},
  J.I., {Adibekyan}, V.Z., {Delgado Mena}, E., {Santos}, N.C., {Sousa}, S.G.:
  {CNO behaviour in planet-harbouring stars. I. Nitrogen abundances in stars
  with planets}.
\newblock \aap \textbf{591}, A69 (2016).
\newblock \doi{10.1051/0004-6361/201628455}

\bibitem{suarezandres2017}
{Su{\'a}rez-Andr{\'e}s}, L., {Israelian}, G., {Gonz{\'a}lez Hern{\'a}ndez},
  J.I., {Adibekyan}, V.Z., {Delgado Mena}, E., {Santos}, N.C., {Sousa}, S.G.:
  {CNO behaviour in planet-harbouring stars. II. Carbon abundances in stars
  with and without planets using the CH band}.
\newblock \aap \textbf{599}, A96 (2017).
\newblock \doi{10.1051/0004-6361/201629434}

\bibitem{thiabaud+2014}
{Thiabaud}, A., {Marboeuf}, U., {Alibert}, Y., {Cabral}, N., {Leya}, I.,
  {Mezger}, K.: {From stellar nebula to planets: The refractory components}.
\newblock \aap \textbf{562}, A27 (2014).
\newblock \doi{10.1051/0004-6361/201322208}

\bibitem{Tinetti2018}
{Tinetti}, G., {Drossart}, P., {Eccleston}, P., {Hartogh}, P., {Heske}, A.,
  {Leconte}, J., {Micela}, G., {Ollivier}, M., {Pilbratt}, G., {Puig}, L.,
  et~al.: {A chemical survey of exoplanets with ARIEL}.
\newblock Experimental Astronomy \textbf{46}, 135--209 (2018).
\newblock \doi{10.1007/s10686-018-9598-x}

\bibitem{Tsantaki2018}
{Tsantaki}, M., {Andreasen}, D.T., {Teixeira}, G.D.C., {Sousa}, S.G., {Santos},
  N.C., {Delgado-Mena}, E., {Bruzual}, G.: {Atmospheric stellar parameters for
  large surveys using FASMA, a new spectral synthesis package}.
\newblock \mnras \textbf{473}(4), 5066--5097 (2018).
\newblock \doi{10.1093/mnras/stx2564}

\bibitem{Turrini2020}
{Turrini}, D., {Codella}, C., {Danielski}, C., {Fedele}, D., {Fonte}, S.,
  {Garudi}, e.a.: {Exploring the link between star and planet formation with
  Ariel}.
\newblock Experimental Astronomy, this issue  (2020)

\bibitem{Turrini2018}
{Turrini}, D., {Miguel}, Y., {Zingales}, T., {Piccialli}, A., {Helled}, R.,
  {Vazan}, A., {Oliva}, F., {Sindoni}, G., {Pani{\'c}}, O., {Leconte}, J.,
  {Min}, M., {Pirani}, S., {Selsis}, F., {Coud{\'e} du Foresto}, V., {Mura},
  A., {Wolkenberg}, P.: {The contribution of the ARIEL space mission to the
  study of planetary formation}.
\newblock Experimental Astronomy \textbf{46}(1), 45--65 (2018).
\newblock \doi{10.1007/s10686-017-9570-1}

\bibitem{turrini+2015}
{Turrini}, D., {Nelson}, R.P., {Barbieri}, M.: {The role of planetary formation
  and evolution in shaping the composition of exoplanetary atmospheres}.
\newblock Experimental Astronomy \textbf{40}(2-3), 501--522 (2015).
\newblock \doi{10.1007/s10686-014-9401-6}

\bibitem{Turriniinprep}
{Turrini}, D., {Schisano}, E., {Fonte}, S., {Molinari}, S., {Politi}, R.,
  {Fedele}, D., {Pani{\'c}}, O., {Kama}, M., {Changeat}, Q., {Tinetti}, G.:
  {Tracing the Formation History of Giant Planets in Protoplanetary Disks with
  Carbon, Oxygen, Nitrogen, and Sulfur}.
\newblock \apj \textbf{909}(1), 40 (2021).
\newblock \doi{10.3847/1538-4357/abd6e5}

\bibitem{ValentiFischer2005}
{Valenti}, J.A., {Fischer}, D.A.: {Spectroscopic Properties of Cool Stars
  (SPOCS). I. 1040 F, G, and K Dwarfs from Keck, Lick, and AAT Planet Search
  Programs}.
\newblock \apjs \textbf{159}(1), 141--166 (2005).
\newblock \doi{10.1086/430500}

\bibitem{vaughan1980}
{Vaughan}, A.H., {Preston}, G.W.: {A survey of chromospheric CA II H and K
  emission in field stars of the solar neighborhood.}
\newblock \pasp \textbf{92}, 385--391 (1980).
\newblock \doi{10.1086/130683}

\bibitem{Vergely2021}
{Vergely}, J.L.S., {Lallement}, R., {Babusiaux}, C., {et al.}: {}.
\newblock to be submitted to \aap

\bibitem{wenger2000}
{Wenger}, M., {Ochsenbein}, F., {Egret}, D., {Dubois}, P., {Bonnarel}, F.,
  {Borde}, S., {Genova}, F., {Jasniewicz}, G., {Lalo{\"e}}, S., {Lesteven}, S.,
  {Monier}, R.: {The SIMBAD astronomical database. The CDS reference database
  for astronomical objects}.
\newblock Astronomy and Astrophysics Supplement \textbf{143}, 9--22 (2000).
\newblock \doi{10.1051/aas:2000332}

\bibitem{Wilson1963}
{Wilson}, O.C.: {A Probable Correlation Between Chromospheric Activity and Age
  in Main-Sequence Stars.}
\newblock \apj \textbf{138}, 832 (1963).
\newblock \doi{10.1086/147689}

\bibitem{wilson1968}
{Wilson}, O.C.: {Flux Measurements at the Centers of Stellar H- and K-Lines}.
\newblock \apj \textbf{153}, 221 (1968).
\newblock \doi{10.1086/149652}

\bibitem{Wilson2018}
{Wilson}, R.F., {Teske}, J., {Majewski}, S.R., {Cunha}, K., {Smith}, V.,
  {Souto}, D., {Bender}, C., {Mahadevan}, S., {Troup}, N., {Allende Prieto},
  C., {Stassun}, K.G., {Skrutskie}, M.F., {Almeida}, A.,
  {Garc{\'\i}a-Hern{\'a}ndez}, D.A., {Zamora}, O., {Brinkmann}, J.: {Elemental
  Abundances of Kepler Objects of Interest in APOGEE. I. Two Distinct Orbital
  Period Regimes Inferred from Host Star Iron Abundances}.
\newblock \aj \textbf{155}(2), 68 (2018).
\newblock \doi{10.3847/1538-3881/aa9f27}

\bibitem{Zechmeister2009}
{Zechmeister}, M., {K{\"u}rster}, M.: {The generalised Lomb-Scargle
  periodogram. A new formalism for the floating-mean and Keplerian
  periodograms}.
\newblock \aap \textbf{496}(2), 577--584 (2009).
\newblock \doi{10.1051/0004-6361:200811296}

\bibitem{2017ApJ...844...27Z}
{Zellem}, R.T., {Swain}, M.R., {Roudier}, G., {Shkolnik}, E.L.,
  {Creech-Eakman}, M.J., {Ciardi}, D.R., {Line}, M.R., {Iyer}, A.R., {Bryden},
  G., {Llama}, J., {Fahy}, K.A.: {Forecasting the Impact of Stellar Activity on
  Transiting Exoplanet Spectra}.
\newblock \apj \textbf{844}(1), 27 (2017).
\newblock \doi{10.3847/1538-4357/aa79f5}

\end{thebibliography}

\end{document}